\documentclass[preprintnumbers,article,amsmath,amssymb,floatfix,10pt,prd,onecolumn, superscriptaddress,nofootinbib]{revtex4}
\usepackage[colorlinks=true, pdfstartview=FitV, linkcolor=blue, citecolor=red, urlcolor=magenta]{hyperref}
\usepackage{bbm}
\usepackage{amsfonts}
\usepackage{amsfonts,graphicx,amsmath,amsthm,amssymb,epsfig}
\usepackage{float}
\allowdisplaybreaks
\usepackage{mathrsfs}
\usepackage{latexsym}
\usepackage{epsfig}
\usepackage{epstopdf}
\usepackage{epstopdf}
\usepackage{graphicx}
\usepackage{amssymb}
\usepackage{amsmath}
\usepackage{dcolumn}
\usepackage{bm}
\usepackage{color}
\usepackage{comment}
\usepackage{xcolor}
\begin{document}
\title{\bf Non-singular anisotropic solutions for strange star model in $f(\mathcal{R},\mathcal{T},\mathcal{R}_{\zeta\gamma}\mathcal{T}^{\zeta\gamma})$ gravity theory}

\author{Yihu Feng}
\email{fengyihubzxy@163.com}\affiliation{Department of Electronics and Information Engineering, Bozhou University, Bozhou, 236800, Anhui, China}

\author{Tayyab Naseer}
\email{tayyab.naseer@math.uol.edu.pk; tayyab.naseer@khazar.org;
tayyabnaseer48@yahoo.com (Corresponding author)}
\affiliation{Department of Mathematics and Statistics, The University of Lahore,\\
1-KM Defence Road Lahore-54000, Pakistan}\affiliation{Research
Center of Astrophysics and Cosmology, Khazar University,\\ Baku,
AZ1096, 41 Mehseti Street, Azerbaijan}

\author{G. Mustafa}
\email{gmustafa3828@gmail.com}\affiliation{Department of Physics, Zhejiang Normal University, Jinhua 321004, People's Republic of China}

\author{S. K. Maurya}
\email{sunil@unizwa.edu.om (Corresponding
author)}\affiliation{Department of Mathematical and Physical
Sciences, College of Arts and Sciences, University of Nizwa, Nizwa,
Sultanate of Oman}

\begin{abstract}
This article focuses on different anisotropic models within the
framework of a specific modified
$f(\mathcal{R},\mathcal{T},\mathcal{R}_{\zeta\gamma}\mathcal{T}^{\zeta\gamma})$
gravity theory. The study adopts a static spherically symmetric
spacetime to determine the field equations for two different
modified models: (i)
$f(\mathcal{R},\mathcal{T},\mathcal{R}_{\zeta\gamma}\mathcal{T}^{\zeta\gamma})=\mathcal{R}+\eta\mathcal{R}_{\zeta\gamma}\mathcal{T}^{\zeta\gamma}$,
and (ii)
$f(\mathcal{R},\mathcal{T},\mathcal{R}_{\zeta\gamma}\mathcal{T}^{\zeta\gamma})=\mathcal{R}(1+\eta\mathcal{R}_{\zeta\gamma}\mathcal{T}^{\zeta\gamma})$,
where $\eta$ is a constant parameter. To address the additional
degrees of freedom in the field equations and obtain their
corresponding unique solution, the Durgapal-Fuloria spacetime
geometry and MIT bag model are utilized. Matching conditions are
applied to determine unknown constants within the chosen spacetime
geometry. We adopt a certain range of model parameters to analyze
the physical characteristics of the developed models in the interior
distribution of a particular compact star candidate 4U 1820-30.
Energy conditions and some other tests are also implemented to
ensure their viability and stability. Additionally, the disappearing
radial pressure constraint is employed to find the values of the
model parameter, aligning with the observed information of an array
of stars. The study concludes that both of our models are
well-behaved and satisfy all necessary conditions, and thus we
observe them suitable for the modeling of astrophysical objects.
\end{abstract}

\maketitle

\date{\today}

\section{Introduction}

The scientific community has long regarded general theory of
relativity (GR) as the dominant gravitational theory, successfully
tackling numerous challenges. However, it has limitations in fully
explaining the rapid cosmic expansion. Recent observations have
hinted at a repulsive force, known as `dark energy', which is
believed to drive galaxies apart and contribute to the cosmic
acceleration. In response to various cosmic mysteries, scientists
have been investigating multiple extensions to GR. The
$f(\mathcal{R})$ theory represents the pioneering extension of GR
achieved through modifying the action, where the Ricci scalar
$\mathcal{R}$ is supplanted with the generalized function \cite{1}.
Researchers have extensively utilized various approaches in this
context to investigate the feasibility of self-gravitating
structures \cite{2}-\cite{2f}. Moreover, $f(\mathcal{R})$ gravity
models have been instrumental in addressing diverse cosmological
issues, such as the late-time cosmic evolution, the inflationary
epoch in which our universe grown at an exponential rate, and more
\cite{3}-\cite{5a}, extending their applications beyond celestial
bodies.

In a research endeavor led by Bertolami and colleagues \cite{10},
the quest to unveil the enigmatic aspects of the cosmos prompted an
investigation into a novel concept concerning the coupling between
matter and geometry. This exploration involved unifying the
influence of geometry and the fluid Lagrangian in $f(\mathcal{R})$
framework. The innovative nature of this approach sparked the
interest of numerous astronomers, who subsequently shifted their
focus towards comprehending the accelerated expansion. Building upon
these initial insights, recent advancements have given rise to
modified theories, garnering significant attention in the scientific
field. Harko et al. \cite{20} proposed the first-ever theory based
on this idea, termed $f(\mathcal{R},\mathcal{T})$ gravity, where
$\mathcal{T}$ signifies the trace of the energy-momentum tensor
(EMT) whose incorporation leads to non-conservation effects,
prompting thorough analysis of the self-gravitating bodies, resulted
in numerous significant discoveries \cite{21}-\cite{21h}. In a
related development, Haghani et al. \cite{22} proposed another
theory whose functional dependent on
$\mathcal{R}_{\zeta\gamma}\mathcal{T}^{\zeta\gamma}$ along with
previous two entities. They explored an epoch of cosmic expansion
characterized by rapid growth and analyzed multiple models to
evaluate their validity in this context. They further enhanced their
analysis by employing the Lagrange multiplier method and computed
conserved EMT even in this theoretical framework.

In this theory, the presence of a non-conserved EMT introduces an
extra force that disrupts the motion of test particles along
geodesic paths. Researchers \cite{22a} investigated two models,
namely
$\mathcal{R}+\eta\mathcal{R}_{\zeta\gamma}\mathcal{T}^{\zeta\gamma}$
and
$\mathcal{R}(1+\eta\mathcal{R}_{\zeta\gamma}\mathcal{T}^{\zeta\gamma})$,
along with different fluid Lagrangians. Their study delved into
various properties related to black holes and the corresponding laws
to discuss their thermodynamics. Odintsov and S\'{a}ez-G\'{o}mez
\cite{23} discussed the effects of varying fluid configurations,
demonstrating that such alternations can result in a pure de Sitter
model within this modified framework. Additionally, Ayuso et al.
\cite{24} analyzed the field equations in this extended theory
through the incorporation of some fields (either scalar or vector),
revealing the presence of non-linear terms arising from such
couplings between fluid and geometry. The work presented in
\cite{25} extensively delves into the stability checks for various
models through the incorporation of perturbation functions. A
crucial observation was made regarding the impact of the matter
Lagrangian density, particularly in relation to the radial and
tangential components of pressure \cite{25a}. Additionally, the
decomposition of the curvature tensor resulted in scalar functions
relevant to the fluids possessing charge/uncharge properties,
bearing importance in studying celestial systems
\cite{26}-\cite{26e}. Through diverse techniques, solutions to
modified field equations were extracted, leading to the modeling of
several anisotropic systems that proved consistent and physically
valid results \cite{27a,27aaa}.

The reigning factors which define the self-gravitating interiors,
does not matter which properties they exhibit, are commonly the
energy density and pressure (same or different in different
directions depending on the nature of the fluid). These elements are
interconnected in specific ways, one of them being the MIT equation
of state (EoS) \cite{30}. It is worth noting that this model
effectively grabs the features of the objects composed of quark-like
elements such as RXJ 185635-3754, 4U 1820-30, PSR 0943+10, 4U
1728-34, SAX J 1808.4-3658, Her X-1, among others. Conversely, such
characterization is not achievable using a neutron star EoS
\cite{33a}. Researchers have utilized the same model to explore the
internal configuration of strange stars \cite{33b}-\cite{34aa}.
Demorest with his colleagues \cite{34b} conducted a comprehensive
exploration of the massive system PSR J1614-2230, determining that
only the MIT model can account for such extraordinarily dense
bodies. Rahaman et al. \cite{35} described in more depth some
particular stellar interiors using the same model. Similarly,
various researchers expanded upon this research by investigating
various modified gravity scenarios, leading to the identification of
physically stable models \cite{38,38c}.

This paper is focused on assessing the viability of the
Durgapal-Fuloria models coupled with anisotropic pressure in the
$f(\mathcal{R},\mathcal{T},\mathcal{R}_{\zeta\gamma}\mathcal{T}^{\zeta\gamma})$
framework. The following lines explain the organization of the
current paper. In the next section, we introduce some fundamental
definitions and calculate the field equations for a couple of
modified models. Furthermore, we utilize the MIT bag model to
describe the inner structure of the quark-like structures. Section III employs matching criteria to determine a doublet $(d_1, d_2)$
appearing due to the consideration of Durgapal-Fuloria metric.
Moving to section IV, we conduct a graphical analysis of different
properties of the obtained fluid determinants by fixing the model
parameter. After this, we identify the model parameter that align
with the calculated data of various stars in section V. Finally, the
last section presents a brief summary regarding our results.

\section{Modified Theory and Field Equations}

The modified Einstein-Hilbert action with $\kappa=8\pi$ takes the
form \cite{23}
\begin{equation}\label{g1}
S=\int\sqrt{-g}
\left\{\frac{f(\mathcal{R},\mathcal{T},\mathcal{R}_{\zeta\gamma}\mathcal{T}^{\zeta\gamma})}{16\pi}
+\L_{m}\right\}d^{4}x,
\end{equation}
where $\L_{m}$ being the matter's Lagrangian density. We apply the
least-action principle on the above equation, resulting in
\begin{equation}\label{g2}
\mathcal{G}_{\zeta\gamma}=8\pi\mathcal{T}_{\zeta\gamma}^{(\mathrm{ef})}=\frac{8\pi\mathcal{T}_{\zeta\gamma}}
{f_{\mathcal{R}}-\L_{m}f_{\mathcal{Q}}}+\mathcal{T}_{\zeta\gamma}^{(\mathrm{cr})},
\end{equation}
which relates the effective matter distribution and geometry
expressed by the EMT $\mathcal{T}_{\zeta\gamma}^{(\mathrm{eff})}$
and the Einstein tensor $\mathcal{G}_{\zeta\gamma}$, respectively. Also, $\mathcal{Q}=\mathcal{R}_{\zeta\gamma}\mathcal{T}^{\zeta\gamma}$.
The insertion of generalized functional $f$ in Eq.\eqref{g1}
produces some additional terms along with the fluid distribution of
GR, and we represent it by
$\mathcal{T}_{\zeta\gamma}^{(\mathrm{cr})}$. Its expression is
\begin{eqnarray}\nonumber
\mathcal{T}_{\zeta\gamma}^{(\mathrm{cor})}&=&-\frac{1}{\bigg(\L_{m}f_{\mathcal{Q}}-f_{\mathcal{R}}\bigg)}
\left[\left(f_{\mathcal{T}}+\frac{1}{2}\mathcal{R}f_{\mathcal{Q}}\right)\mathcal{T}_{\zeta\gamma}
-\left\{\L_{m}f_{\mathcal{T}}-\frac{\mathcal{R}}{2}\bigg(\frac{f}{\mathcal{R}}-f_{\mathcal{R}}\bigg)\right.\right.\\\nonumber
&+&\left.\frac{1}{2}\nabla_{\varrho}\nabla_{\sigma}(f_{\mathcal{Q}}\mathcal{T}^{\varrho\sigma})\right\}g_{\zeta\gamma}
-\frac{1}{2}\Box(f_{\mathcal{Q}}\mathcal{T}_{\zeta\gamma})-(g_{\zeta\gamma}\Box-
\nabla_{\zeta}\nabla_{\gamma})f_{\mathcal{R}}\\\label{g4}
&-&2f_{\mathcal{Q}}\mathcal{R}_{\varrho(\zeta}\mathcal{T}_{\gamma)}^{\varrho}
+\nabla_{\varrho}\nabla_{(\zeta}[\mathcal{T}_{\gamma)}^{\varrho}
f_{\mathcal{Q}}]+2(f_{\mathcal{Q}}\mathcal{R}^{\varrho\sigma}+\left.f_{\mathcal{T}}g^{\varrho\sigma})\frac{\partial^2\L_{m}}
{\partial g^{\zeta\gamma}\partial g^{\varrho\sigma}}\right],
\end{eqnarray}
where the partial derivatives $f_{\mathcal{R}},~f_{\mathcal{T}}$ and
$f_{\mathcal{Q}}$ are involved that differentiate the functional
with respect to its arguments. Two other mathematical symbols are
used, known as the covariant derivative ($\nabla_\varrho$) and and
D'Alambert operator \big($\Box\equiv
\frac{1}{\sqrt{-g}}\partial_\zeta(\sqrt{-g}g^{\zeta\gamma}\partial_{\gamma})$\big).
It has been observed in the literature that the fluid Lagrangian
takes the value $\L_{m}=-\mu$ ($\mu$ being the energy density) in
this theory to produce well-behaving results. On the other hand, we
also observe that the equivalence principle is not satisfied in the
current scenario (i.e., $\nabla_\zeta \mathcal{T}^{\zeta\gamma}\neq
0$), so an extra force must be required to make the system in stable
equilibrium. Such kind of force alters the geodesic motion of the
particles in spacetime regions. Mathematical, this force becomes
\begin{align}\nonumber
\nabla^\zeta
\mathcal{T}_{\zeta\gamma}&=\frac{2}{2f_\mathcal{T}+\mathcal{R}f_\mathcal{Q}+16\pi}\bigg[\nabla_\zeta
\big(f_\mathcal{Q}\mathcal{R}^{\varrho\zeta}\mathcal{T}_{\varrho\gamma}\big)-\mathcal{G}_{\zeta\gamma}\nabla^\zeta
\big(f_\mathcal{Q}\L_m\big)+\nabla_\gamma
\big(\L_mf_\mathcal{T}\big)\\\label{g4a}
&-\frac{1}{2}\nabla_\gamma\mathcal{T}^{\varrho\sigma}\big(f_\mathcal{T}g_{\varrho\sigma}+f_\mathcal{Q}\mathcal{R}_{\varrho\sigma}\big)
-\frac{1}{2}\big\{\nabla^{\zeta}(\mathcal{R}f_{\mathcal{Q}})+2\nabla^{\zeta}f_{\mathcal{T}}\big\}\mathcal{T}_{\zeta\gamma}\bigg].
\end{align}

The EMT plays a crucial role in examining the internal properties of
self-gravitating systems, which is vital in the realm of
astrophysics. In this field, a diverse array of cosmic entities is
thought to display the anisotropy in pressure. This makes the EMT an
indispensable tool for studying the complex processes of stellar
evolution. The mathematical expression of the anisotropic EMT is as
follows
\begin{equation}\label{g5}
\mathcal{T}_{\zeta\gamma}=(\mu+P_\bot) \mathrm{U}_{\zeta}
\mathrm{U}_{\gamma}+P_\bot
g_{\zeta\gamma}+\left(P_r-P_\bot\right)\mathrm{V}_\zeta\mathrm{V}_\gamma,
\end{equation}
where $P_\bot$ is the tangential pressure, $\mathrm{V}_{\zeta}$
indicates the four-vector, $P_r$ being the radial pressure and
$\mathrm{U}_\zeta$ represents the four-velocity. The trace of
Eq.\eqref{g2} along with \eqref{g4} is
\begin{align}\nonumber
&3\nabla^{\varrho}\nabla_{\varrho}
f_\mathcal{R}-\mathcal{T}(8\pi+f_\mathcal{T})-\mathcal{R}\left(\frac{\mathcal{T}}{2}f_\mathcal{Q}-f_\mathcal{R}\right)+\frac{1}{2}
\nabla^{\varrho}\nabla_{\varrho}(f_\mathcal{Q}\mathcal{T})\\\nonumber
&+\nabla_\zeta\nabla_\varrho(f_\mathcal{Q}\mathcal{T}^{\zeta\varrho})
-2f+(\mathcal{R}f_\mathcal{Q}+4f_\mathcal{T})\L_m+2\mathcal{R}_{\zeta\varrho}\mathcal{T}^{\zeta\varrho}f_\mathcal{Q}\\\nonumber
&-2g^{\gamma\sigma} \frac{\partial^2\L_m}{\partial
g^{\gamma\sigma}\partial
g^{\zeta\varrho}}\left(f_\mathcal{T}g^{\zeta\varrho}+f_\mathcal{Q}R^{\zeta\varrho}\right)=0.
\end{align}

The metric or spacetime is a fundamental concept that enables us to
investigate the gravitational field and its effects on the curvature
of spacetime within heavily celestial objects. In this context, we
focus on the interior spherical spacetime that is described by the
line element given as
\begin{equation}\label{g6}
ds^2=e^{{A_1}} dr^2+r^2d\theta^2+r^2\sin^2\theta d\phi^2-e^{A_0}
dt^2,
\end{equation}
where $A_0=A_0(r)$ and ${{A_1}}={{A_1}}(r)$. The quantities used in
Eq.\eqref{g5} are now become
\begin{equation}\label{g7}
\mathrm{V}^\zeta=\delta^\zeta_1 e^{-\frac{{{A_1}}}{2}}, \quad
\mathrm{U}^\zeta=\delta^\zeta_0 e^{-\frac{A_0}{2}},
\end{equation}
satisfying $\mathrm{V}^\zeta \mathrm{U}_{\zeta}=0$ and
$\mathrm{U}^\zeta \mathrm{U}_{\zeta}=-1$.

Our cosmos is currently experiencing a period of rapid expansion and
is filled with a multitude of stars that exist within a non-linear
context. Despite this non-linearity, conducting analyzes using
linear methods offer a better understanding of the formation and
behavior of these heavily structures. To delve into this, we explore
following two models as \cite{22}.
$$\textbf{Model 1:}\quad f(\mathcal{R},\mathcal{T},\mathcal{R}_{\zeta\gamma}\mathcal{T}^{\zeta\gamma})=\mathcal{R}+\eta
\mathcal{R}_{\zeta\gamma}\mathcal{T}^{\zeta\gamma},$$
$$\textbf{Model 2:}\quad f(\mathcal{R},\mathcal{T},\mathcal{R}_{\zeta\gamma}\mathcal{T}^{\zeta\gamma})=\mathcal{R}(1+\eta
\mathcal{R}_{\zeta\gamma}\mathcal{T}^{\zeta\gamma}),$$ where $\eta$
being the arbitrary parameter. It is important to highlight that
different parametric values, all falling within the estimated range,
guarantee the validity of stars. Haghani et al. \cite{22} conducted
a thorough examination of these models, delving into the evolution
of the scale factor and deceleration parameter. Similarly, Sharif
and Zubair \cite{22a} focused on isotropic configurations within the
same context, deriving acceptable values for the respective
parameters. The expression for the last term in above models can be expressed as
\begin{eqnarray}\nonumber
\mathcal{R}_{\zeta\gamma}\mathcal{T}^{\zeta\gamma}&=&e^{-{A_1}}\bigg[\frac{\mu}{4}\left(A_0'^2-A_0'{A_1}'+2A_0''+\frac{4A_0'}{r}\right)
+P_\bot\left(\frac{{A_1}'}{r}-\frac{A_0'}{r}+\frac{2e^{A_1}}{r^2}-\frac{2}{r^2}\right)\\\nonumber
&-&\frac{P_r}{4}\left(A_0'^2-A_0'{A_1}'+2A_0''+\frac{4{A_1}'}{r}\right)\bigg],
\end{eqnarray}
where $'=\frac{\partial}{\partial r}$.

The field equations \eqref{g2} characterizing the anisotropic
interior \eqref{g5} are given for $\textbf{Model 1}$ as

\begin{align}\nonumber
8\pi\mu&=e^{-{A_1}}\bigg[\frac{{A_1}'}{r}+\frac{e^{A_1}}{r^2}-\frac{1}{r^2}+\eta\bigg\{\mu\bigg(\frac{3A_0'{A_1}'}{8}-\frac{A_0'^2}{8}
+\frac{{A_1}'}{r}+\frac{e^{A_1}}{r^2}-\frac{3A_0''}{4}-\frac{3A_0'}{2r}\\\nonumber
&-\frac{1}{r^2}\bigg)-\mu'\bigg(\frac{{A_1}'}{4}-\frac{1}{r}-A_0'\bigg)+\frac{\mu''}{2}+P_r\bigg(\frac{A_0'{A_1}'}{8}
-\frac{A_0'^2}{8}-\frac{A_0''}{4}+\frac{{A_1}'}{2r}+\frac{{A_1}''}{2}\\\label{g8}
&-\frac{3{A_1}'^2}{4}\bigg)+\frac{5{A_1}'P'_r}{4}-\frac{P''_r}{2}+P_\bot\bigg(\frac{{A_1}'}{2r}-\frac{A_0'}{2r}+\frac{3e^{A_1}}{r^2}
-\frac{1}{r^2}\bigg)-\frac{P'_\bot}{r}\bigg\}\bigg],\\\nonumber 8\pi
P_r&=e^{-{A_1}}\bigg[\frac{A_0'}{r}-\frac{e^{A_1}}{r^2}+\frac{1}{r^2}+\eta\bigg\{\mu\bigg(\frac{A_0'{A_1}'}{8}+\frac{A_0'^2}{8}
-\frac{A_0''}{4}-\frac{A_0'}{2r}\bigg)-\frac{A_0'\mu'}{4}-P_r\\\nonumber
&\times\bigg(\frac{5A_0'^2}{8}-\frac{7A_0'{A_1}'}{8}+\frac{5A_0''}{4}-\frac{7{A_1}'}{2r}+\frac{A_0'}{r}-{A_1}'^2
-\frac{e^{A_1}}{r^2}+\frac{1}{r^2}\bigg)+P'_r\bigg(\frac{A_0'}{4}+\frac{1}{r}\bigg)\\\label{g8a}
&-P_\bot\bigg(\frac{{A_1}'}{2r}-\frac{A_0'}{2r}+\frac{3e^{A_1}}{r^2}-\frac{1}{r^2}\bigg)+\frac{P'_\bot}{r}\bigg\}\bigg],\\\nonumber
8\pi
P_\bot&=e^{-{A_1}}\bigg[\frac{A_0'^2}{4}-\frac{A_0'{A_1}'}{4}+\frac{A_0''}{2}-\frac{{A_1}'}{2r}+\frac{A_0'}{2r}
+\eta\bigg\{\mu\bigg(\frac{A_0'^2}{8}+\frac{A_0'{A_1}'}{8}-\frac{A_0''}{4}-\frac{A_0'}{2r}\bigg)\\\nonumber
&-\frac{A_0'\mu'}{4}+P_r\bigg(\frac{A_0'^2}{8}-\frac{A_0'{A_1}'}{8}+\frac{A_0''}{4}-\frac{{A_1}'}{2r}-\frac{{A_1}''}{2}
+\frac{3{A_1}'^2}{4}\bigg)-\frac{5{A_1}'P'_r}{4}+\frac{P''_r}{2}\\\label{g8b}
&-P_\bot\bigg(\frac{A_0'^2}{4}-\frac{A_0'{A_1}'}{4}+\frac{A_0''}{2}-\frac{{A_1}'}{r}+\frac{A_0'}{r}\bigg)
-P'_\bot\bigg(\frac{{A_1}'}{4}-\frac{A_0'}{4}-\frac{3}{r}\bigg)+\frac{P''_\bot}{2}\bigg\}\bigg].
\end{align}

Similarly, expressions for matter variables corresponding to $\textbf{Model 2}$ as
\begin{align}\nonumber
8\pi\mu&=e^{-{A_1}}\bigg[\frac{{A_1}'}{r}+\frac{e^{A_1}}{r^2}-\frac{1}{r^2}+\eta\bigg\{\mu\bigg(\bigg(\frac{{A_1}'}{r}+\frac{e^{A_1}}{r^2}
-\frac{1}{r^2}\bigg)\alpha_1-\mathcal{R}\bigg(\frac{1}{r^2}-\frac{{A_1}'}{r}-\frac{e^{A_1}}{r^2}\\\nonumber
&+\frac{\mathcal{R}e^{A_1}}{2}-\frac{3A_0'^2}{8}-\frac{3A_0'}{2r}+\frac{5A_0'{A_1}'}{8}-\frac{3A_0''}{4}\bigg)
+\mathcal{R}'\bigg(\frac{{A_1}'}{2}-\frac{1}{r}\bigg)-\frac{\mathcal{R}''}{2}-\alpha_4\bigg(\frac{2}{r}\\\nonumber
&-\frac{{A_1}'}{2}\bigg)-\alpha_7\bigg)+\mu'\bigg(\alpha_1\bigg(\frac{{A_1}'}{2}-\frac{2}{r}\bigg)-\mathcal{R}\bigg(\frac{1}{r}-\frac{{A_1}'}{4}\bigg)
-\mathcal{R}'-2\alpha_4\bigg)-\mu''\bigg(\alpha_1\\\nonumber
&+\frac{\mathcal{R}}{2}\bigg)-P_r\bigg(\bigg(\frac{1}{r^2}-\frac{{A_1}'}{r}-\frac{e^{A_1}}{r^2}\bigg)\alpha_2+\mathcal{R}\bigg(\frac{A_0'^2}{8}
-\frac{1}{r^2}-\frac{A_0'{A_1}'}{8}+\frac{{A_1}'}{2r}+\frac{A_0''}{4}\bigg)\\\nonumber
&-\mathcal{R}'\bigg(\frac{2}{r}-\frac{{A_1}'}{2}\bigg)-\frac{\mathcal{R}''}{2}+\alpha_5\bigg(\frac{2}{r}-\frac{{A_1}'}{2}\bigg)+\alpha_8\bigg)
-P'_r\bigg(\alpha_2\bigg(\frac{2}{r}-\frac{{A_1}'}{2}\bigg)+\mathcal{R}\\\nonumber
&\times\bigg(\frac{{A_1}'}{4}-\frac{2}{r}\bigg)-\mathcal{R}'+2\alpha_5\bigg)-P''_r\bigg(\alpha_2-\frac{\mathcal{R}}{2}\bigg)
-P_\bot\bigg(\bigg(\frac{1}{r^2}-\frac{{A_1}'}{r}-\frac{e^{A_1}}{r^2}\bigg)\alpha_3\\\nonumber
&+\mathcal{R}\bigg(\frac{A_0'}{2r}+\frac{1}{r^2}-\frac{{A_1}'}{2r}\bigg)+\frac{\mathcal{R}'}{r}+\alpha_6\bigg(\frac{2}{r}-\frac{{A_1}'}{2}\bigg)
+\alpha_9\bigg)-P'_\bot\bigg(\alpha_3\bigg(\frac{2}{r}-\frac{{A_1}'}{2}\bigg)\\\label{g8c}
&+\frac{\mathcal{R}}{2}+2\alpha_6\bigg)-P''_\bot\alpha_3\bigg\}\bigg], \\\nonumber
8\pi
P_r&=e^{-{A_1}}\bigg[\frac{A_0'}{r}-\frac{e^{A_1}}{r^2}+\frac{1}{r^2}+\eta\bigg\{\mu\bigg(\bigg(\frac{A_0'}{r}-\frac{e^{A_1}}{r^2}
+\frac{1}{r^2}\bigg)\alpha_1-\mathcal{R}\bigg(\frac{e^{A_1}}{r^2}-\frac{1}{r^2}-\frac{A_0'}{r}\\\nonumber
&+\frac{A_0'^2}{8}+\frac{A_0'}{2r}-\frac{A_0'{A_1}'}{8}+\frac{A_0''}{4}\bigg)
+\frac{\mathcal{R}'A_0'}{4}+\alpha_4\bigg(\frac{2}{r}+\frac{A_0'}{2}\bigg)\bigg)+\mu'\bigg(\alpha_1\bigg(\frac{2}{r}
+\frac{A_0'}{2}\bigg)\\\nonumber
&+\frac{\mathcal{R}A_0'}{4}\bigg)+P_r\bigg(\bigg(\frac{1}{r^2}+\frac{A_0'}{r}-\frac{e^{A_1}}{r^2}\bigg)\alpha_2
-\mathcal{R}\bigg(\frac{\mathcal{R}e^{A_1}}{2}-\frac{3A_0'^2}{8}+\frac{1}{r^2}+\frac{A_0'}{r}+\frac{3{A_1}'}{2r}\\\nonumber
&+\frac{3A_0'{A_1}'}{8}-\frac{3A_0''}{4}\bigg)-\mathcal{R}'\bigg(\frac{1}{r}+\frac{A_0'}{4}\bigg)
+\alpha_5\bigg(\frac{2}{r}+\frac{A_0'}{2}\bigg)\bigg)+P'_r\bigg(\alpha_2\bigg(\frac{2}{r}+\frac{A_0'}{2}\bigg)\\\nonumber
&-\mathcal{R}\bigg(\frac{A_0'}{4}+\frac{1}{r}\bigg)\bigg)-P_\bot\bigg(\bigg(\frac{A_0'}{r}+\frac{e^{A_1}}{r^2}-\frac{1}{r^2}\bigg)\alpha_3
+\mathcal{R}\bigg(\frac{{A_1}'}{2r}-\frac{1}{r^2}-\frac{A_0'}{2r}\bigg)+\frac{\mathcal{R}'}{r}\\\label{g8d}
&-\alpha_6\bigg(\frac{2}{r}+\frac{A_0'}{2}\bigg)\bigg)+P'_\bot\bigg(\alpha_3\bigg(\frac{2}{r}
+\frac{A_0'}{2}\bigg)-\frac{\mathcal{R}}{2}\bigg)\bigg\}\bigg],\\\nonumber
8\pi
P_\bot&=e^{-{A_1}}\bigg[\frac{A_0''}{2}-\frac{A_0'{A_1}'}{4}+\frac{A_0'^2}{4}+\frac{A_0'}{2r}-\frac{{A_1}}{2r}
+\eta\bigg\{\mu\bigg(\alpha_1\bigg(\frac{A_0''}{2}-\frac{A_0'{A_1}'}{4}+\frac{A_0'^2}{4}\\\nonumber
&+\frac{A_0'}{2r}-\frac{{A_1}}{2r}\bigg)-\mathcal{R}\bigg(\frac{A_0'{A_1}'}{8}-\frac{A_0''}{4}-\frac{A_0'^2}{8}+\frac{{A_1}'}{2r}\bigg)
-\bigg(\frac{{A_1}'}{2}-\frac{1}{r}-\frac{A_0'}{2}\bigg)\alpha_4+\alpha_7\\\nonumber
&-\frac{\mathcal{R}'A_0'}{4}\bigg)-\mu'\bigg(\alpha_1\bigg(\frac{{A_1}'}{2}-\frac{1}{r}-\frac{A_0'}{2}\bigg)+\frac{\mathcal{R}A_0'}{4}-2\alpha_4\bigg)
+\mu''\alpha_1-P_r\bigg(\bigg(\frac{{A_1}'}{2r}\\\nonumber
&-\frac{A_0''}{2}-\frac{A_0'^2}{4}-\frac{A_0'}{2r}+\frac{A_0'{A_1}'}{4}\bigg)\alpha_2+\mathcal{R}\bigg(\frac{A_0'^2}{8}+\frac{A_0'}{2r}-\frac{A_0'{A_1}'}{8}
+\frac{A_0''}{4}\bigg)+\mathcal{R}'\\\nonumber
&\times\bigg(\frac{A_0'}{2}+\frac{1}{r}-\frac{{A_1}'}{4}\bigg)+\frac{\mathcal{R}''}{2}+\alpha_5\bigg(\frac{{A_1}'}{2}-\frac{1}{r}-\frac{A_0'}{2}\bigg)
-\alpha_8\bigg)-P'_r\bigg(\alpha_2\bigg(\frac{{A_1}'}{2}\\\nonumber
&-\frac{1}{r}-\frac{A_0'}{2}\bigg)+\mathcal{R}\bigg(\frac{A_0'}{2}+\frac{1}{r}-\frac{{A_1}'}{4}\bigg)+\mathcal{R}'-2\alpha_5\bigg)+P''_r\bigg(\alpha_2-\frac{\mathcal{R}}{2}\bigg)
+P_\bot\bigg(\alpha_3\\\nonumber
&\times\bigg(\frac{A_0''}{2}-\frac{A_0'{A_1}'}{4}+\frac{A_0'^2}{4}+\frac{A_0'}{2r}-\frac{{A_1}}{2r}\bigg)-\mathcal{R}\bigg(\frac{\mathcal{R}e^{A_1}}{2}-\frac{2}{r^2}+\frac{{A_1}'}{r}-\frac{A_0'}{r}
+\frac{2e^{A_1}}{r^2}\bigg)\\\nonumber
&-\mathcal{R}'\bigg(\frac{A_0'}{4}-\frac{{A_1}'}{4}\bigg)-\frac{\mathcal{R}''}{2}-\alpha_6\bigg(\frac{{A_1}'}{2}-\frac{1}{r}-\frac{A_0'}{2}\bigg)+\alpha_9\bigg)-P'_\bot\bigg(\alpha_3\bigg(\frac{{A_1}'}{2}-\frac{1}{r}\\\label{g8e}
&-\frac{A_0'}{2}\bigg)+\mathcal{R}\bigg(\frac{A_0'}{4}-\frac{{A_1}'}{4}+\frac{2}{r}\bigg)+\mathcal{R}'-2\alpha_6\bigg)+P''_\bot\bigg(\alpha_3-\frac{\mathcal{R}}{2}\bigg)\bigg\}\bigg],
\end{align}
where $\alpha_j^{'s}$, ($j=1$ to $9$) are presented in Appendix
$\mathbf{A}$. A formula for calculating the spherical mass
distribution has been calculated \cite{41b}, represented as
\begin{equation}\nonumber
m(r)=\frac{r}{2}\big(1-g^{\zeta\gamma}r_{,\zeta}r_{,\gamma}\big),
\end{equation}
simplifies to
\begin{equation}\label{g12a}
m(r)=\frac{r}{2}\big(1-e^{-{A_1}}\big).
\end{equation}

The two sets of differential equations \eqref{g8}-\eqref{g8b} and
\eqref{g8c}-\eqref{g8e} involve matter determinants and their
derivatives, making it a challenging task to obtain their solutions.
Therefore, it becomes imperative to impose some constraints in order
to derive the required solutions. In this context, we consider the
MIT bag model to quark's interior \cite{30}. The corresponding EoS
is defined as
\begin{equation}\label{g14a}
P_r=\frac{1}{3}\left(\mu-4B_c\right).
\end{equation}
Researchers have conducted investigations into various quark
interiors using the aforementioned equation and calculated values
for $B_c$ that are consistent with observed data \cite{41f,41h}.

After combining Eqs.\eqref{g8} and \eqref{g8a} with \eqref{g14a},
the explicit form of matter variables $\mu$, $P_r$ and $P_t$ for $\textbf{Model 1}$ can be expressed by
\begin{align}\nonumber
\mu&=\bigg[\eta\bigg(\frac{9A_0''}{8}-\frac{e^{{A_1}}}{r^2}+\frac{1}{r^2}-\frac{{A_1}''}{8}-\frac{5A_0'{A_1}'}{8}-\frac{{A_1}'^2}{16}
-\frac{7{A_1}'}{2r}+\frac{3A_0'^2}{16}+\frac{7A_0'}{4r}\bigg)\\\nonumber
&+8\pi
e^{{A_1}}\bigg]^{-1}\bigg[\frac{3}{4}\bigg(\frac{{A_1}'}{r}+\frac{A_0'}{r}\bigg)+
B_c\bigg\{8\pi
e^{A_1}-\eta\bigg(\frac{4{A_1}'}{r}-\frac{3A_0'^2}{4}-\frac{3A_0''}{2}+\frac{{A_1}''}{2}\\\label{g14b}
&+\frac{{A_1}'^2}{4}+A_0'{A_1}'-\frac{A_0'}{r}+\frac{e^{A_1}}{r^2}-\frac{1}{r^2}\bigg)\bigg\}\bigg],\\\nonumber
P_r&=\bigg[\eta\bigg(\frac{9A_0''}{8}-\frac{e^{{A_1}}}{r^2}+\frac{1}{r^2}-\frac{{A_1}''}{8}-\frac{5A_0'{A_1}'}{8}-\frac{{A_1}'^2}{16}
-\frac{7{A_1}'}{2r}+\frac{3A_0'^2}{16}+\frac{7A_0'}{4r}\bigg)\\\nonumber
&+8\pi
e^{{A_1}}\bigg]^{-1}\bigg[\frac{1}{4}\bigg(\frac{{A_1}'}{r}+\frac{A_0'}{r}\bigg)-
B_c\bigg\{8\pi e^{A_1}-\eta\bigg(\frac{A_0'{A_1}'}{2}
+\frac{{A_1}'}{r}-\frac{2A_0'}{r}\\\label{g14c}
&+\frac{e^{A_1}}{r^2}-A_0''-\frac{1}{r^2}\bigg)\bigg\}\bigg].
\end{align}

On the other hand, Eqs.\eqref{g8c} and \eqref{g8d} along with EOS ({g14a}) give the expressions for $\textbf{Model 2}$
\begin{align}\nonumber
\mu&=\bigg[\eta\bigg\{\frac{3}{4}\bigg(\frac{{A_1}'}{r}+\frac{A_0'}{r}\bigg)\bigg(\alpha_1+\frac{\alpha_2}{3}\bigg)
+\frac{3}{8}\big({A_1}'+A_0'\big)\bigg(\alpha_4+\frac{\alpha_5}{3}\bigg)
+\mathcal{R}\bigg(\frac{A_0''}{2}+\frac{{A_1}'}{4r}\\\nonumber
&-\frac{7A_0'{A_1}'}{16}+\frac{A_0'^2}{16}+\frac{5A_0'}{4r}-\frac{\mathcal{R}e^{A_1}}{2}\bigg)+\mathcal{R}'\bigg(\frac{A_0'}{8}-\frac{1}{2r}
+\frac{{A_1}'}{8}\bigg)-\frac{\mathcal{R''}}{4}-\frac{\alpha_8}{4}\bigg\}\\\nonumber
&-8\pi
e^{{A_1}}\bigg]^{-1}\bigg[-\frac{3}{4}\bigg(\frac{{A_1}'}{r}+\frac{A_0'}{r}\bigg)-8\pi
B_ce^{A_1}-\eta B_c\bigg\{\bigg(\frac{{A_1}'}{r}
+\frac{A_0'}{r}\bigg)\alpha_2-\mathcal{R}\bigg(\frac{A_0'^2}{2}\\\nonumber
&+\frac{\mathcal{R}e^{A_1}}{2}+\frac{A_0'}{r}+\frac{A_0'{A_1}'}{4}+\frac{2{A_1}'}{r}-\frac{A_0''}{2}\bigg)-\mathcal{R}'\bigg(\frac{A_0'}{4}-\frac{1}{r}
+\frac{{A_1}'}{4}\bigg)+\frac{\mathcal{R''}}{2}-\alpha_8\\\label{g14e}
&+\alpha_5\bigg(\frac{{A_1}'}{2}+\frac{A_0'}{2}\bigg)\bigg\}\bigg],\\\nonumber
P_r&=\bigg[\eta\bigg\{\frac{3}{4}\bigg(\frac{{A_1}'}{r}+\frac{A_0'}{r}\bigg)\bigg(\alpha_1+\frac{\alpha_2}{3}\bigg)+\frac{3}{8}\big({A_1}'
+A_0'\big)\bigg(\alpha_4+\frac{\alpha_5}{3}\bigg)+\mathcal{R}\bigg(\frac{A_0''}{2}+\frac{{A_1}'}{4r}\\\nonumber
&-\frac{7A_0'{A_1}'}{16}+\frac{A_0'^2}{16}+\frac{5A_0'}{4r}-\frac{\mathcal{R}e^{A_1}}{2}\bigg)+\mathcal{R}'\bigg(\frac{A_0'}{8}-\frac{1}{2r}
+\frac{{A_1}'}{8}\bigg)-\frac{\mathcal{R''}}{4}-\frac{\alpha_8}{4}\bigg\}\\\nonumber
&-8\pi
e^{{A_1}}\bigg]^{-1}\bigg[-\frac{1}{4}\bigg(\frac{{A_1}'}{r}+\frac{A_0'}{r}\bigg)+8\pi
B_ce^{A_1}-\eta B_c\bigg\{\bigg(\frac{{A_1}'}{r}
+\frac{A_0'}{r}\bigg)\alpha_1+\mathcal{R}\bigg(\frac{A_0'^2}{4}\\\nonumber
&-\frac{\mathcal{R}e^{A_1}}{2}+\frac{{A_1}'}{r}-\frac{A_0'{A_1}'}{2}+\frac{A_0''}{2}+\frac{2A_0'}{r}\bigg)+\mathcal{R}'\bigg(\frac{A_0'}{4}-\frac{1}{r}
+\frac{{A_1}'}{4}\bigg)-\frac{\mathcal{R''}}{2}-\alpha_7\\\label{g14f}
&+\alpha_4\bigg(\frac{{A_1}'}{2}+\frac{A_0'}{2}\bigg)\bigg\}\bigg].
\end{align}
Moreover, we avoid writing an expression for tangential pressure ($P_t$) for both models due to the very long expression. But, we can determine the tangential pressure corresponding to $\textbf{Model 1}$, one
can put Eqs.\eqref{g14b} and \eqref{g14c} into \eqref{g8b}. For the
$\textbf{Model 2}$, it can be achieved using
Eqs.\eqref{g8e}, \eqref{g14e} and \eqref{g14f}.

\section{A Particular Ansatz and Implementation of Matching Criteria}

Given that the equations of motion possess additional unknowns till
now, we shift our focus to a particular metric, i.e.,
Durgapal-Fuloria spacetime, that has attracted significant interest
among astrophysicists. The components of this metric are provided as
\cite{38ca}
\begin{equation}\label{g15}
e^{A_0}=d_1\big(d_2 r^2+1\big)^4, \quad\quad
e^{{A_1}}=\frac{7\big(d_2 r^2+1\big)^2}{7-d_2^2 r^4-10 d_2 r^2},
\end{equation}
involving a doublet constant ($d_1,d_2$) and we need to calculate
its values. In our analysis, we shall utilize matching criteria to
ascertain these values. Given the variety of metrics proposed in the
literature, it is essential to verify the physical acceptability of
the ansatz under consideration. For this purpose, a specific
criterion has been suggested \cite{41j,41ja}, which involves taking
derivatives of both the time and radial components to validate its
suitability, as outlined below
\begin{align}\nonumber
A_0'(r)&=\frac{8 d_2 r}{d_2 r^2+1}, \quad A_0''(r)=\frac{8 d_2}{d_2
r^2+1}-\frac{16 d_2^2 r^2}{\big(d_2 r^2+1\big)^2},\\\nonumber
{A_1}'(r)&=\frac{16 d_2 r \big(d_2 r^2-3\big)}{\big(d_2 r^2+1\big)
\big(d_2^2 r^4+10 d_2 r^2-7\big)},\\\nonumber {A_1}''(r)&=\frac{16
d_2 \big(21-3 d_2^4 r^8+4 d_2^3 r^6+102 d_2^2 r^4-12 d_2
r^2\big)}{\big(d_2 r^2+1\big)^2\big(d_2^2 r^4+10 d_2 r^2-7\big)^2},
\end{align}
from which we notice that $A_0'(0)=0={A_1}'(0)$ and $A_0''(0)$,
${A_1}''(0)$$>$$0$ in the whole domain ($r=0$ is the star's core).
Hence, the considered ansatz is found to be acceptable.

The matching of the inner and outer sectors at the hypersurface
provides multiple conditions that serve as a valuable tool for
comprehending the structure of massive bodies. In this context, the
Schwarzschild metric describing the exterior geometry is considered
that has the form
\begin{equation}\label{g20}
ds^2=\frac{dr^2}{\left(1-\frac{2\mathrm{M}}{r}\right)}
+r^2d\theta^2+r^2\sin^2\theta
d\phi^2-\bigg(1-\frac{2\mathrm{M}}{r}\bigg)dt^2,
\end{equation}
where $\mathrm{M}$ being the total exterior mass. We use only the
first fundamental form of these constraints, ensuring that metric
components of both geometries are continuous across the boundary.
This leads to
\begin{eqnarray}\label{g21}
g_{tt}&{_{=}^{\Sigma}}&e^{A_0(\emph{R})}=d_1\big(d_2
\emph{R}^2+1\big)^4=1-\frac{2\mathrm{M}}{\emph{R}},\\\label{g21a}
g_{rr}&{_{=}^{\Sigma}}&e^{{A_1}(\emph{R})}=\frac{7\big(d_2
\emph{R}^2+1\big)^2}{7-d_2^2 \emph{R}^4-10 d_2 \emph{R}^2}
=\bigg(1-\frac{2\mathrm{M}}{\emph{R}}\bigg)^{-1}.
\end{eqnarray}
Equations \eqref{g21} and \eqref{g21a} simultaneously provide these
two constants as
\begin{align}\label{g23}
d_1&=\frac{\emph{R}-2\mathrm{M}}{\emph{R}\big(d_2\emph{R}^2+1\big)^4},\\\label{g24}
d_2&=\frac{6\emph{R}^3-7\mathrm{M}\emph{R}^2-2
\sqrt{9\emph{R}^6-14\mathrm{M}\emph{R}^5}}{7\mathrm{M}\emph{R}^4-4\emph{R}^5}.
\end{align}

\section{Exploring Stellar Solutions through Graphical Analysis}

We devoted this section to graphically analyzing the resulting
models for a particular object, namely 4U 1820-30. The initial data
for the considered candidate includes a mass of $\mathrm{M}=1.58 \pm
0.06~\mathrm{M}_{\bigodot}$ and radius $\emph{R}=9.1 \pm 0.4~km$
\cite{41k}. For this star, the constants \eqref{g23} and \eqref{g24} are calculated as $d_1=0.235009$ and
$d_2=0.00243166~km^{-2}$. Furthermore, a thorough analysis of the model under
consideration is conducted, encompassing various properties to
assess the validity of the outcomes. We also extend this
investigation to assessing the equilibrium as well as stability for
the resulting models by considering the parametric range of $\eta$,
which can either be negative or positive. It must be kept in mind
that the plotting is done for multiple choices of the bag constant
and identify favorable outcomes when $B_c=90~MeV/fm^3$. Therefore,
all the plots presented in the following shall correspond to this
particular value.

It is crucial to highlight that a valid metric ansatz should not
only monotonically increasing function of $r$ but also free from
singularities. This is ensured through the plotting of both
components given in Eq.\eqref{g15}, as depicted in Figure
\textbf{1}.
\begin{figure}[htp!]
\center
\epsfig{file=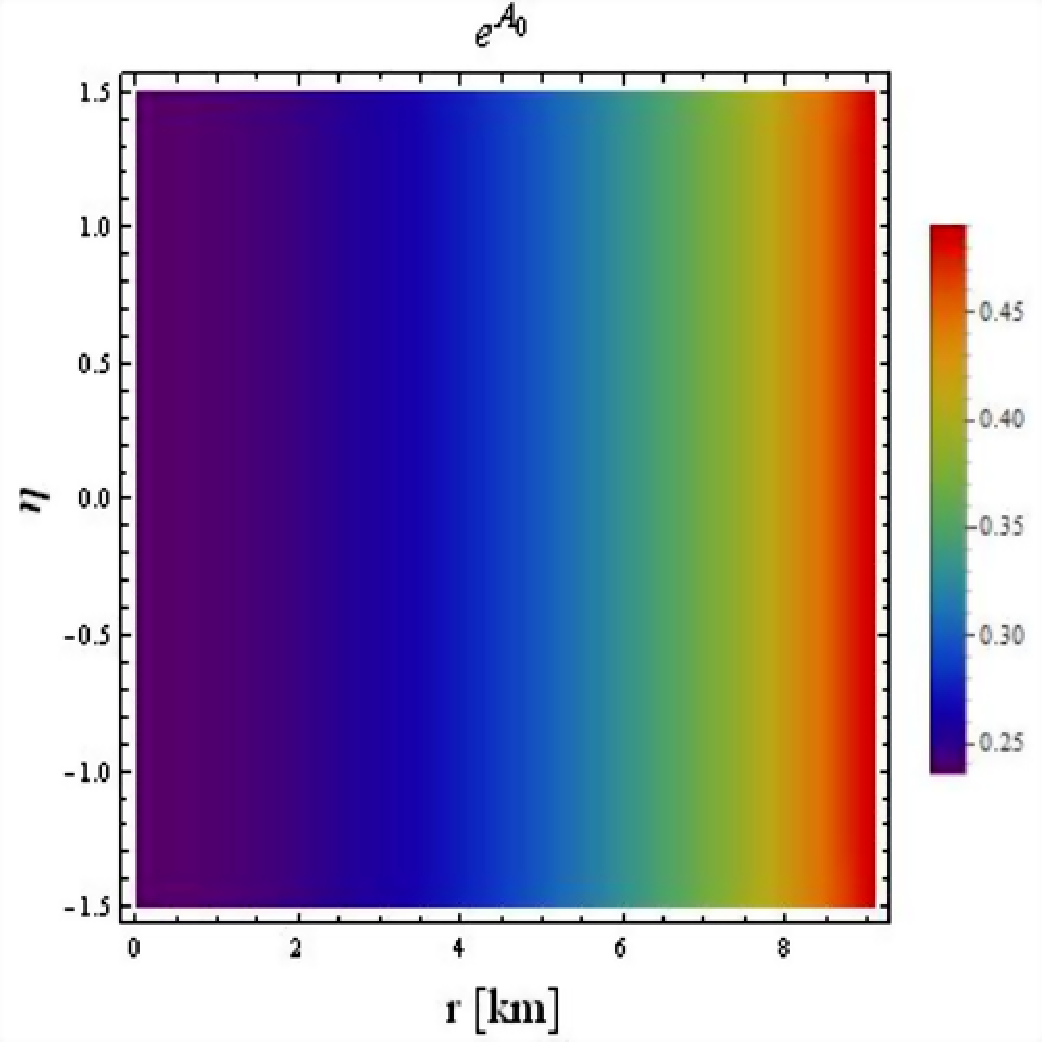,width=0.47\linewidth}\epsfig{file=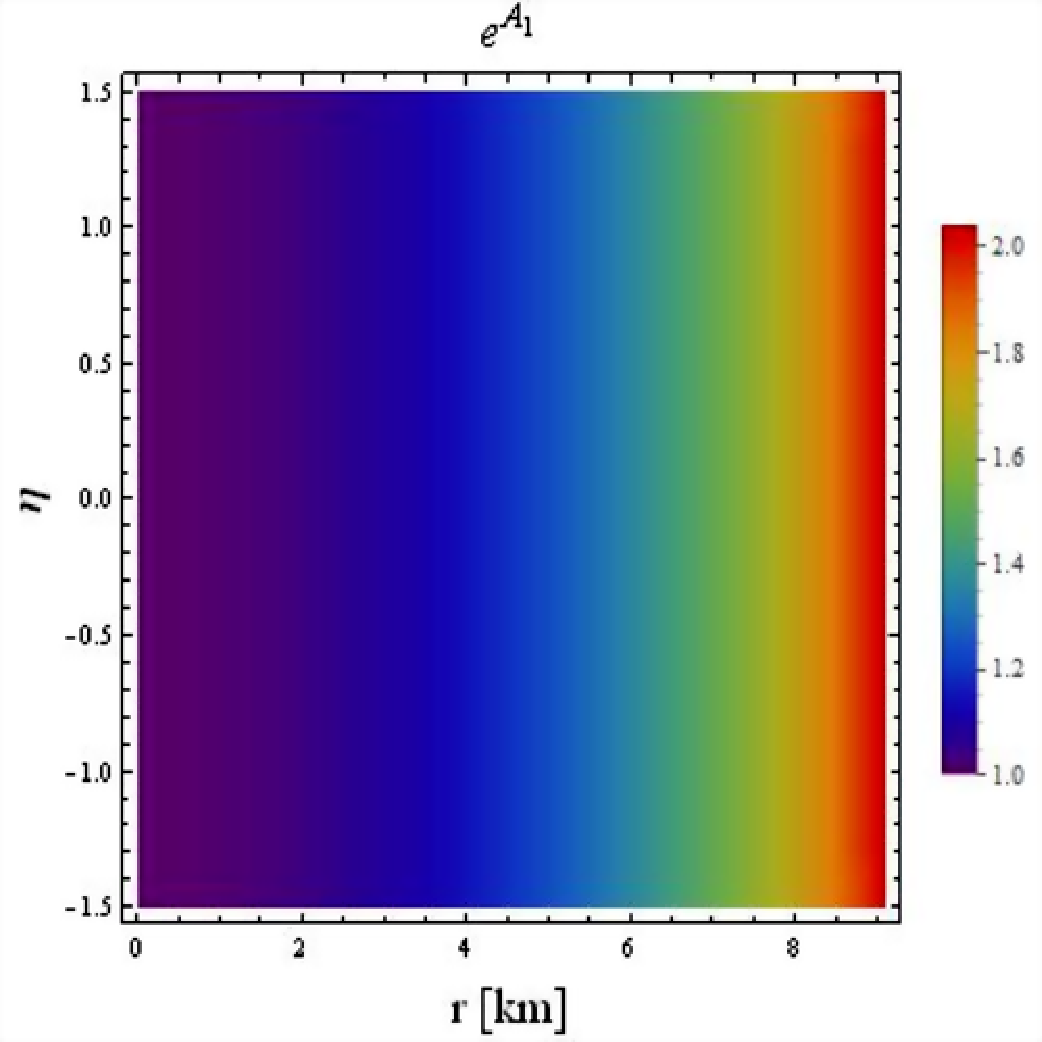,width=0.47\linewidth}
\caption{Metric components \eqref{g15} versus $\eta$ and $r$.}
\end{figure}

\subsection{Matter Variables and Anisotropy}

Ensuring the concentration of fluid within a geometric structure is
crucial for validating the resulting solution. It entails that the
fluid sector must reach their maximum values and be finitely
positive at the center, decreasing outward. To achieve this, we
conduct a comparative analysis for models 1 and 2, as illustrated in
Figure \textbf{2}. Remarkably, the plotting shows that the behavior
of the fluid variables aligns with that of physically acceptable
interiors. Such an observation provides assistance for the existence
of extremely dense interiors within this gravity theory. Upon
exploring the first row of Figure \textbf{2}, it is guaranteed that
the second model produces structures with slightly higher densities. This can also be observed from the numerical values provided in
Tables \textbf{I} and \textbf{II}.
Furthermore, the subsequent graphs in the same Figure show that
model 2 exhibits higher values for pressure components within the
stellar body under discussion. Importantly, we also verify the
regularity constraints by plotting the first and second derivatives
of the fluid variables in Figures \textbf{3} and \textbf{4}.

\begin{table}[H]
\scriptsize \centering \caption{Numerical values of fluid parameters
for 4U 1820-30 corresponding to model 1.} \label{Table1}
\vspace{+0.1in} \setlength{\tabcolsep}{0.95em}
\begin{tabular}{cccccc}
\hline\hline Physical Factors & $\mu_c~(gm/cm^3)$ &
$\mu_s~(gm/cm^3)$ & $P_{c}~(dyne/cm^2)$ & $\beta_s$ & $z_s$
\\\hline $\eta=-1.5$ & 1.7138$\times$10$^{15}$ & 9.4959$\times$10$^{14}$ &
2.4626$\times$10$^{35}$ & 0.248 & 0.909
\\\hline
$\eta=0$ & 1.6736$\times$10$^{15}$ & 8.9461$\times$10$^{14}$ &
2.3267$\times$10$^{35}$ & 0.259 & 1.028
\\\hline
$\eta=1.5$ & 1.6295$\times$10$^{15}$ & 8.5956$\times$10$^{14}$ &
2.2052$\times$10$^{35}$ & 0.269 & 1.147 \\
\hline\hline
\end{tabular}
\end{table}
\begin{table}[H]
\scriptsize \centering \caption{Numerical values of fluid parameters
for 4U 1820-30 corresponding to model 2.} \label{Table2}
\vspace{+0.1in} \setlength{\tabcolsep}{0.95em}
\begin{tabular}{cccccc}
\hline\hline Physical Factors & $\mu_c~(gm/cm^3)$ &
$\mu_s~(gm/cm^3)$ & $P_{c}~(dyne/cm^2)$ & $\beta_s$ & $z_s$
\\\hline $\eta=-1.5$ & 2.0228$\times$10$^{15}$ & 8.4016$\times$10$^{14}$ &
2.8678$\times$10$^{35}$ & 0.278 & 0.744
\\\hline
$\eta=0$ & 2.0094$\times$10$^{15}$ & 1.0727$\times$10$^{15}$ &
2.8089$\times$10$^{35}$ & 0.284 & 1.788
\\\hline
$\eta=1.5$ & 1.9961$\times$10$^{15}$ & 1.4355$\times$10$^{15}$ &
2.7115$\times$10$^{35}$ & 0.291 & 0.833 \\
\hline\hline
\end{tabular}
\end{table}
\begin{figure}[htp!]\center
\epsfig{file=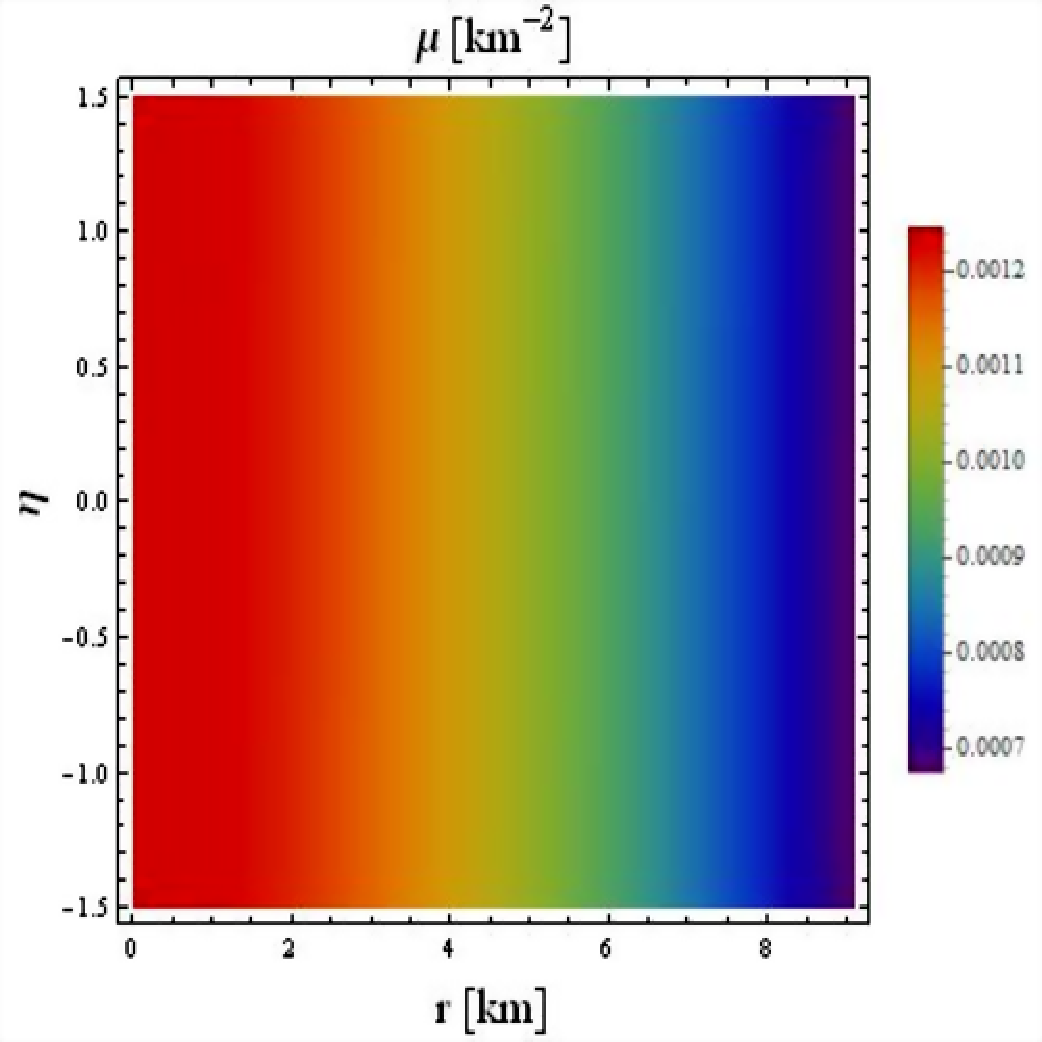,width=0.47\linewidth}\epsfig{file=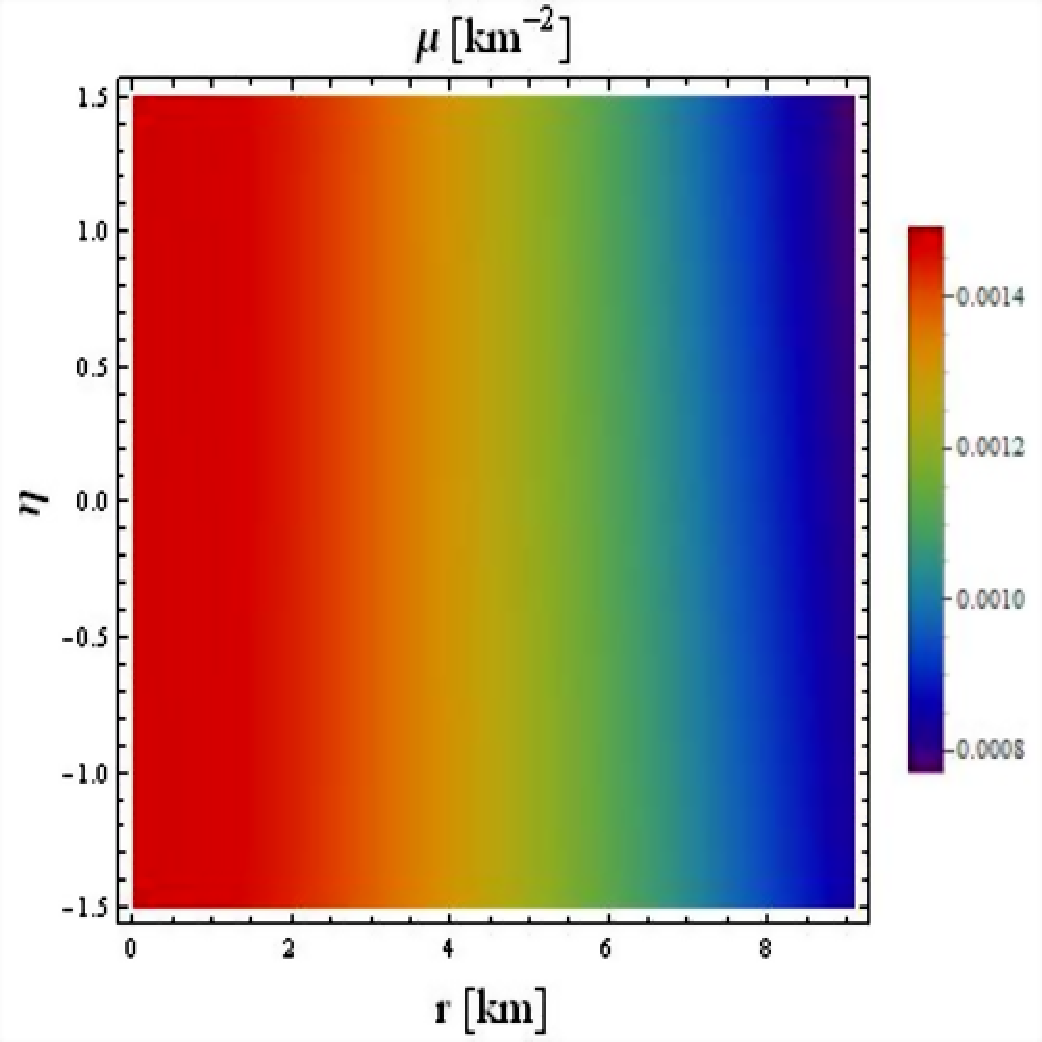,width=0.47\linewidth}
\epsfig{file=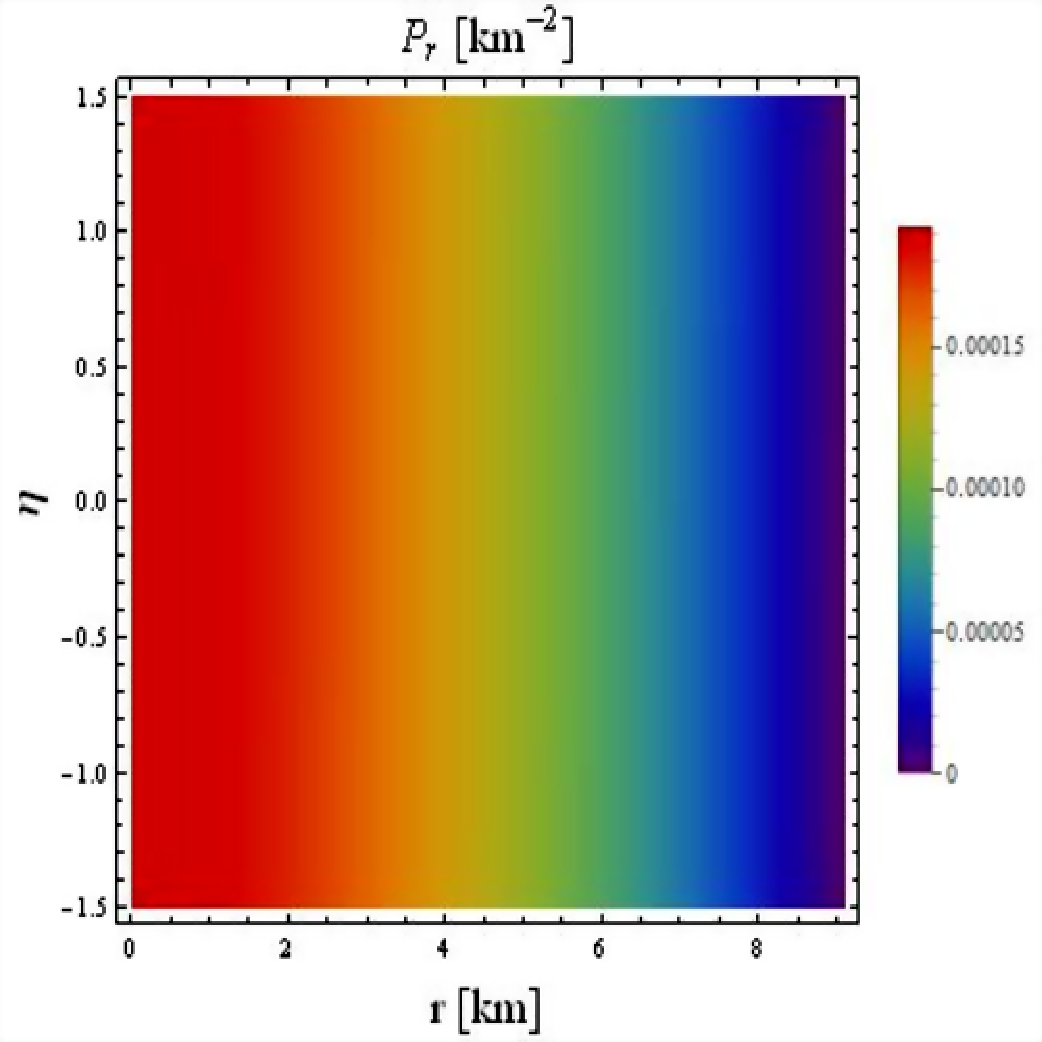,width=0.47\linewidth}\epsfig{file=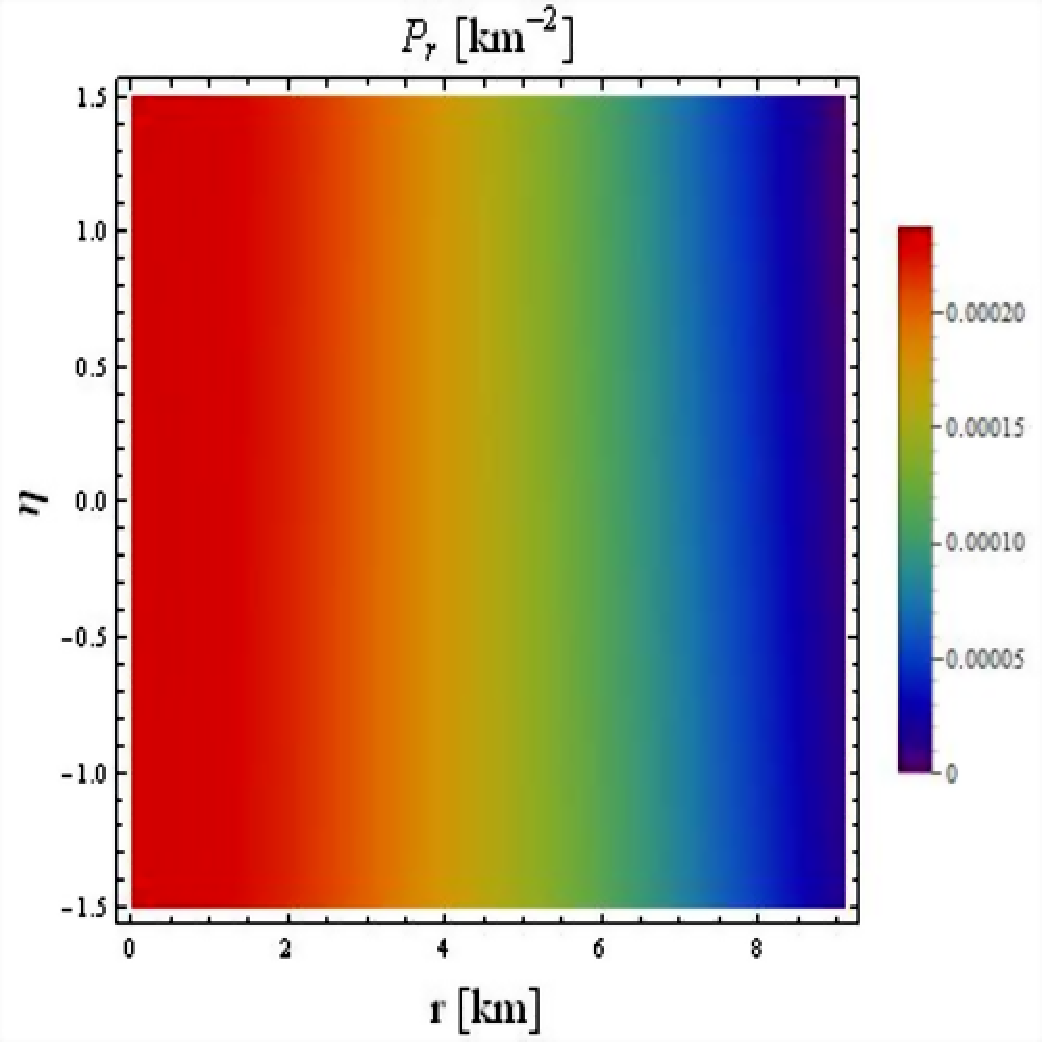,width=0.47\linewidth}
\epsfig{file=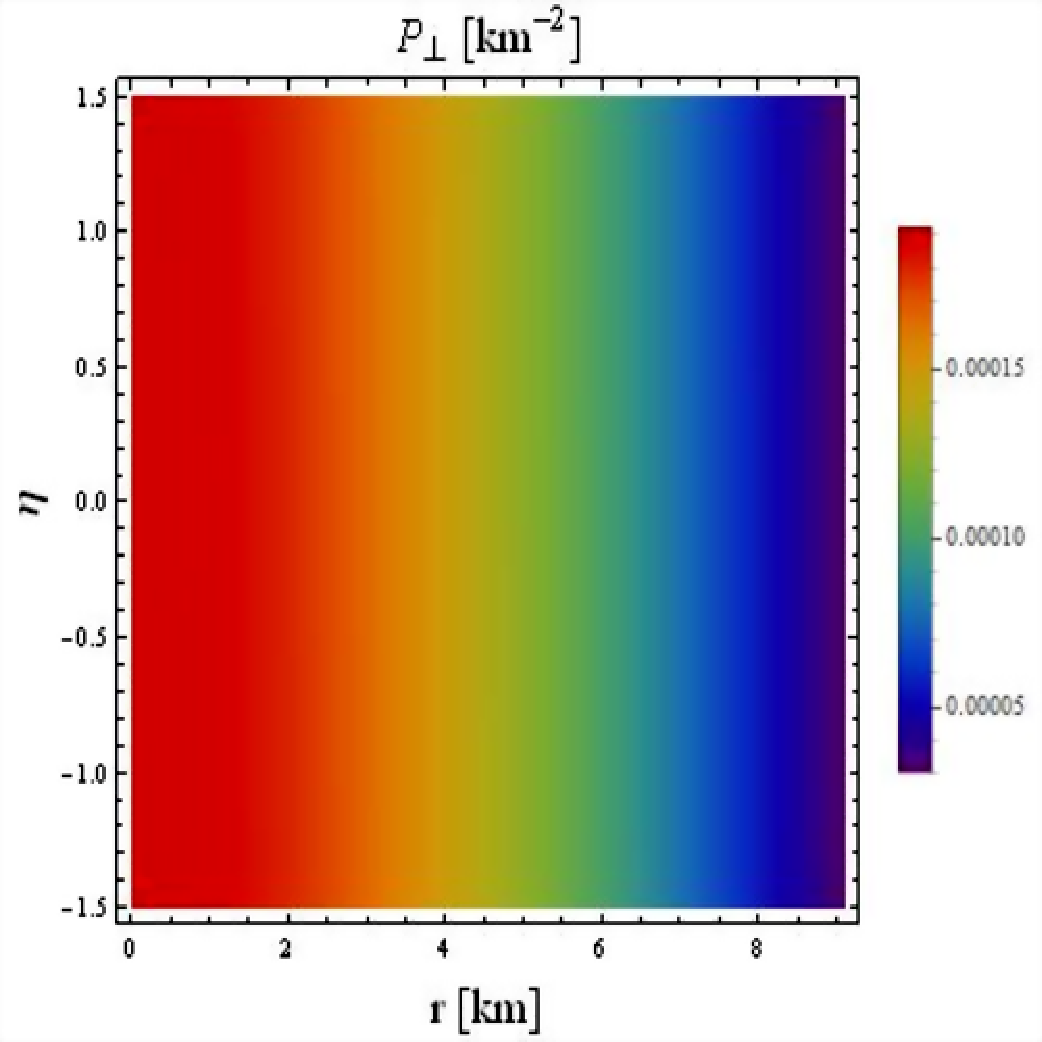,width=0.47\linewidth}\epsfig{file=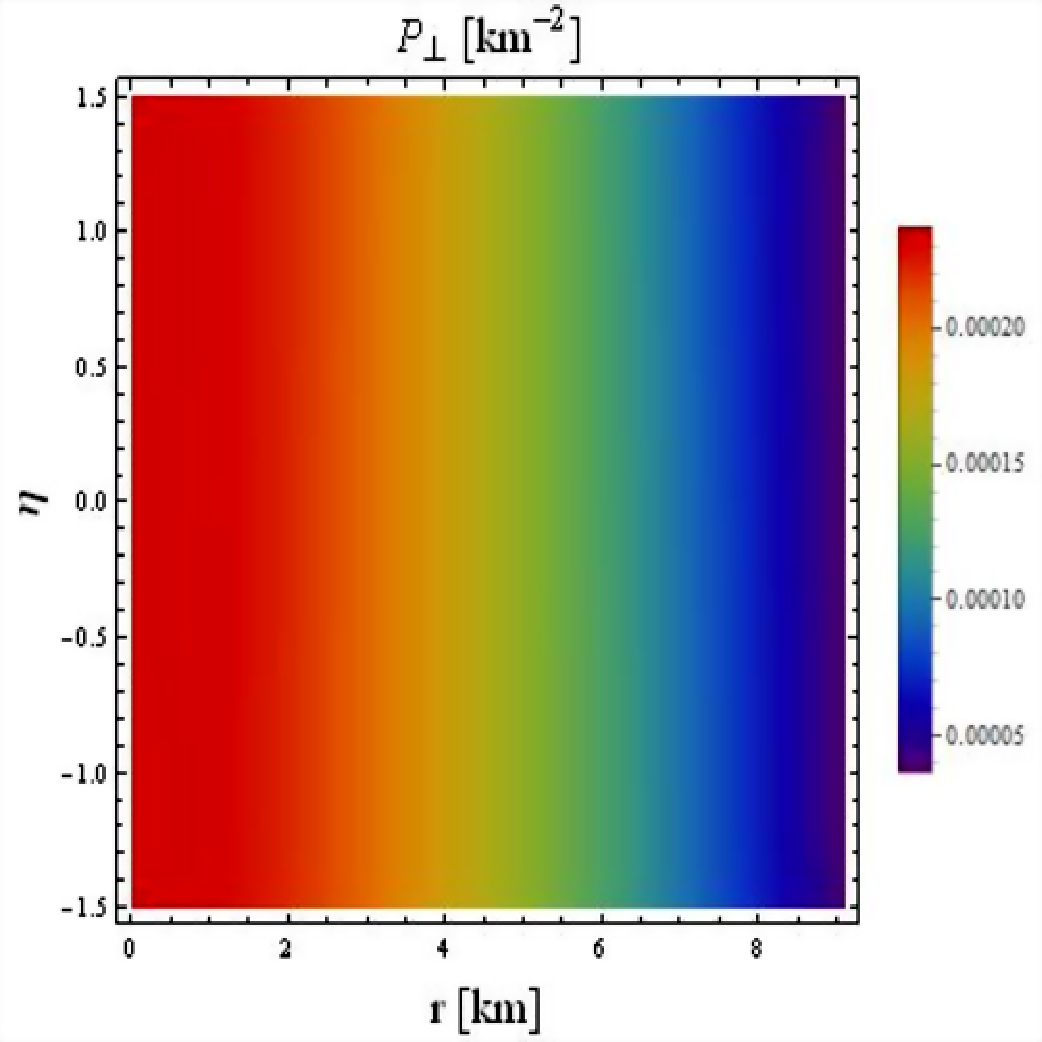,width=0.47\linewidth}
\caption{Matter determinants versus $\eta$ and $r$ for $\textbf{Model 1}$
(left) and $\textbf{Model 2}$ (right).}
\end{figure}
\begin{figure}[htp!]\center
\epsfig{file=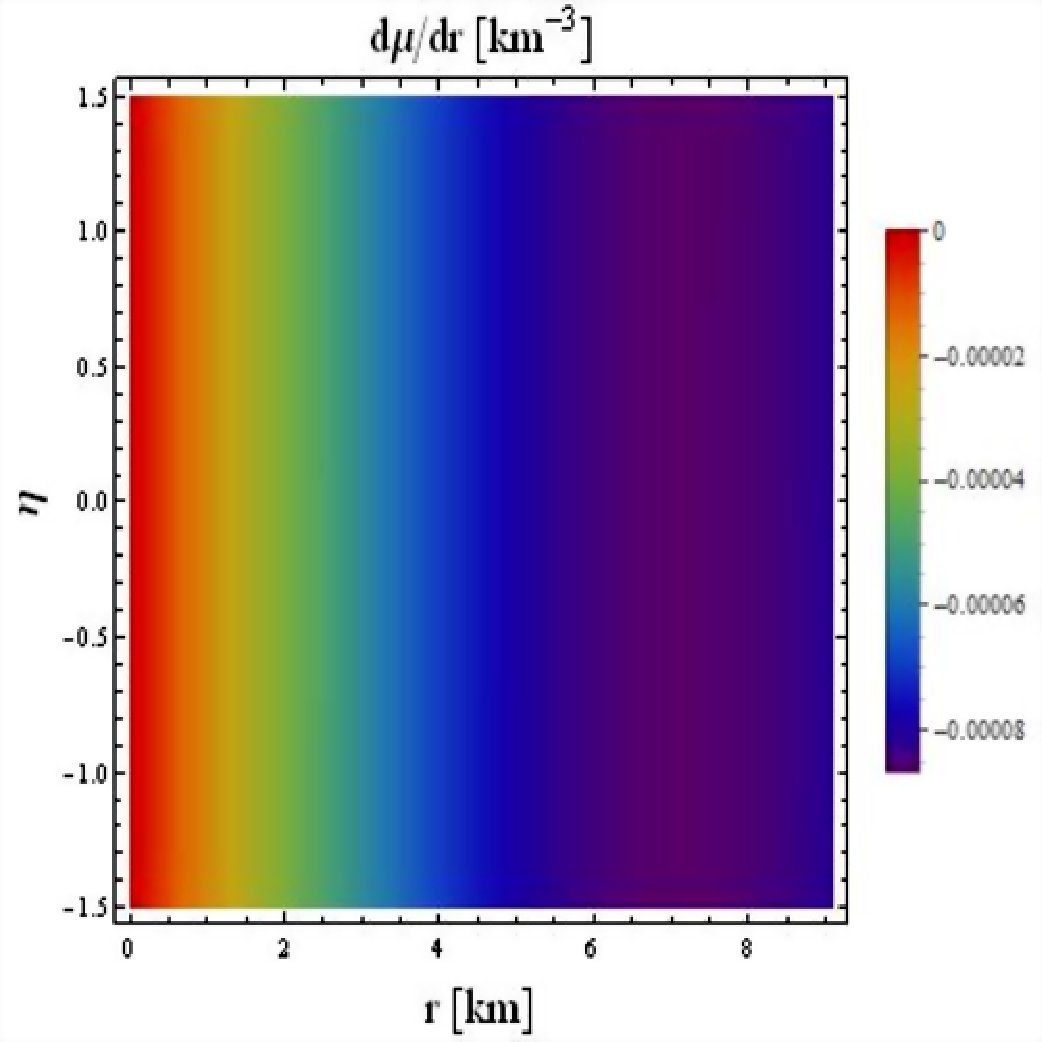,width=0.47\linewidth}\epsfig{file=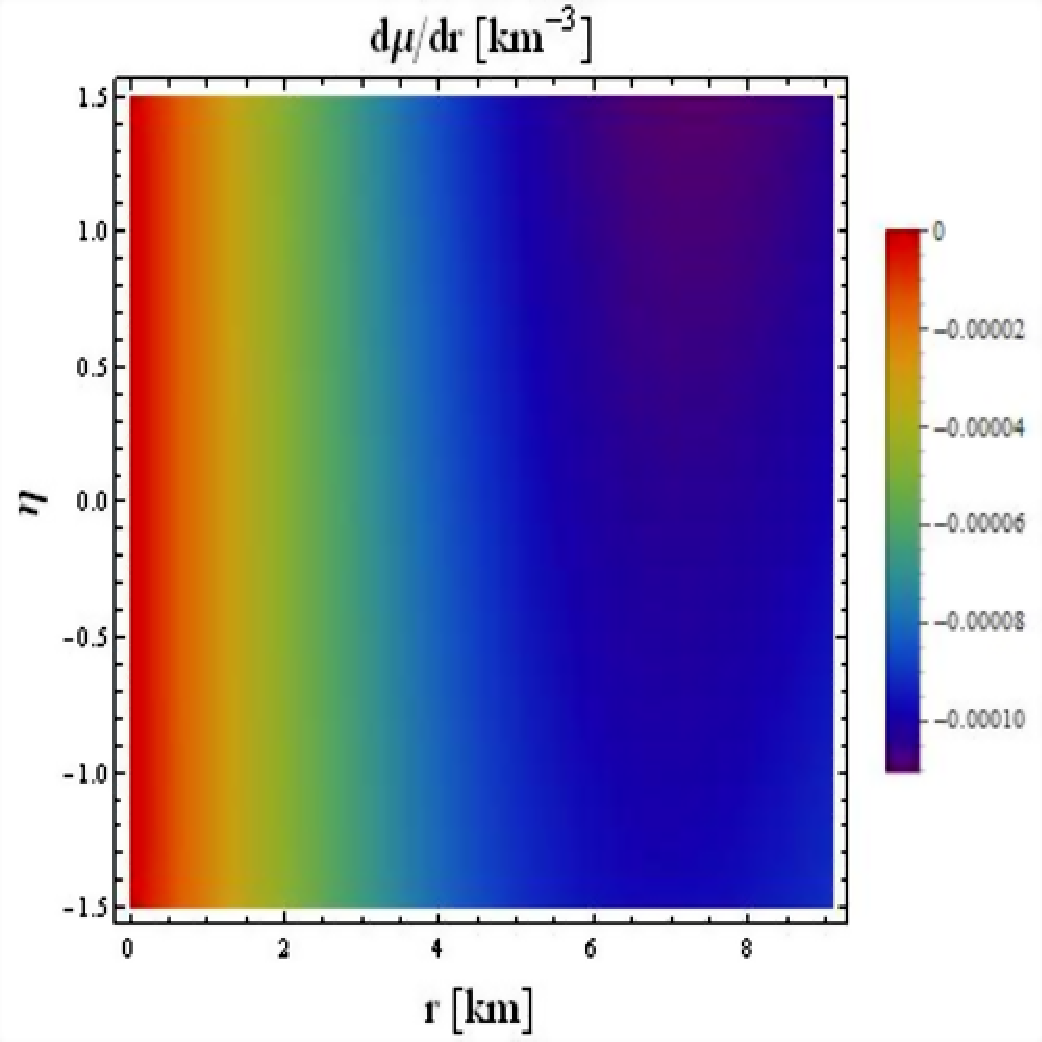,width=0.47\linewidth}
\epsfig{file=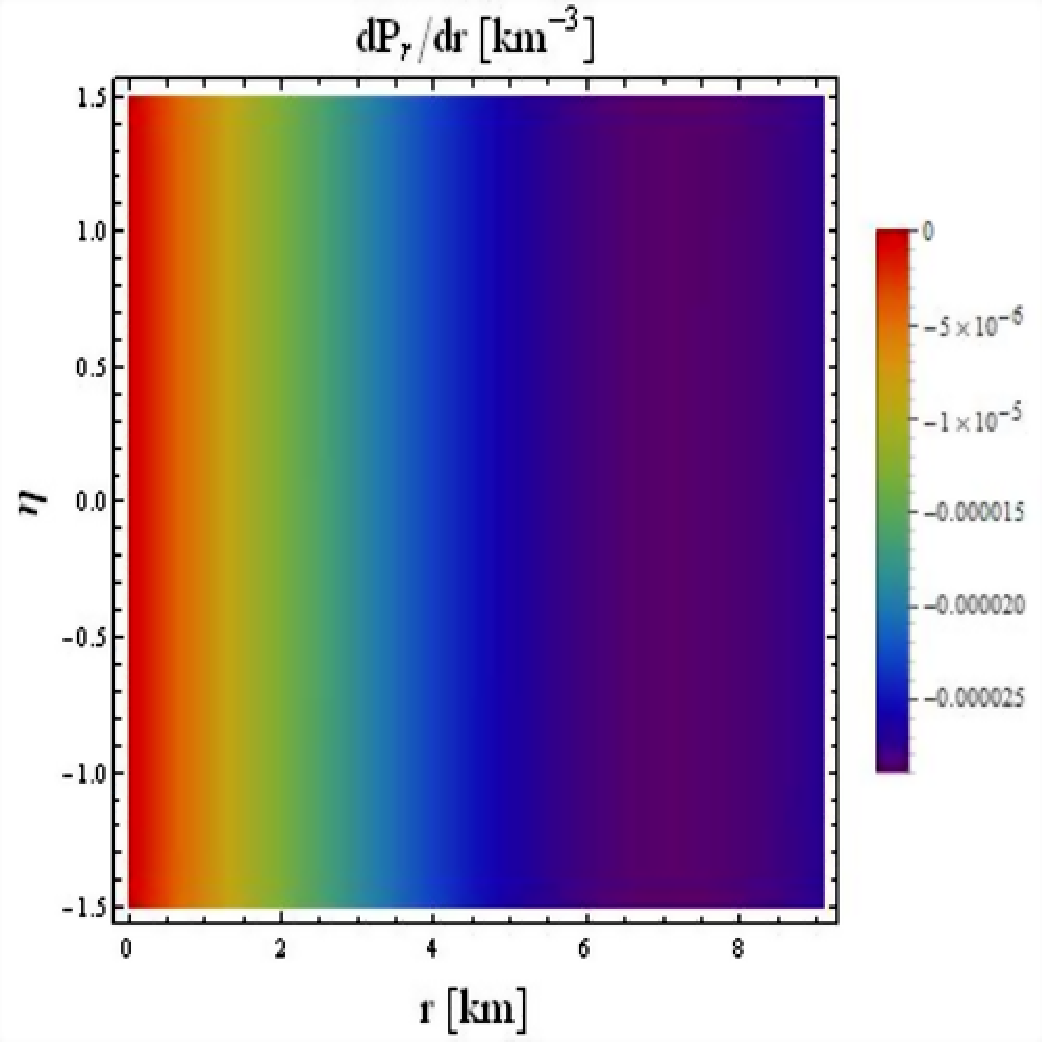,width=0.47\linewidth}\epsfig{file=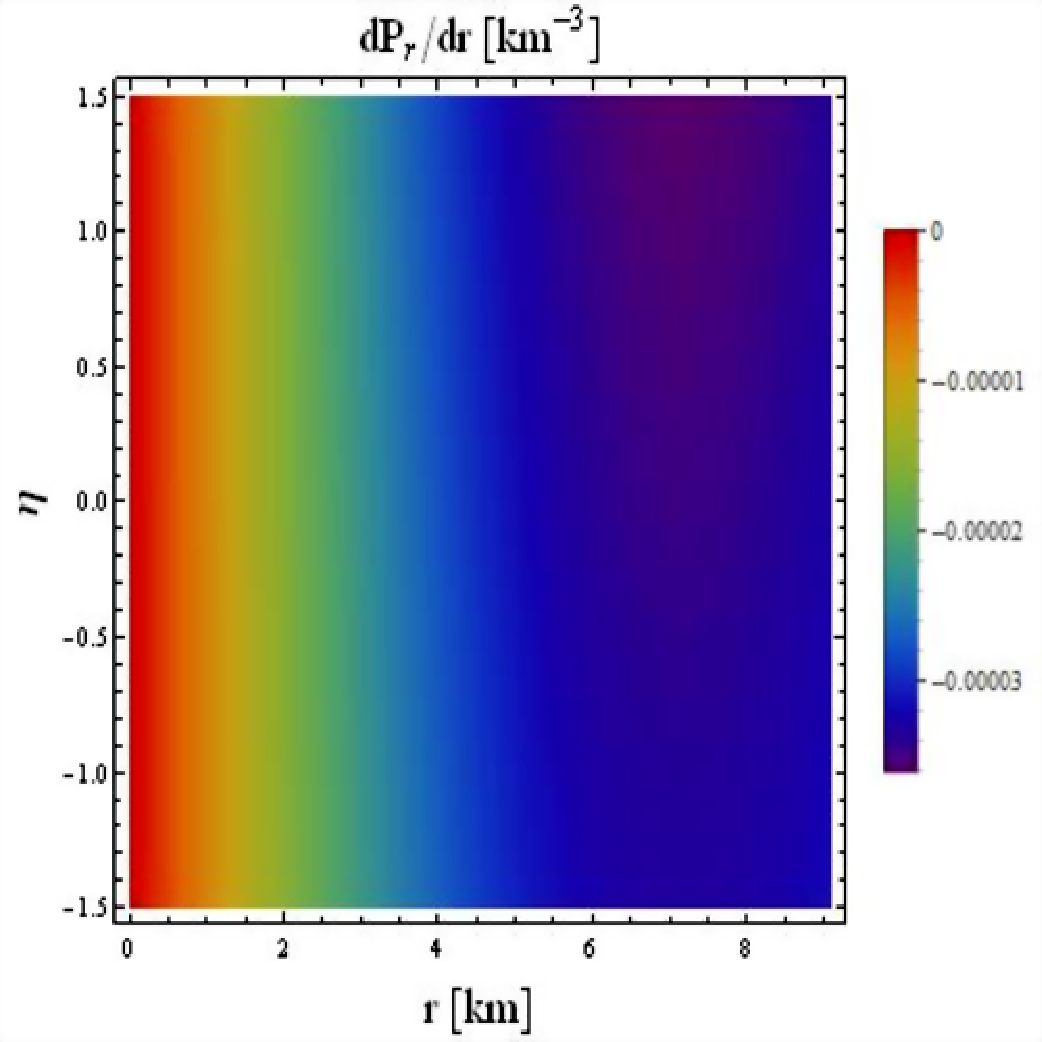,width=0.47\linewidth}
\epsfig{file=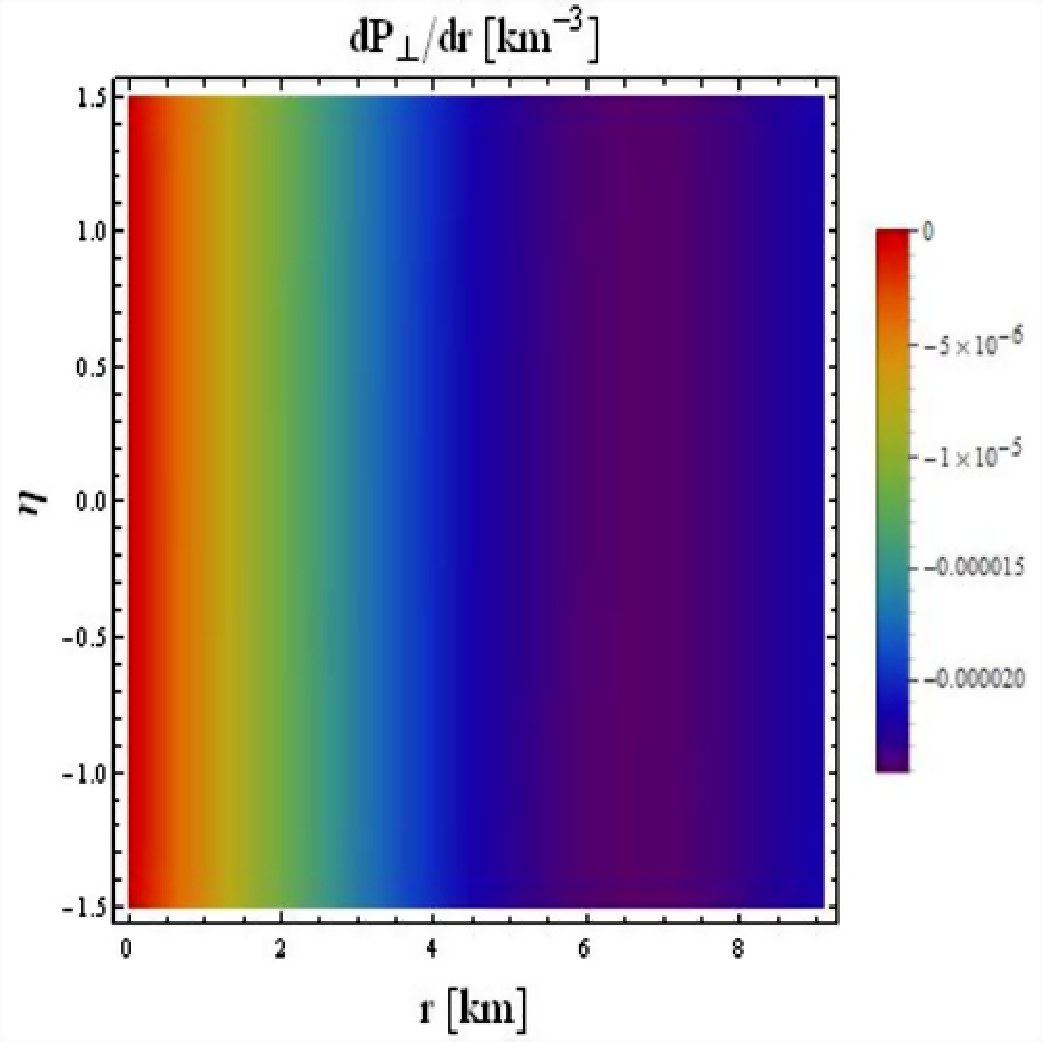,width=0.47\linewidth}\epsfig{file=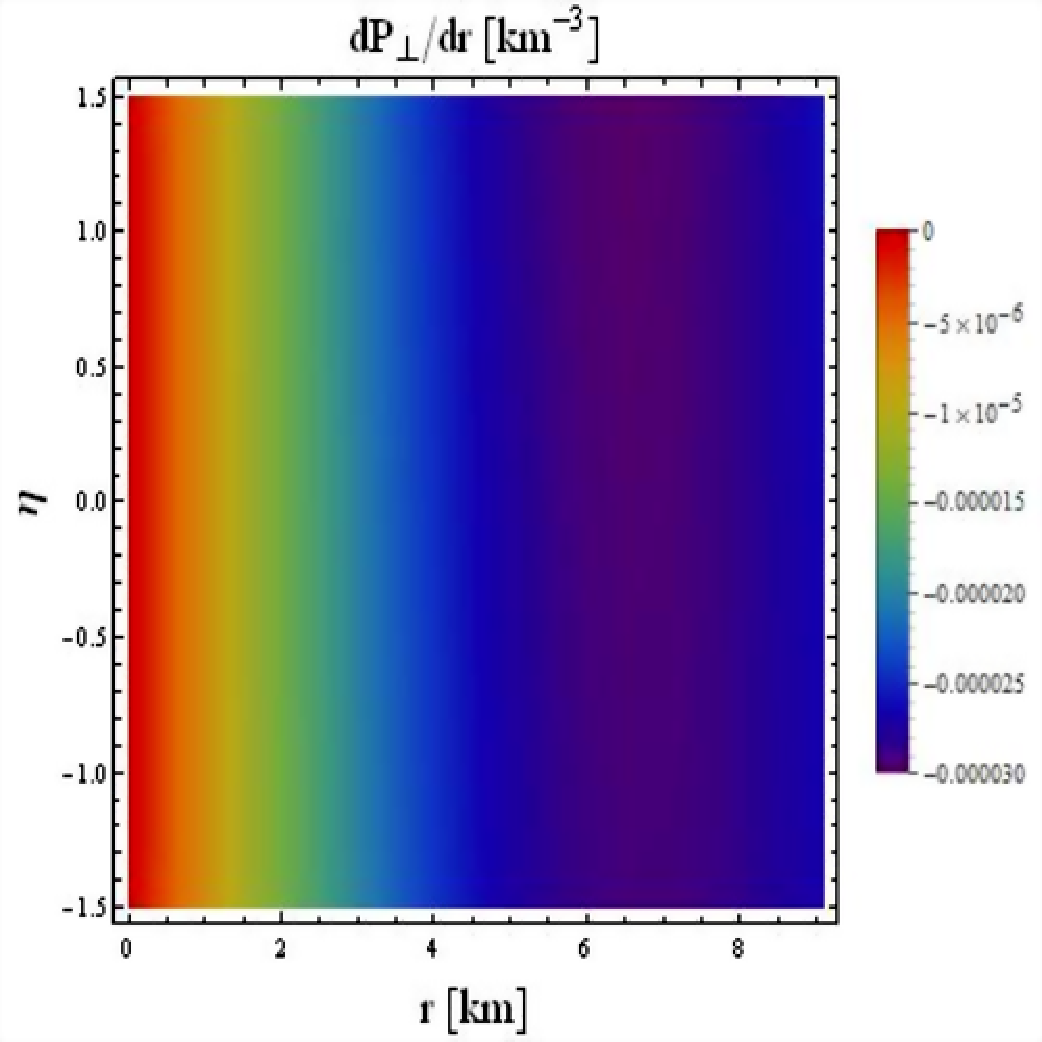,width=0.47\linewidth}
\caption{First-order derivatives of matter determinants versus
$\eta$ and $r$ for $\textbf{Model 1}$ (left) and $\textbf{Model 2}$ (right).}
\end{figure}
\begin{figure}[htp!]\center
\epsfig{file=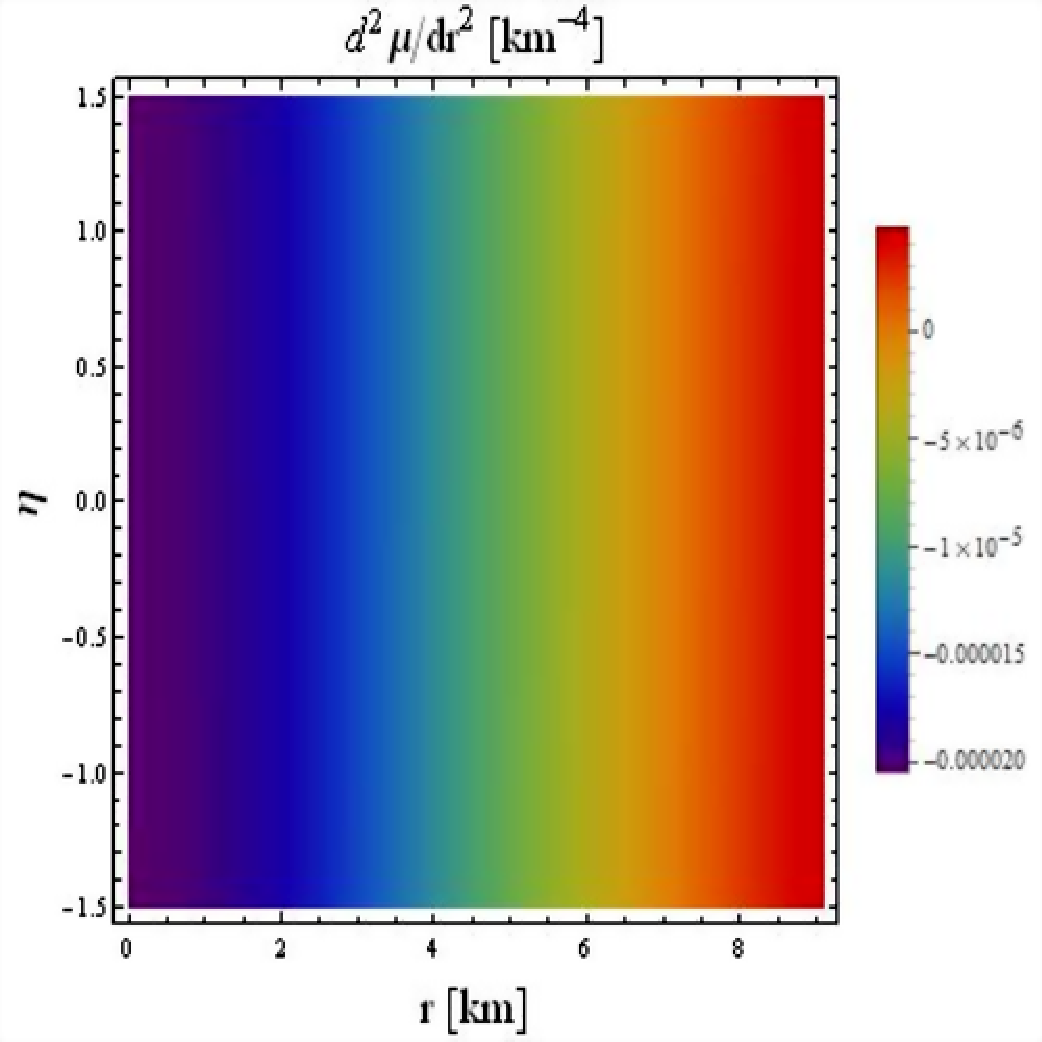,width=0.47\linewidth}\epsfig{file=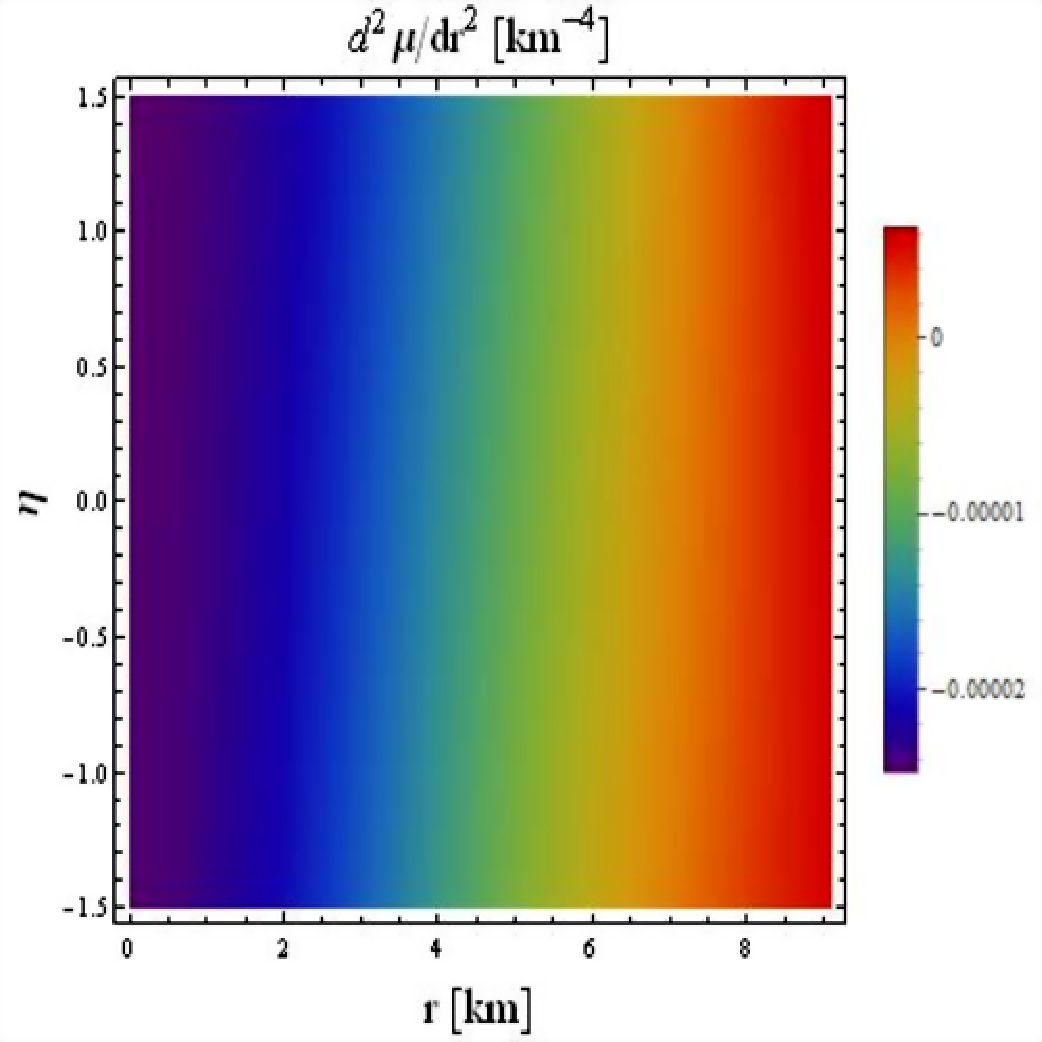,width=0.47\linewidth}
\epsfig{file=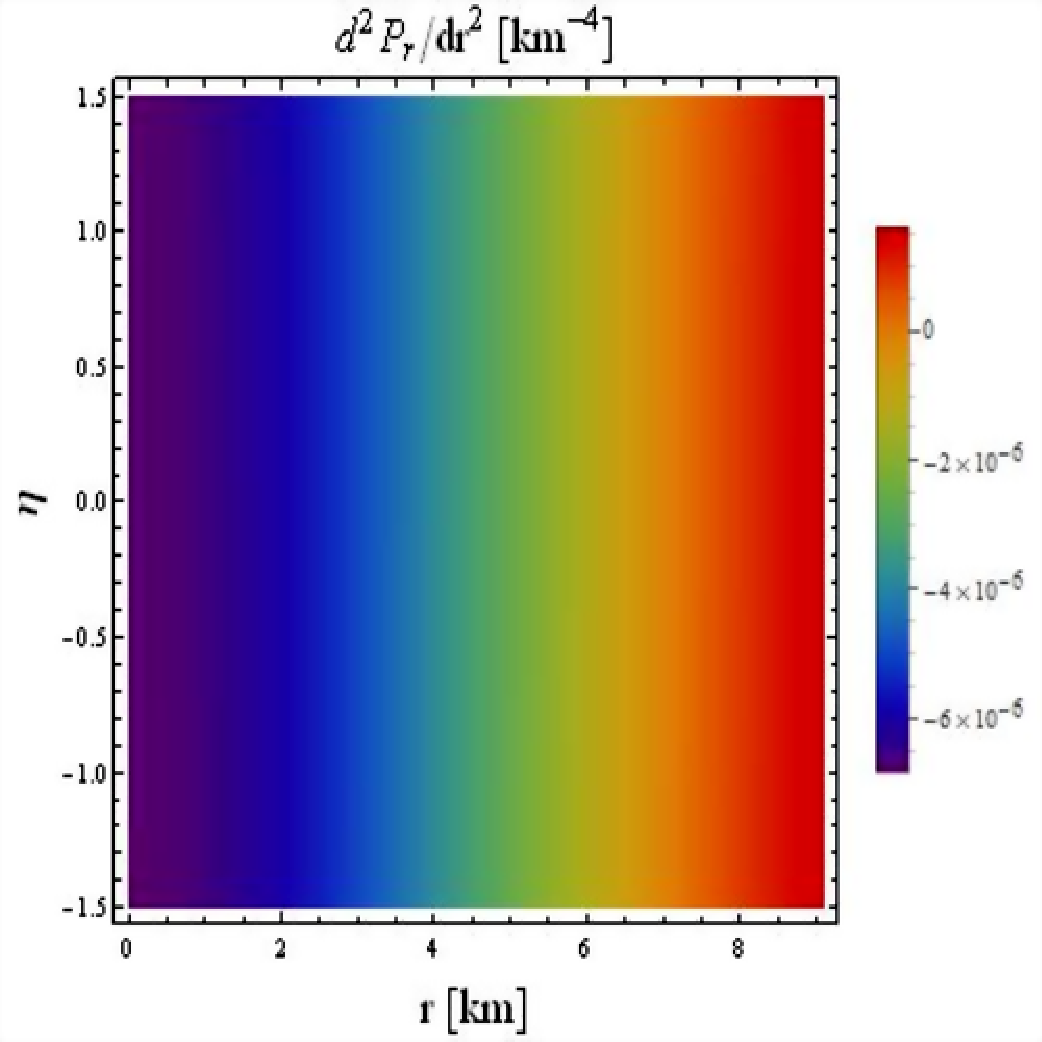,width=0.47\linewidth}\epsfig{file=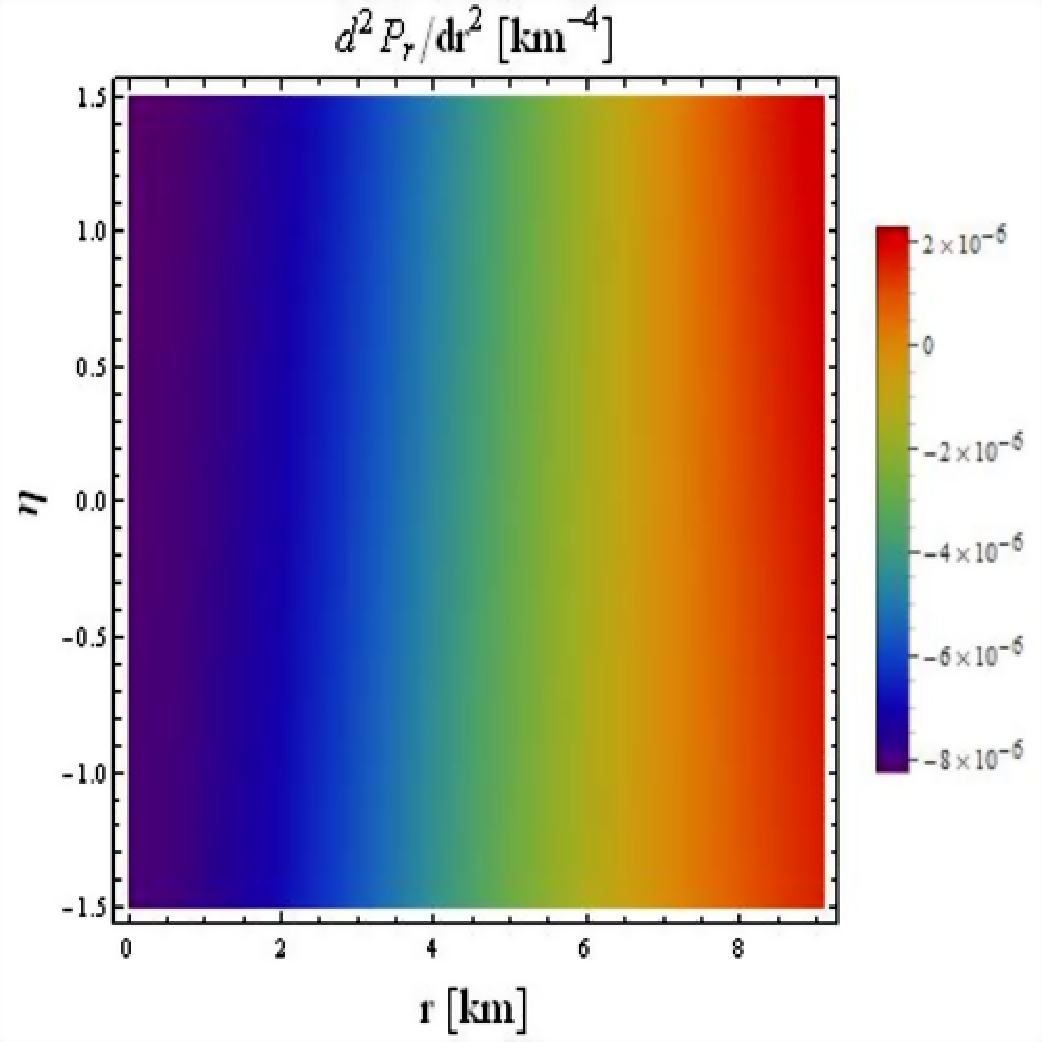,width=0.47\linewidth}
\epsfig{file=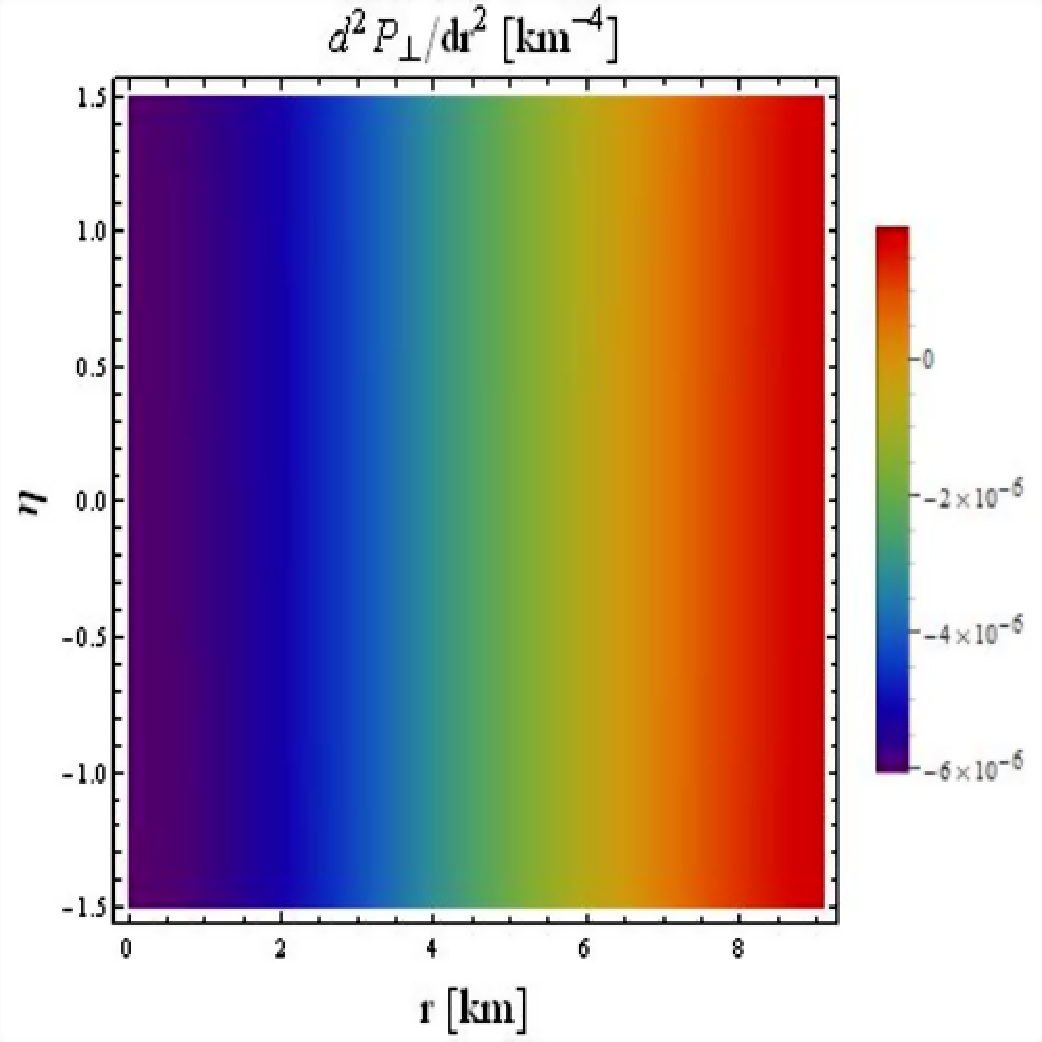,width=0.47\linewidth}\epsfig{file=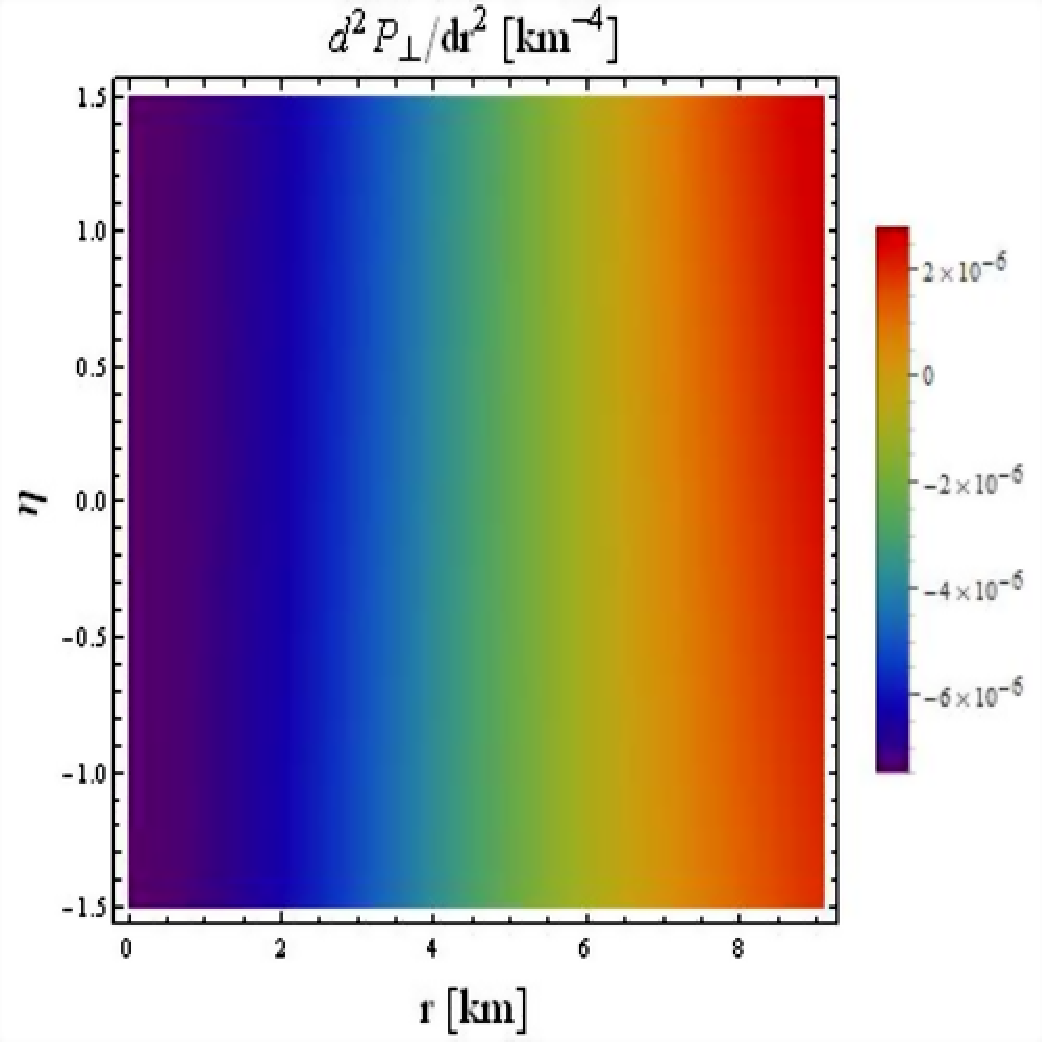,width=0.47\linewidth}
\caption{Second-order derivatives of matter determinants versus
$\eta$ and $r$ for $\textbf{Model 1}$ (left) and $\textbf{Model 2}$ (right).}
\end{figure}

Another important factor in the stellar evolution is the pressure
anisotropy, and we define it by $\Delta=P_\bot-P_r$ in this case.
Here, we shall see how this factor affects the stellar evolutionary
pattern through its graphical representation. If the pressure in the
tangential direction is higher than the other one, an outward force
must occur that prevent the system from collapse. However, the
structure is collapse when the radial pressure is much higher than
the other component. The expressions for anisotropy corresponding to
both models are given in Appendix \textbf{B}. Figure \textbf{5}
exhibits the plots from where we notice its null profile at the
center and consistent increment outwards. We also observe the
presence of little more anisotropy in the interior corresponding to
the second model.
\begin{figure}\center
\epsfig{file=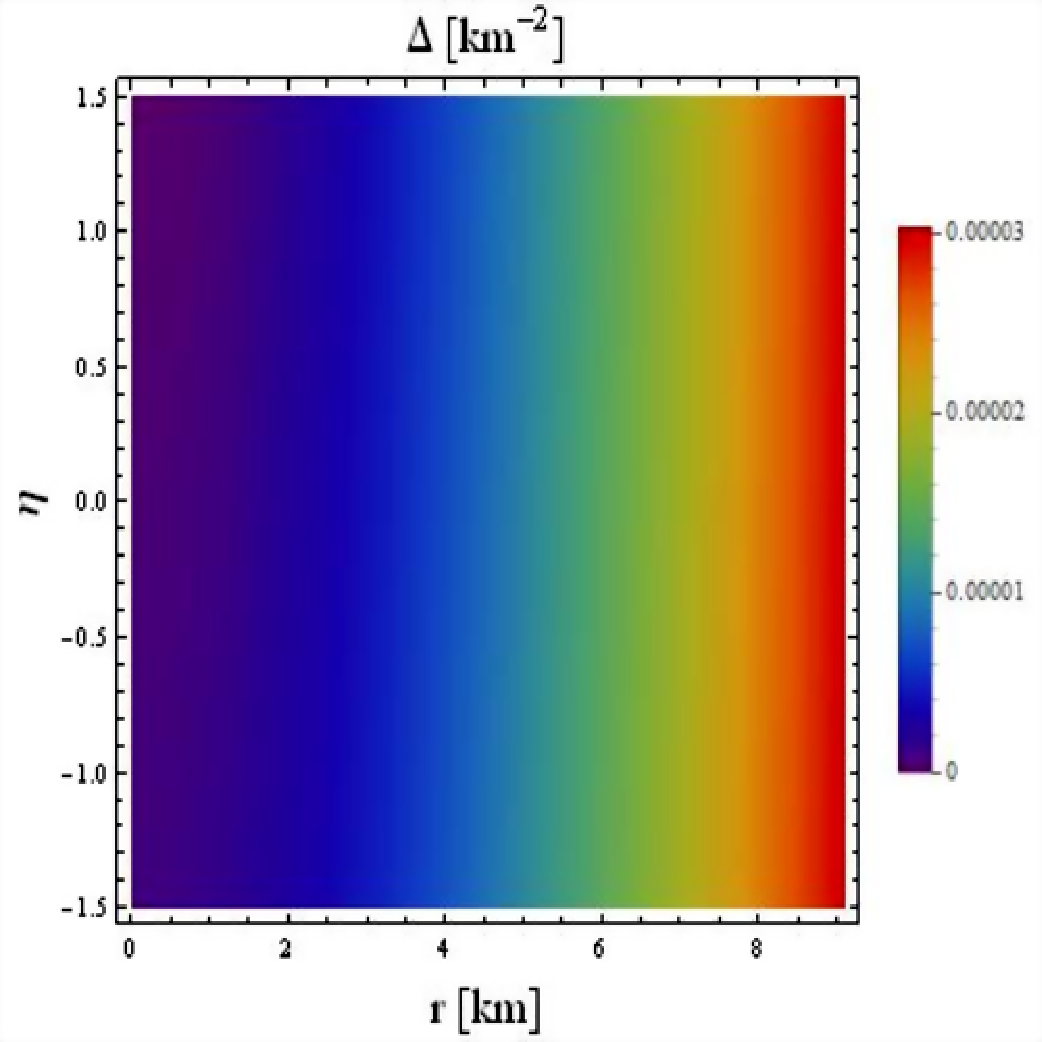,width=0.47\linewidth}\epsfig{file=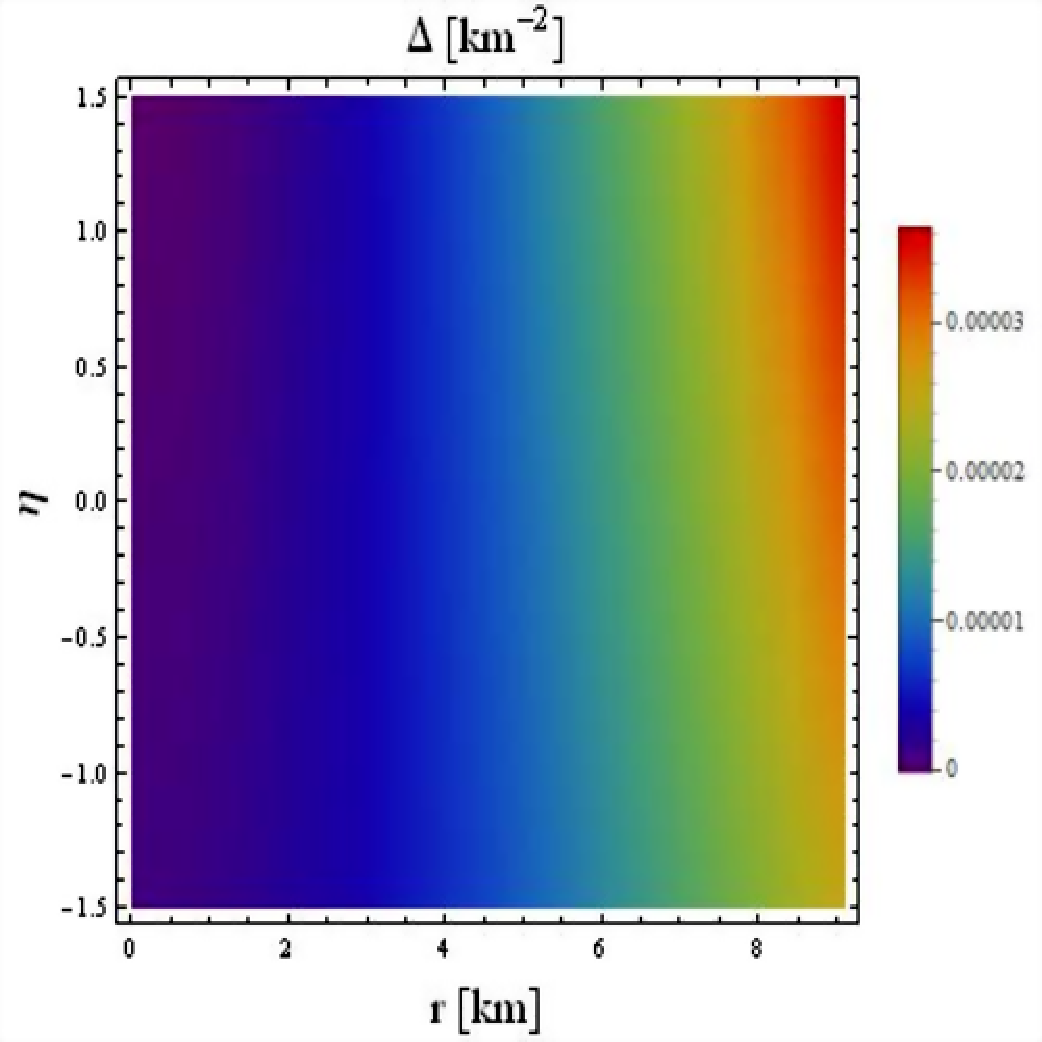,width=0.47\linewidth}
\caption{Anisotropy versus $\eta$ and $r$ for $\textbf{Model 1}$ (left) and $\textbf{Model 2}$
(right).}
\end{figure}

\subsection{Mass Function, Compactness, Redshift and EoS Parameters}

We already defined the mass function in the form of radial metric
function \eqref{g12a} in section 2, and it remains the same in this
theory as well. However, in this subsection, we define this function
in relation with the effective energy density so that the impact of
modification of the action \eqref{g1} can be explored. This is
expressed by
\begin{equation}\label{g32}
m(r)=\frac{1}{2}\int_{0}^{\emph{R}}r^2\mu dr,
\end{equation}
where the values of $\mu$ are presented in Eqs.\eqref{g14b} and
\eqref{g14e} analogous to models 1 and 2, respectively. The first
two plots of Figure $\mathbf{6}$ explains that there is no mass at
the center. However, model 2 generates more massive interior of the
considered compact star as compared to the other model.

Compactness, specifically, refers to the ratio between an object's
mass and its radius. Buchdahl \cite{42a}, in a seminal contribution,
calculated a maximum value for this factor (denoted by $\beta$)
within the context of a celestial structure, establishing it at
$\frac{4}{9}$. A massive object, nestled within a potent gravity
field, emits radiations. Such radiations always travel through
space, and hence, there occur a stretching in their wavelengths,
leading to the redshift phenomenon. Its formula is given as
\begin{equation}\label{g35}
z(r)=\frac{1}{\sqrt{1-2\beta(r)}}-1.
\end{equation}
Theoretical models whose interior is configured with uniform
distribution typically have a maximum limit of 2 for this parameter.
However, an important development came from the contribution of
Ivanov \cite{42b}, who found its value as 5.211 when the anisotropic
star is studied. The remaining plots of Figure \textbf{6} confirm
both of these factors within their proposed limits.
\begin{figure}[htp!]\center
\epsfig{file=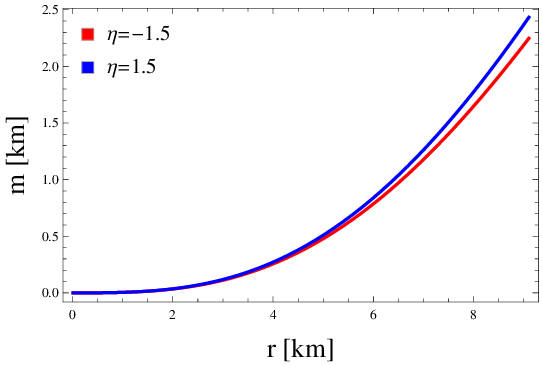,width=0.47\linewidth}\epsfig{file=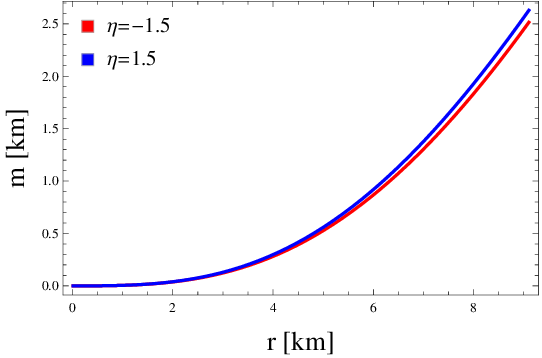,width=0.47\linewidth}
\epsfig{file=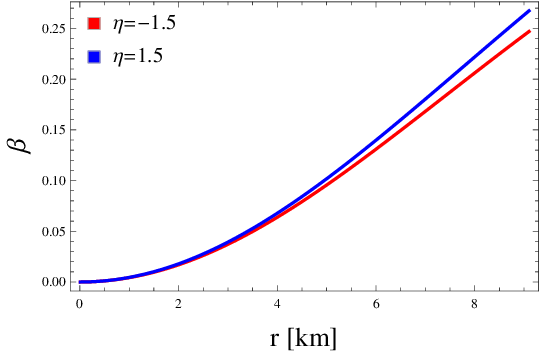,width=0.47\linewidth}\epsfig{file=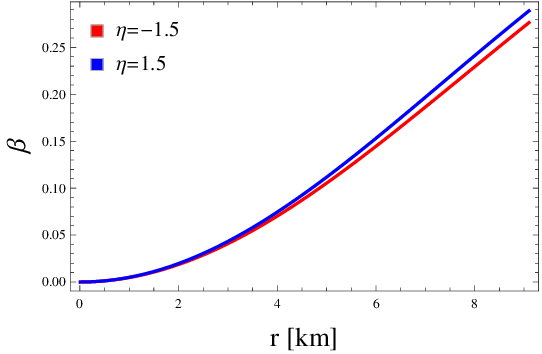,width=0.47\linewidth}
\epsfig{file=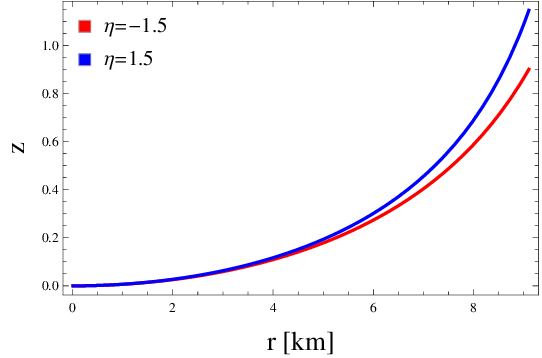,width=0.47\linewidth}\epsfig{file=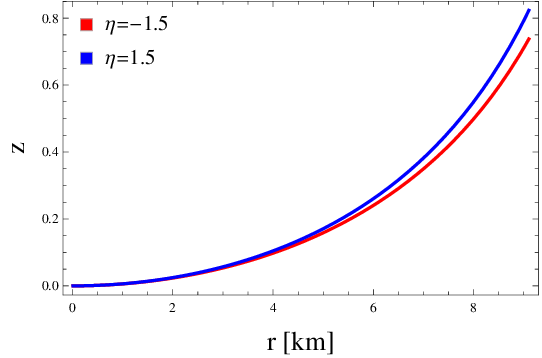,width=0.47\linewidth}
\caption{Physical terms versus $\eta$ and $r$ for $\textbf{Model 1}$ (left)
and $\textbf{Model 2}$ (right).}
\end{figure}

Furthermore, the EoS is divided into two equations when we are
dealing with the anisotropic fluid distribution. They can
mathematically be written as
\begin{equation}\label{g361}
\omega_r=\frac{P_r}{\mu}, \quad \omega_\bot=\frac{P_\bot}{\mu}.
\end{equation}
It must be recalled that both the parameters must remain with the
interval $[0,1]$ to get well-behaved results for the stellar
structures. We plot them in Figure \textbf{7}, ensuring their
required behavior.
\begin{figure}[htp!]\center
\epsfig{file=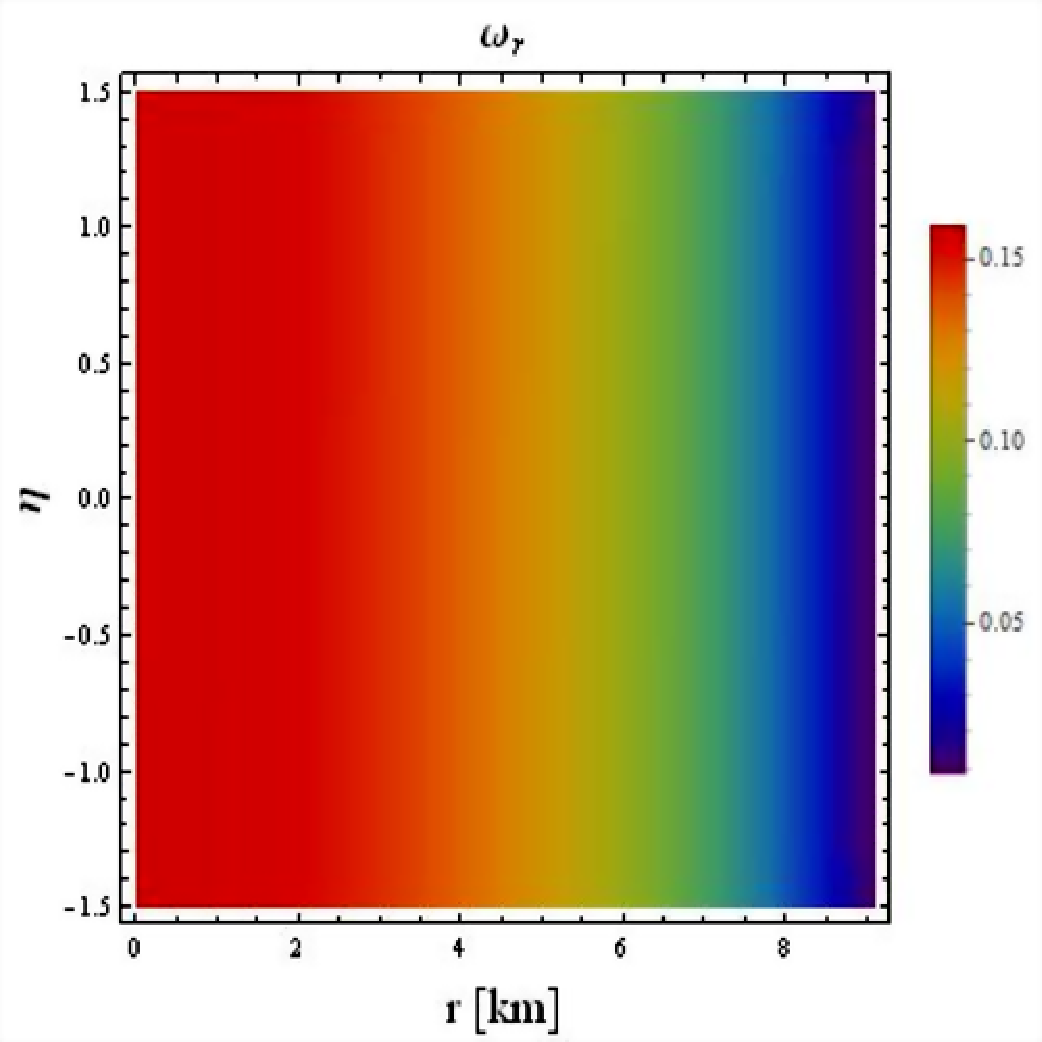,width=0.47\linewidth}\epsfig{file=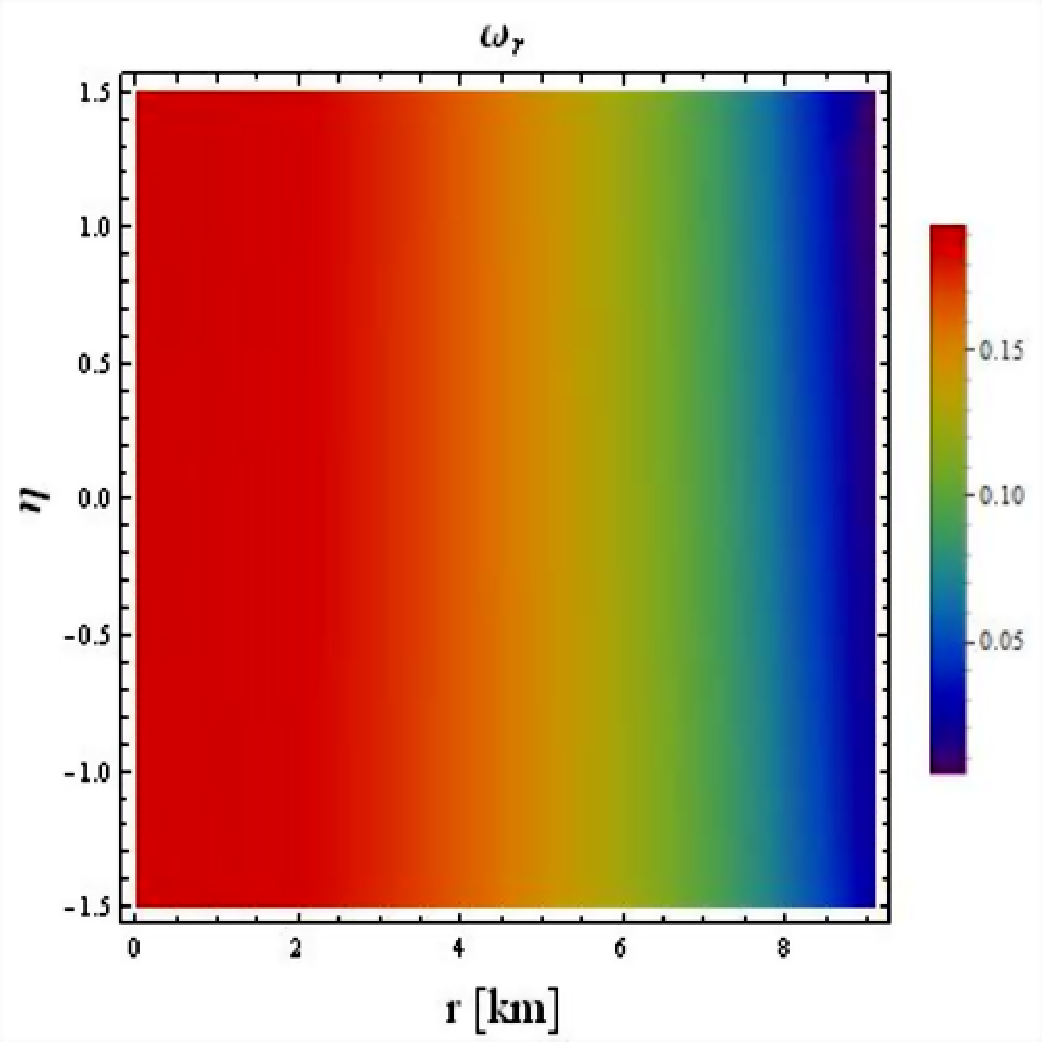,width=0.47\linewidth}
\epsfig{file=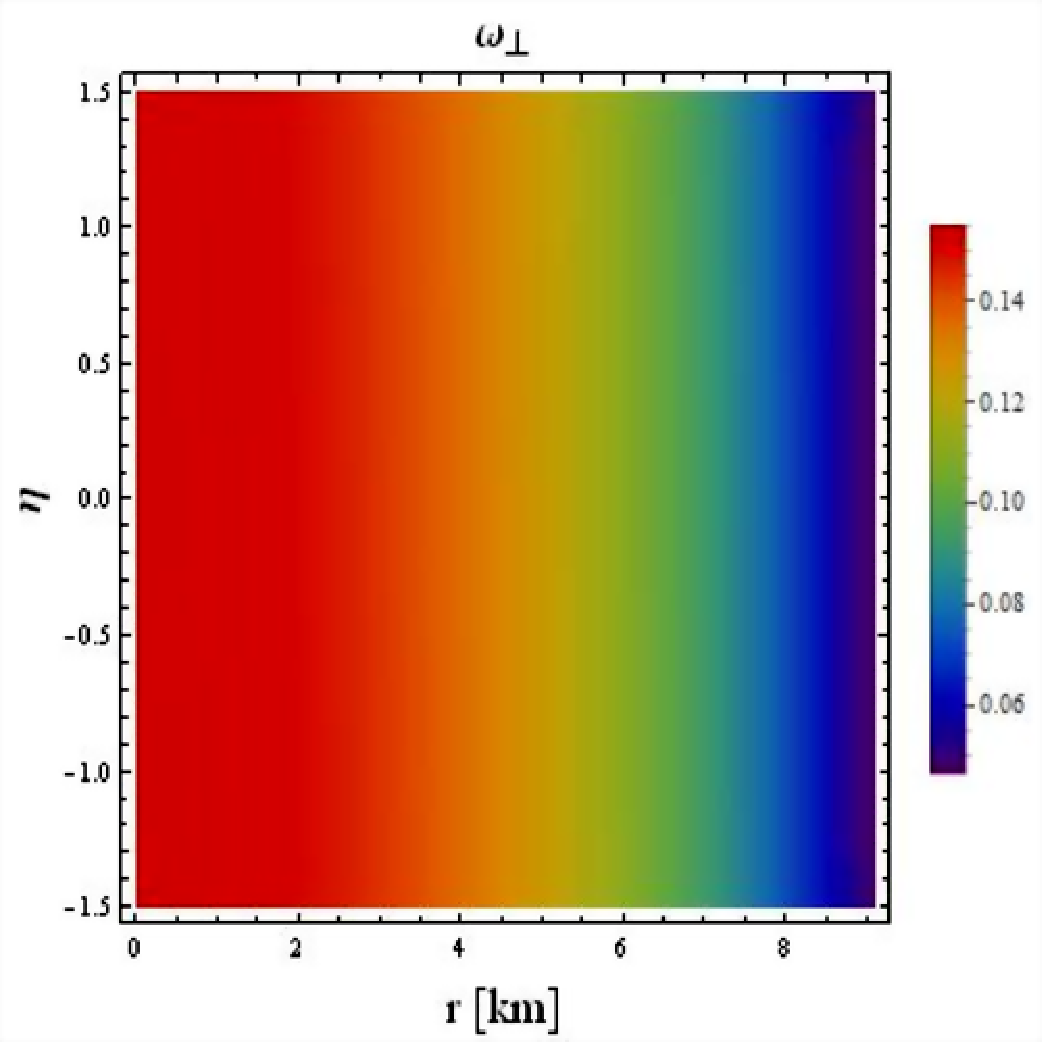,width=0.47\linewidth}\epsfig{file=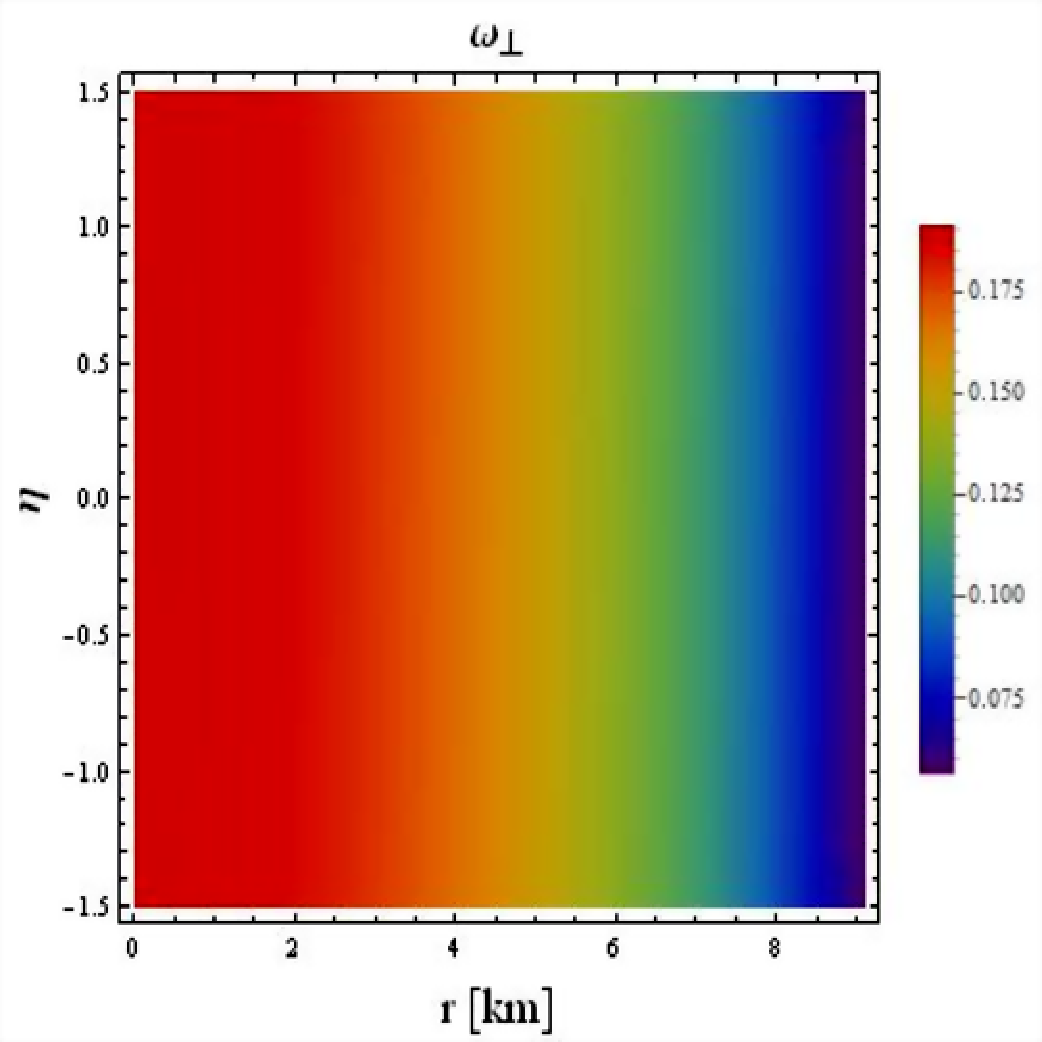,width=0.47\linewidth}
\caption{EoS parameters versus $\eta$ and $r$ for $\textbf{Model 1}$ (left)
and $\textbf{Model 2}$ (right).}
\end{figure}

\subsection{Energy Bounds}

The internal composition of celestial bodies can consist of either
conventional or exotic fluids. Whether a compact star contains a
conventional fluid depends on energy conditions determined by
physical factors reigning the fluid distribution. When examining
astronomical systems within modified theories, it is crucial to
account for these conditions, as correction terms can profoundly
influence their behavior. Therefore, adhering to the following four
types of such conditions ensures the formation of a scientifically
valid configuration
\begin{itemize}
\item Strong: $\mu+P_r+2P_\bot \geq 0$,
\item Weak: $\mu+P_r \geq 0$, \quad $\mu \geq 0$, \quad $\mu+P_\bot \geq 0$,
\item Dominant: $\mu\pm P_\bot \geq 0$, \quad $\mu\pm P_r \geq 0$,
\item Null: $\mu+P_\bot \geq 0$, \quad $\mu+P_r \geq 0$.
\end{itemize}
We observe their graphical depiction Figures \textbf{8} and
\textbf{9}. Each plot showcases consistently positive trend,
indicating that both of derived solutions meet the standards of
physical viability. This suggests the existence of normal matter
within their respective interiors.
\begin{figure}[htp!]\center
\epsfig{file=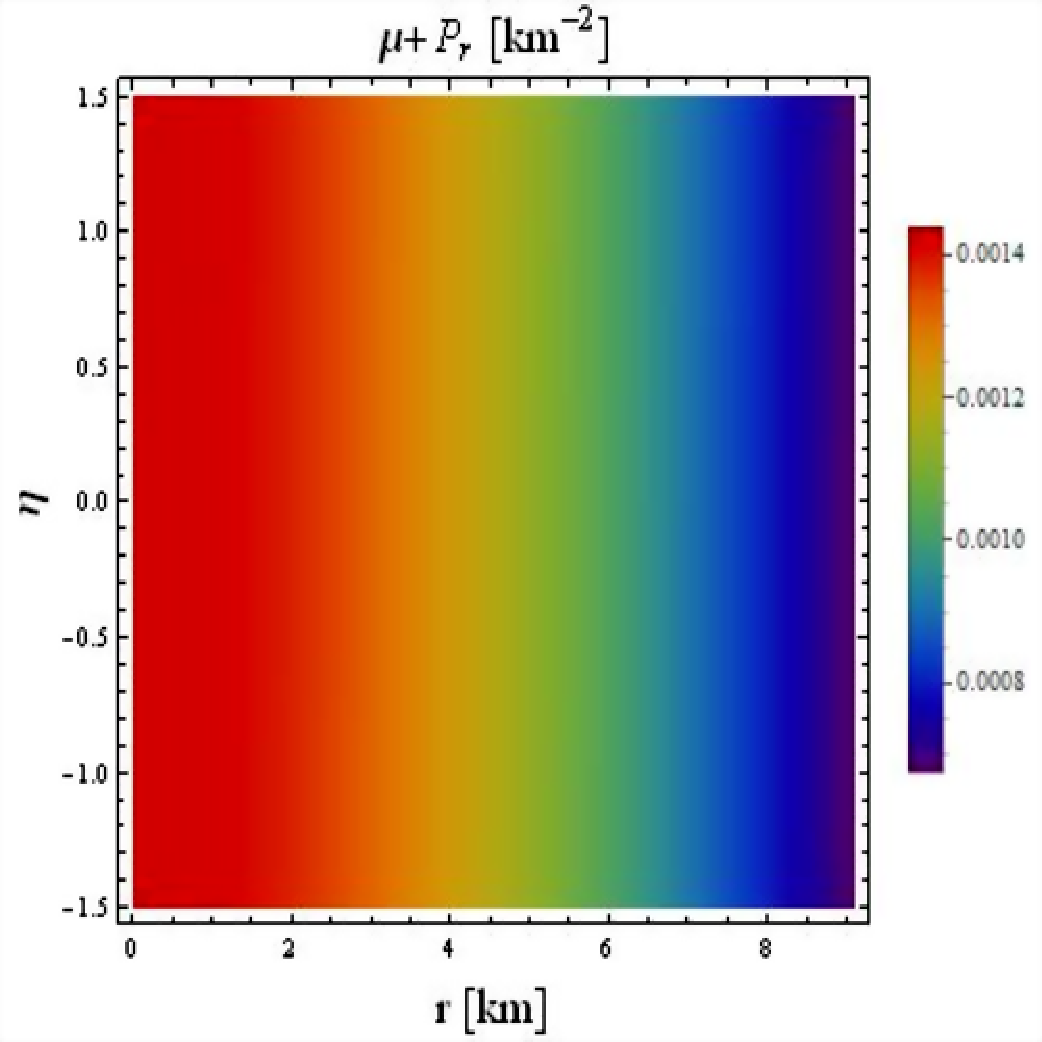,width=0.47\linewidth}\epsfig{file=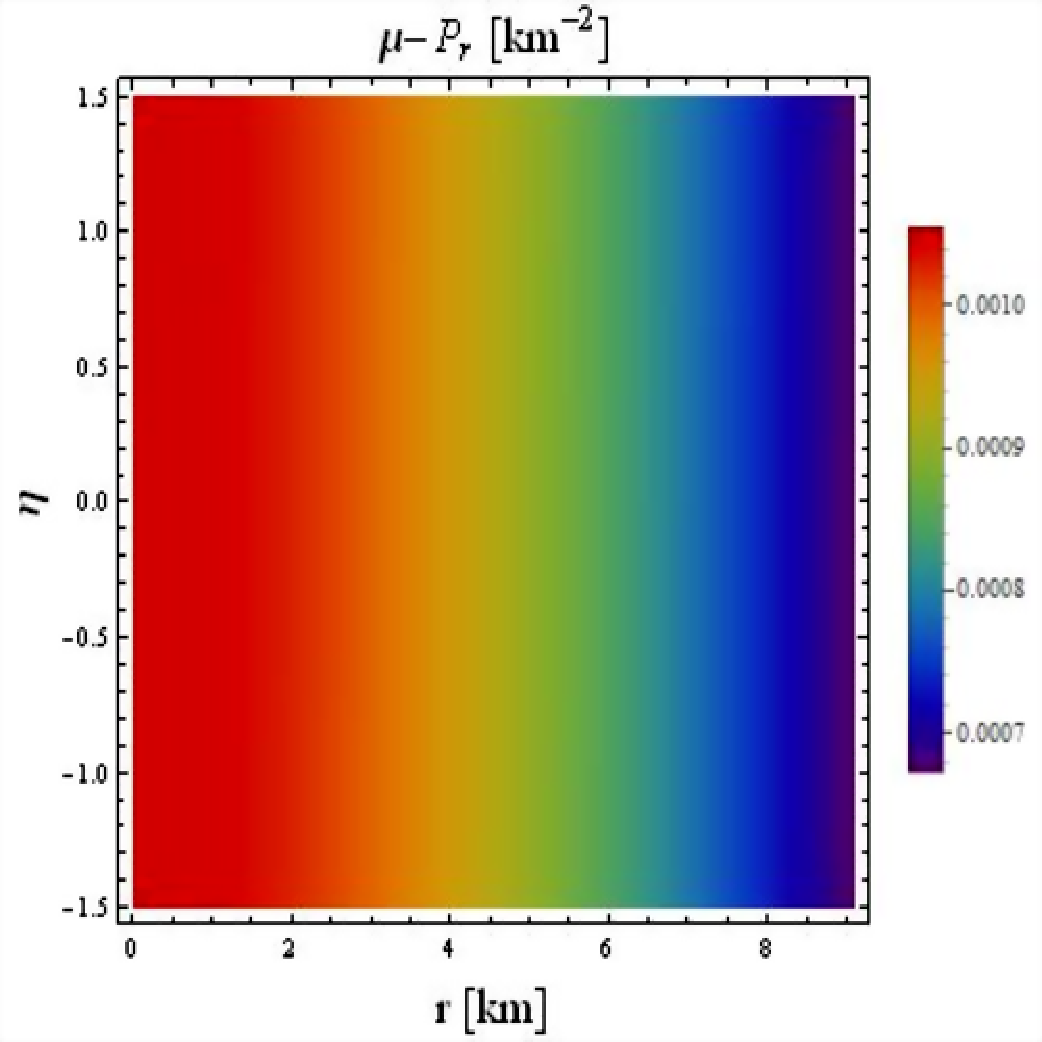,width=0.47\linewidth}
\epsfig{file=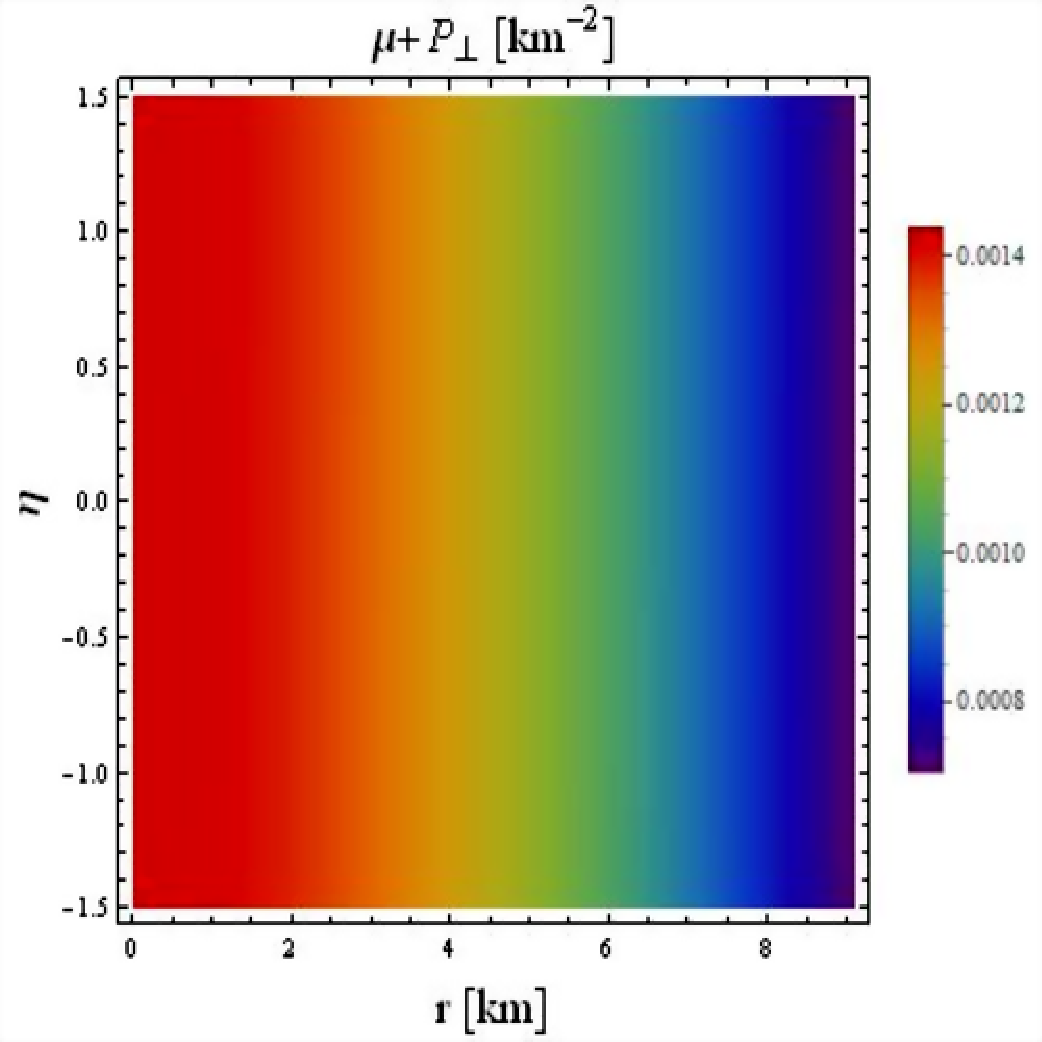,width=0.47\linewidth}\epsfig{file=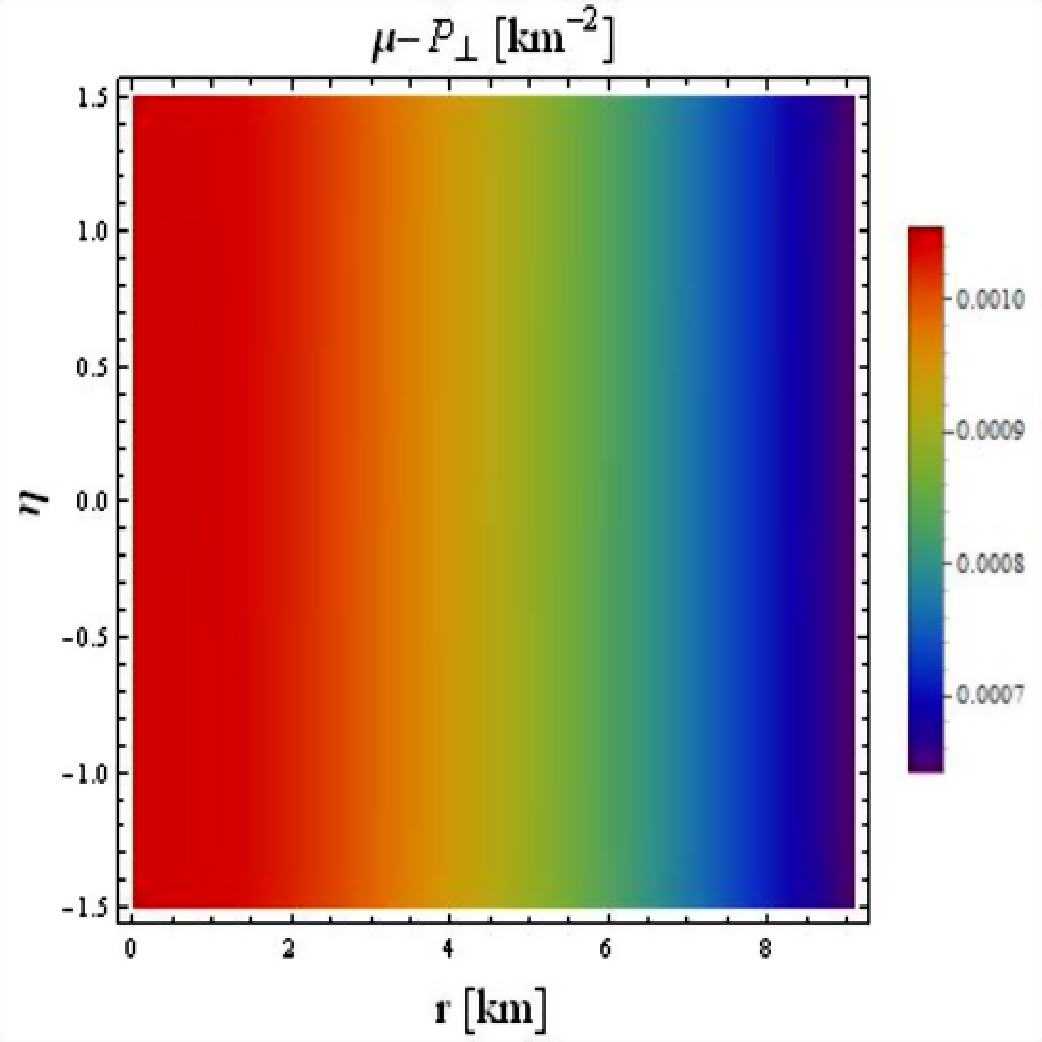,width=0.47\linewidth}
\epsfig{file=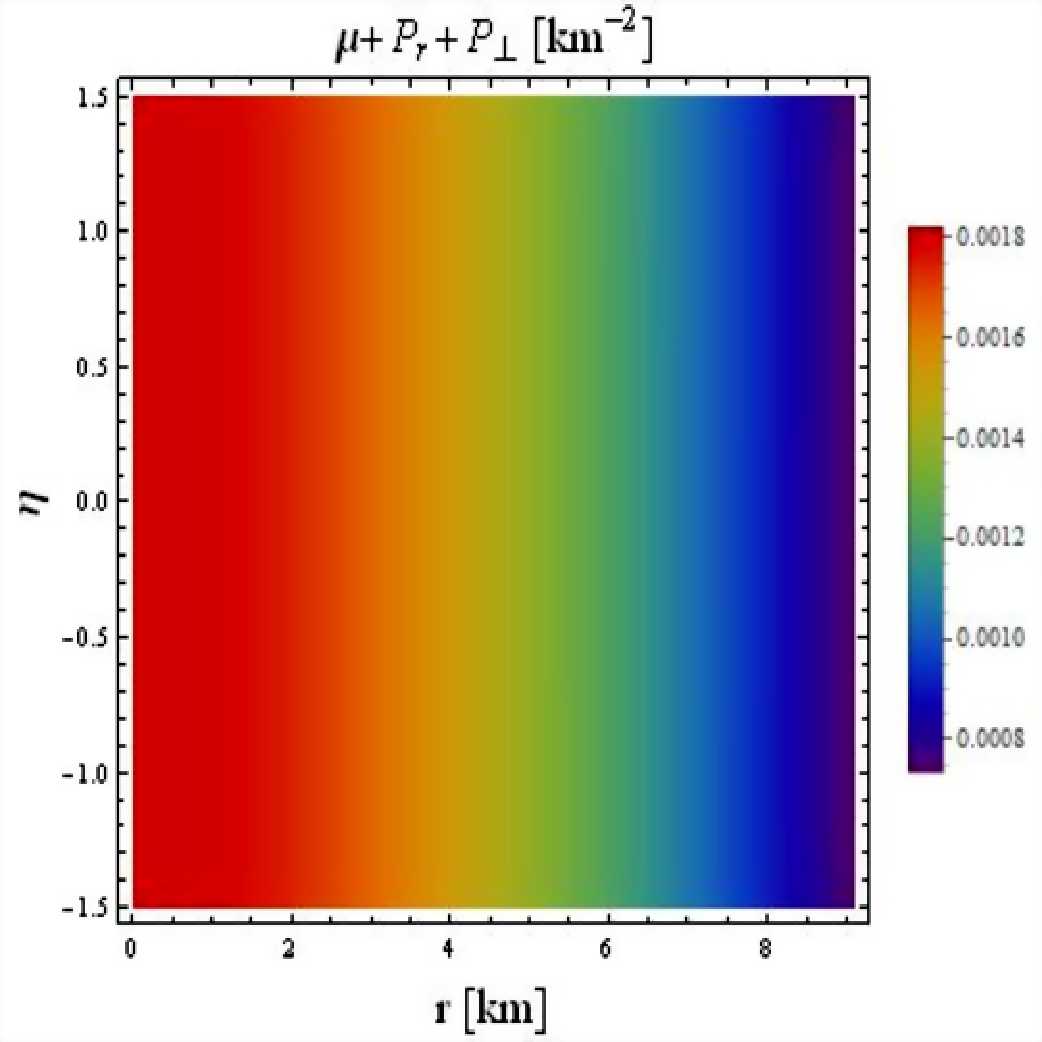,width=0.47\linewidth} \caption{Energy bounds
versus $\eta$ and $r$ for $\textbf{Model 1}$.}
\end{figure}
\begin{figure}[htp!]\center
\epsfig{file=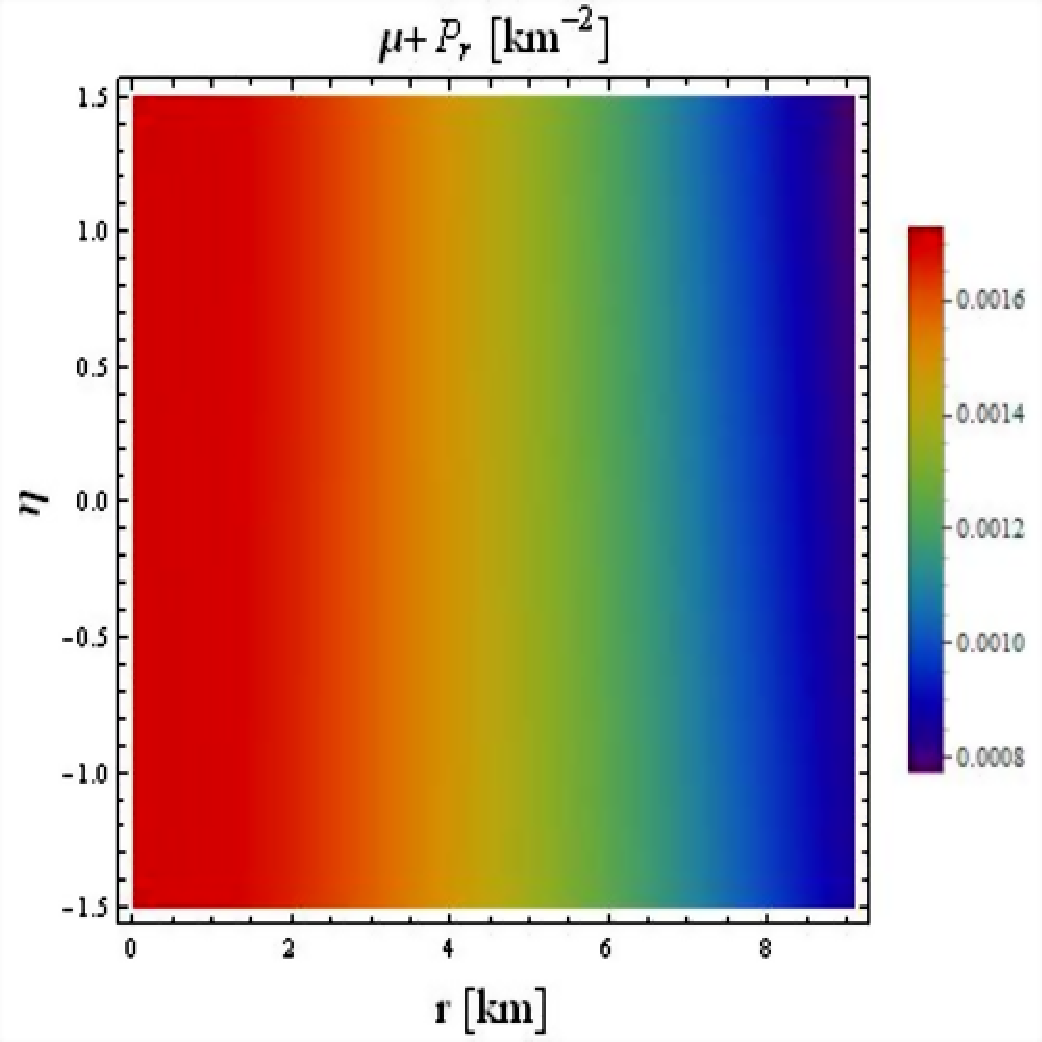,width=0.47\linewidth}\epsfig{file=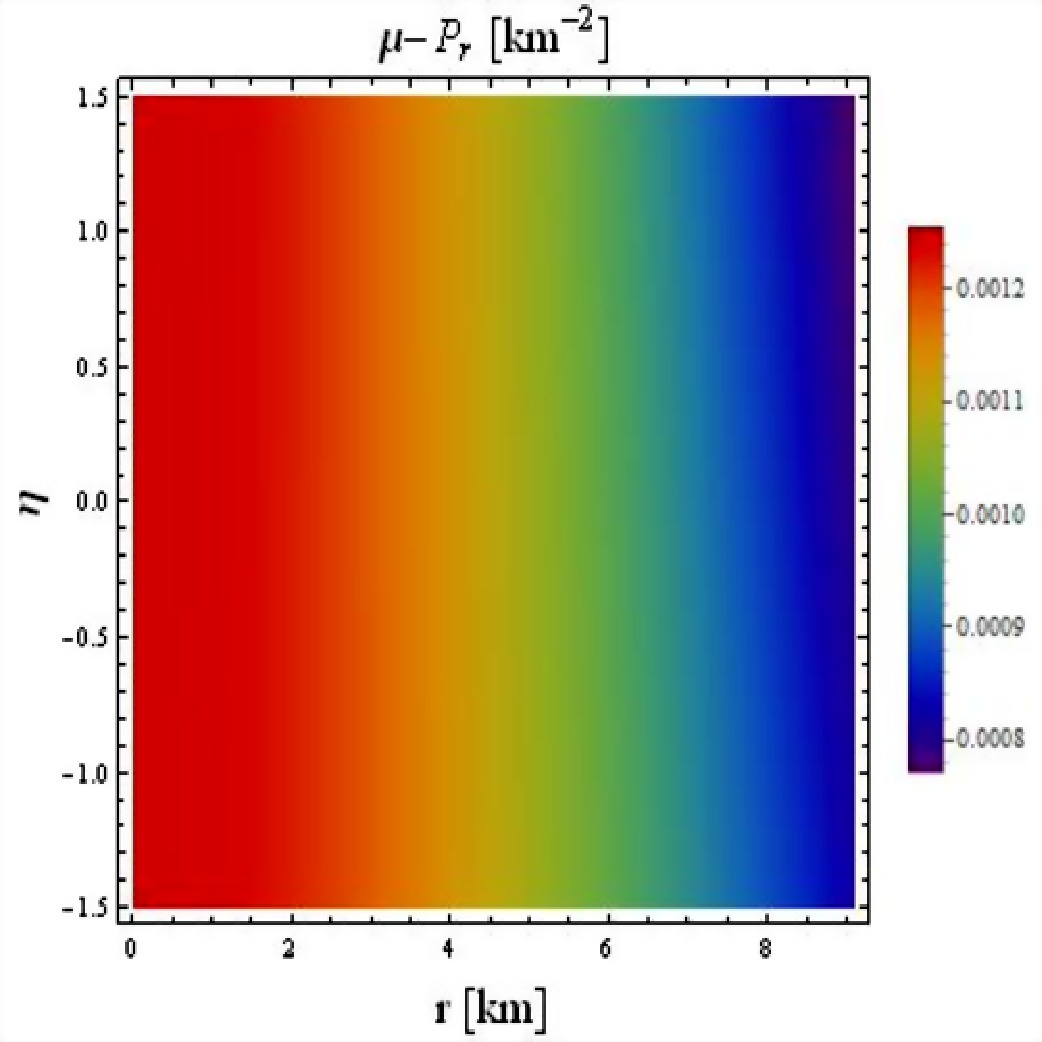,width=0.47\linewidth}
\epsfig{file=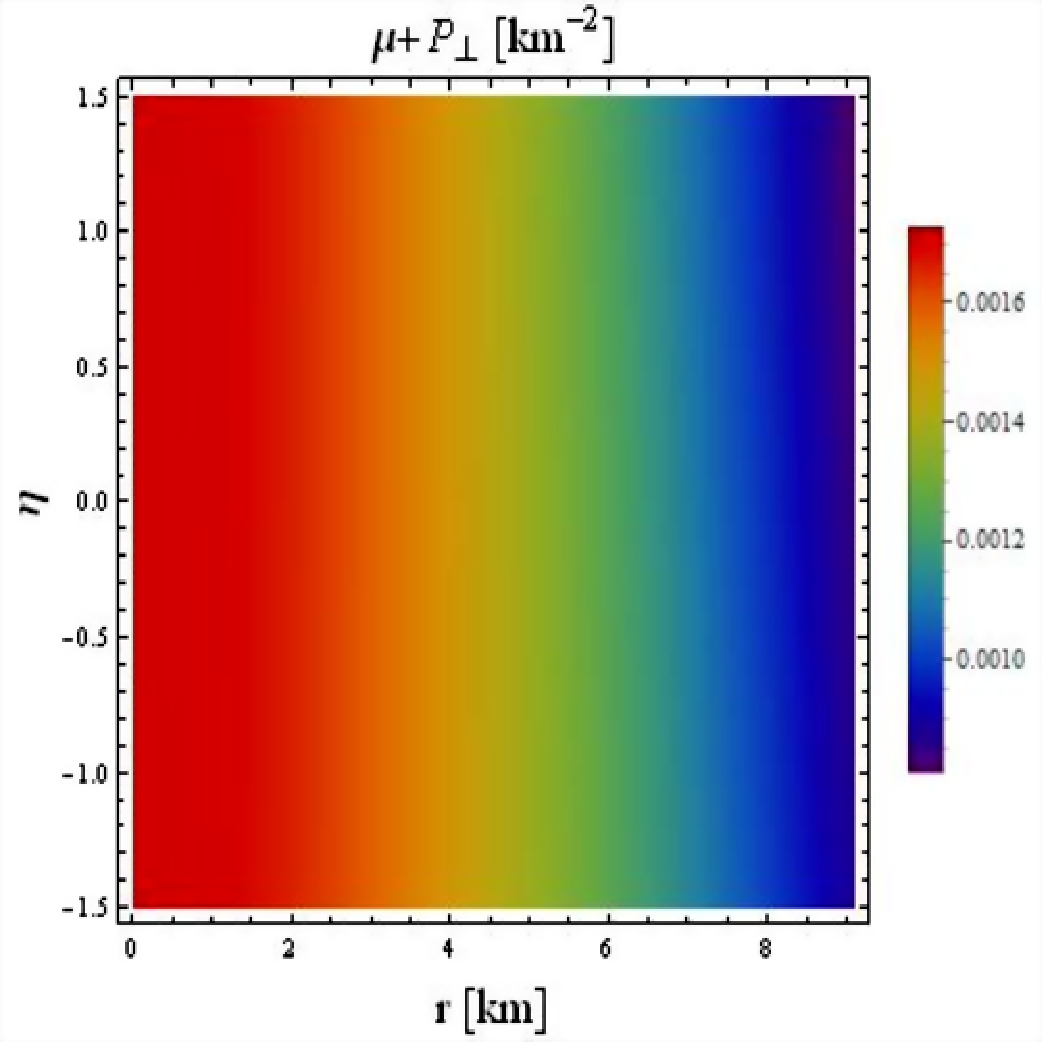,width=0.47\linewidth}\epsfig{file=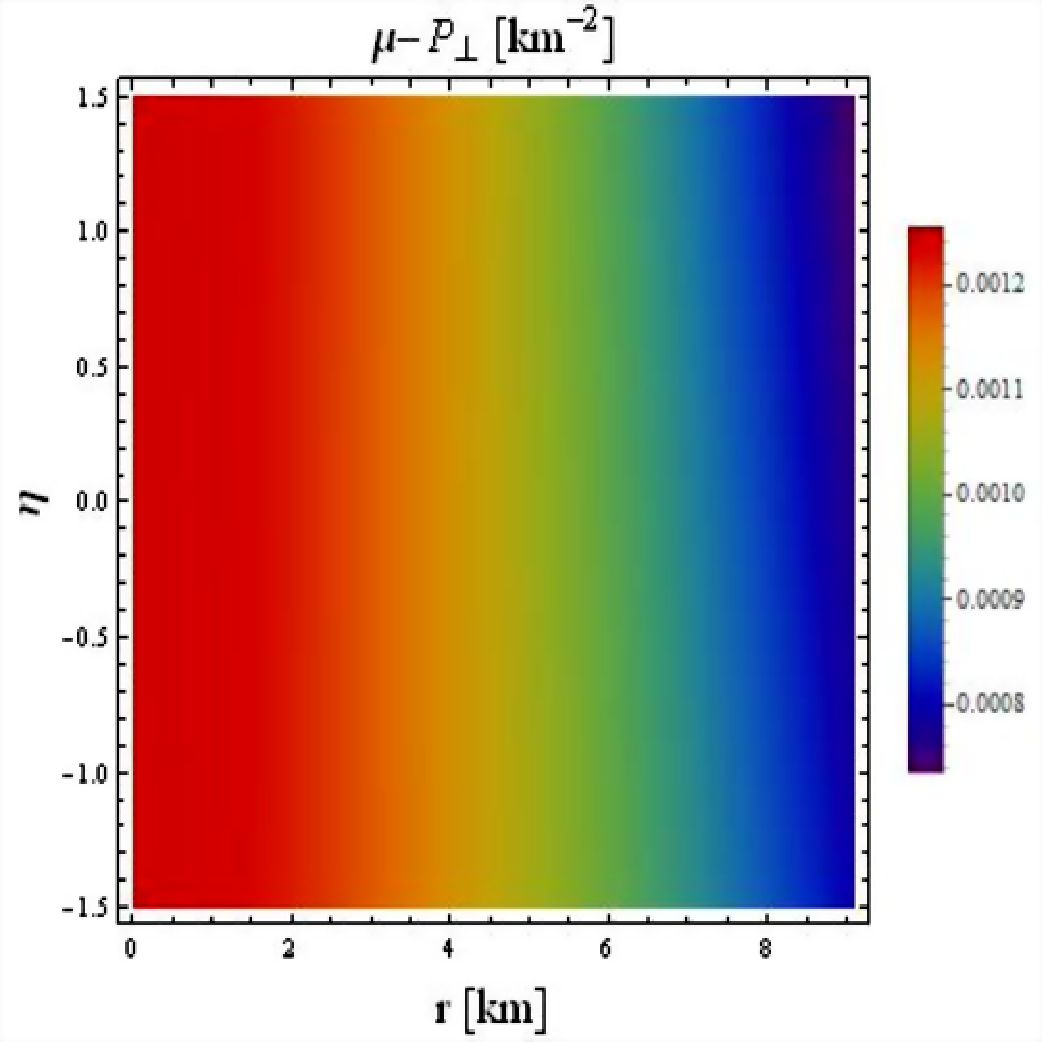,width=0.47\linewidth}
\epsfig{file=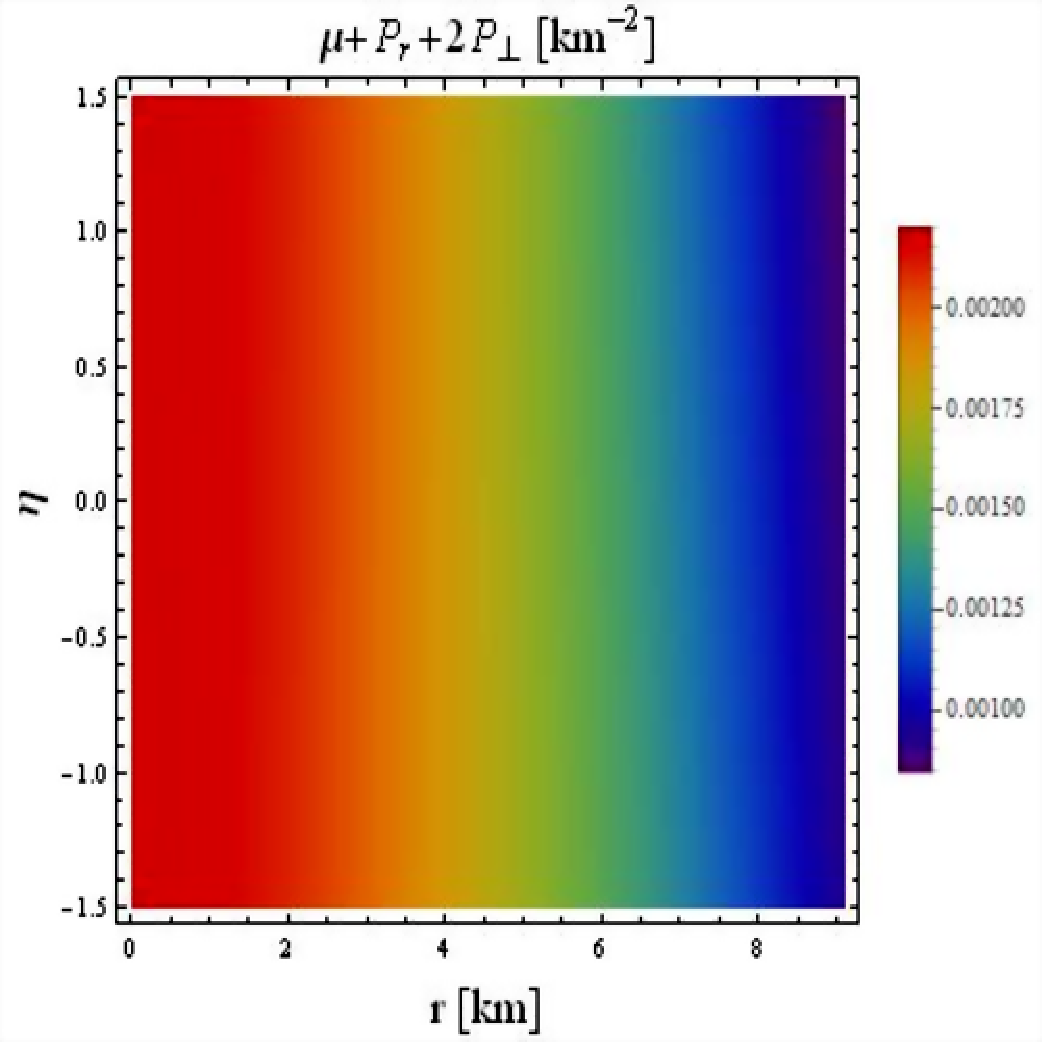,width=0.47\linewidth} \caption{Energy bounds
versus $\eta$ and $r$ for $\textbf{Model 2}$.}
\end{figure}

\subsection{Tolman-Opphenheimer-Volkoff Equation}

Examining various (fundamental) forces is crucial for understanding
the evolution of a self-gravitating object. It is essential to
assess these forces to determine if the system is in a state of
equilibrium or not \cite{37ccc,37ddd}. This can be explored by
formulating the Tolman-Oppenheimer-Volkoff (TOV) equation. In the
subsequent analysis, we compute the corresponding expression using
Eq.\eqref{g4a} for both models as
\begin{align}\nonumber
&\frac{dP_r}{dr}+\frac{A_0'}{2}\left(\mu
+P_r\right)-\frac{2}{r}\left(P_\bot-P_r\right)-\frac{2\eta
e^{-{A_1}}}{\eta\mathcal{R}+16\pi}\bigg[\frac{A_0'\mu}{8}\bigg(A_0'^2-A_0'{A_1}'+2A_0''+\frac{4A_0'}{r}\bigg)\\\nonumber
&-\frac{\mu'}{8}\bigg(A_0'^2-A_0'{A_1}'+2A_0''-\frac{4A_0'}{r}-\frac{8e^{A_1}}{r^2}+\frac{8}{r^2}\bigg)
+P_r\bigg(\frac{5A_0'^2{A_1}'}{8}-\frac{5A_0'{A_1}'^2}{8}+\frac{7A_0''{A_1}'}{4}\\\nonumber
&-A_0'A_0''+\frac{A_0'{A_1}''}{2}-\frac{5{A_1}'^2}{2r}-\frac{A_0'''}{2}+\frac{2{A_1}''}{r}+\frac{A_0'{A_1}'}{r}-\frac{{A_1}'}{r^2}
-\frac{A_0''}{r}+\frac{A_0'}{r^2}+\frac{2e^{A_1}}{r^3}-\frac{2}{r^3}\bigg)\\\nonumber
&-\frac{P'_r}{8}\bigg(A_0'^2-A_0'{A_1}'+2A_0''-\frac{4{A_1}'}{r}\bigg)+\frac{P_\bot}{r^2}\bigg({A_1}'-A_0'+\frac{2e^{A_1}}{r}
-\frac{2}{r}\bigg)-\frac{P'_\bot}{r}\bigg(\frac{{A_1}'}{2}-\frac{A_0'}{2}\\\label{g11}
&+\frac{e^{A_1}}{r}-\frac{1}{r}\bigg)\bigg]=0,
\end{align}
and
\begin{align}\nonumber
&\frac{dP_r}{dr}+\frac{A_0'}{2}\left(\mu
+P_r\right)-\frac{2}{r}\left(P_\bot-P_r\right)-\frac{2\eta}{\eta\mathcal{R}^2+16\pi}
\bigg[\mu\bigg\{\frac{e^{-A_0-{A_1}}A_0'\mathcal{R}\mathcal{R}_{00}}{2}-e^{-2{A_1}}\\\nonumber
&\times\mathcal{R}'\bigg(\frac{A_0'}{r}-\frac{e^{{A_1}}}{r^2}+\frac{1}{r^2}\bigg)\bigg\}-\mu'\bigg\{\frac{e^{-A_0-{A_1}}\mathcal{R}\mathcal{R}_{00}}{2}
-e^{-2{A_1}}\mathcal{R}\bigg(\frac{A_0'}{r}-\frac{e^{{A_1}}}{r^2}+\frac{1}{r^2}\bigg)\bigg\}\\\nonumber
&+P_r\bigg\{\mathcal{R}'\mathcal{R}^{11}+\mathcal{R}(\mathcal{R}^{11})'-e^{-{A_1}}\mathcal{R}\mathcal{R}'+\frac{e^{-2{A_1}}{A_1}'
\mathcal{R}\mathcal{R}_{11}}{2}\bigg\}-\frac{\mathcal{R}\mathcal{R}_{22}}{e^{{A_1}}}\bigg\{\frac{P'_{t}}{r^2}-\frac{2P_{\bot}}{r^3}\bigg\}\\\label{g11a}
&+P'_r\bigg\{\mathcal{R}\mathcal{R}^{11}-\frac{e^{-2{A_1}}\mathcal{R}\mathcal{R}_{11}}{2}\bigg\}\bigg]=0.
\end{align}
The geometric entities ($\mathcal{R}_{00},~\mathcal{R}_{11}$ and
$\mathcal{R}_{22}$) are expressed in Appendix \textbf{A}. We write
the above equations in the concise form as
\begin{equation}\label{g36a}
f_\texttt{a}+f_\texttt{h}+f_\texttt{y}=0,
\end{equation}
with $f_\texttt{a}$ being the anisotropic and $f_\texttt{h}$
indicates the hydrostatic, defined by
\begin{equation}\nonumber
f_\texttt{a}=\frac{2}{r}\big(P_\bot-P_r\big),\quad
f_\texttt{h}=-\frac{dP_r}{dr}.
\end{equation}
Further, the third entity is the sum of gravitational and an extra
force of this gravity theory, i.e.,
$f_\texttt{y}=f_\texttt{g}+f_\texttt{e}$. This force contains all
the remaining terms of \eqref{g11} and \eqref{g11a} along with a
negative sign. These forces are plotted in Figure \textbf{10},
guaranteing the developed interiors to be in hydrostatic
equilibrium.
\begin{figure}\center
\epsfig{file=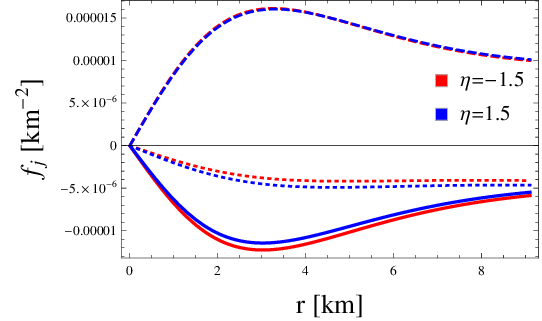,width=0.47\linewidth}\epsfig{file=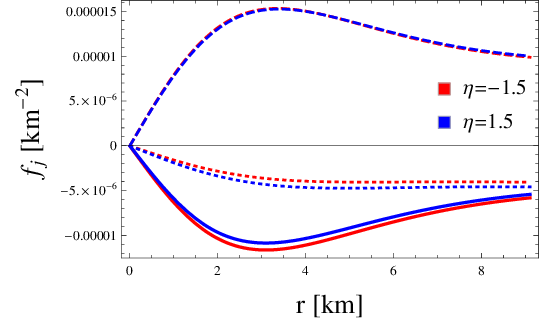,width=0.47\linewidth}
\caption{Variation in $f_\texttt{a}$ (solid), $f_\texttt{h}$
(dotted) and $f_\texttt{y}$ (dashed) versus $r$ for $\textbf{Model 1}$ (left)
and $\textbf{Model 2}$ (right).}
\end{figure}

\subsection{Stability Analysis through Different Tests}

In the expansive domain of cosmic phenomena, significant focus has
been directed towards gravitational models that meet stability
checks. Various methodologies have been documented to analyze the
stability. A crucial approach that often used in such studies is the
causality criterion \cite{42d,42da}, which asserts that the speed of
light in a stable object must be lower than that of light.
Mathematically, we have
\begin{equation}
0 < v_{s\bot}^{2}=\frac{dP_{\bot}}{d\mu}, ~
v_{sr}^{2}=\frac{dP_{r}}{d\mu} < 1.
\end{equation}
Herrera \cite{42e} put forward this approach and merged both sound
speeds into a single expression. He claimed that the absence of
cracking can be assured only if $0 < \mid v_{s\bot}^{2}-v_{sr}^{2}
\mid < 1$ satisfies, leading to the stability of compact stars. In
Figure \textbf{11}, we plot all these factors which are found to be
within their respective ranges. Hence, our models are stable for
every value of $\eta$ within the chosen range. Another test is named
as the adiabatic index, which we denote by $\Gamma$, used in the
literature regarding stability analysis. According to its
definition, both the following components must be greater than
$\frac{4}{3}$ \cite{42f}. These components are
\begin{equation}\label{g62}
\Gamma_r=\frac{\mu+P_{r}}{P_{r}} \left(\frac{dP_{r}}{d\mu}\right),
\quad \Gamma_\bot=\frac{\mu+P_{\bot}}{P_{\bot}}
\left(\frac{dP_{\bot}}{d\mu}\right).
\end{equation}
Both graphs in Figure $\mathbf{12}$ satisfy the above-mentioned
limit, showing the applicability of the considered modified theory
in the framework of anisotropic stellar models.
\begin{figure}[htp!]\center
\epsfig{file=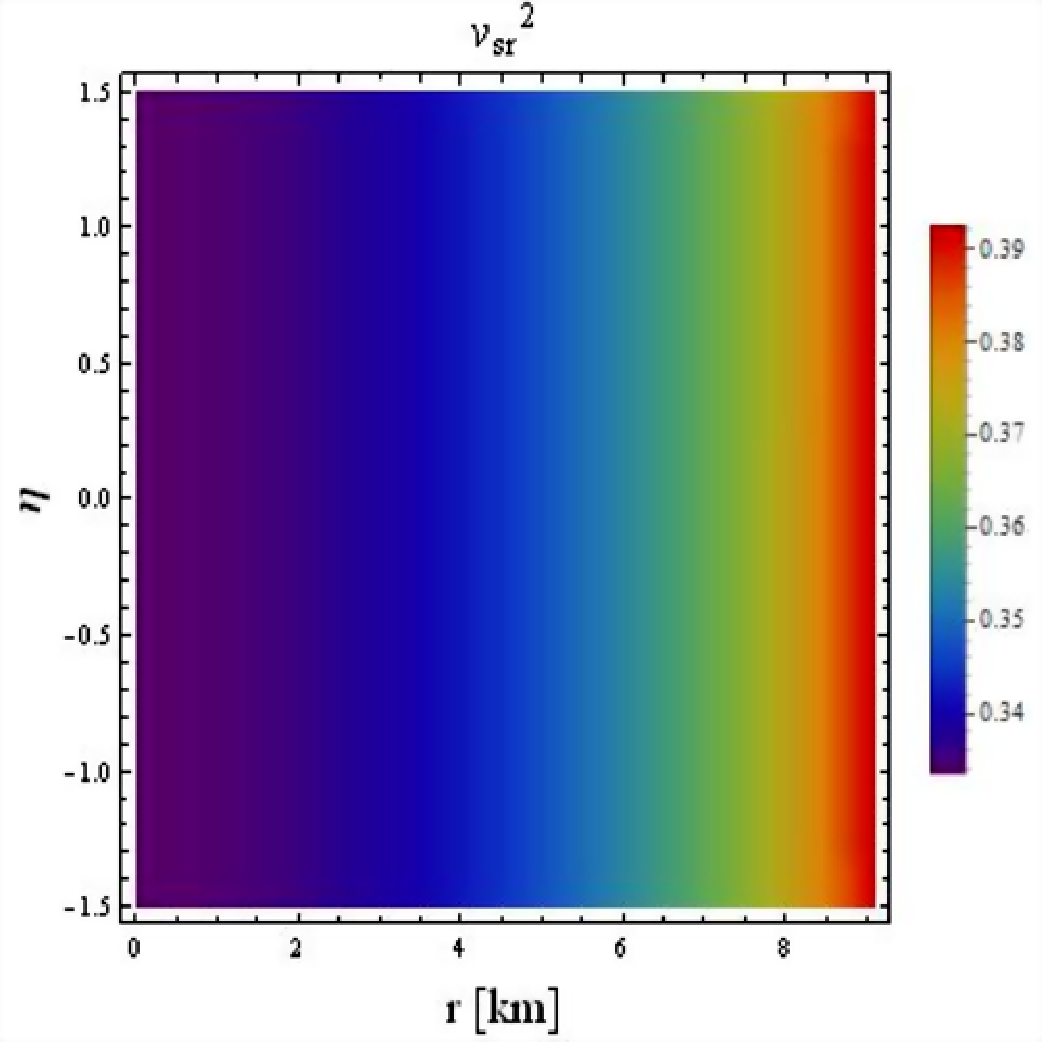,width=0.47\linewidth}\epsfig{file=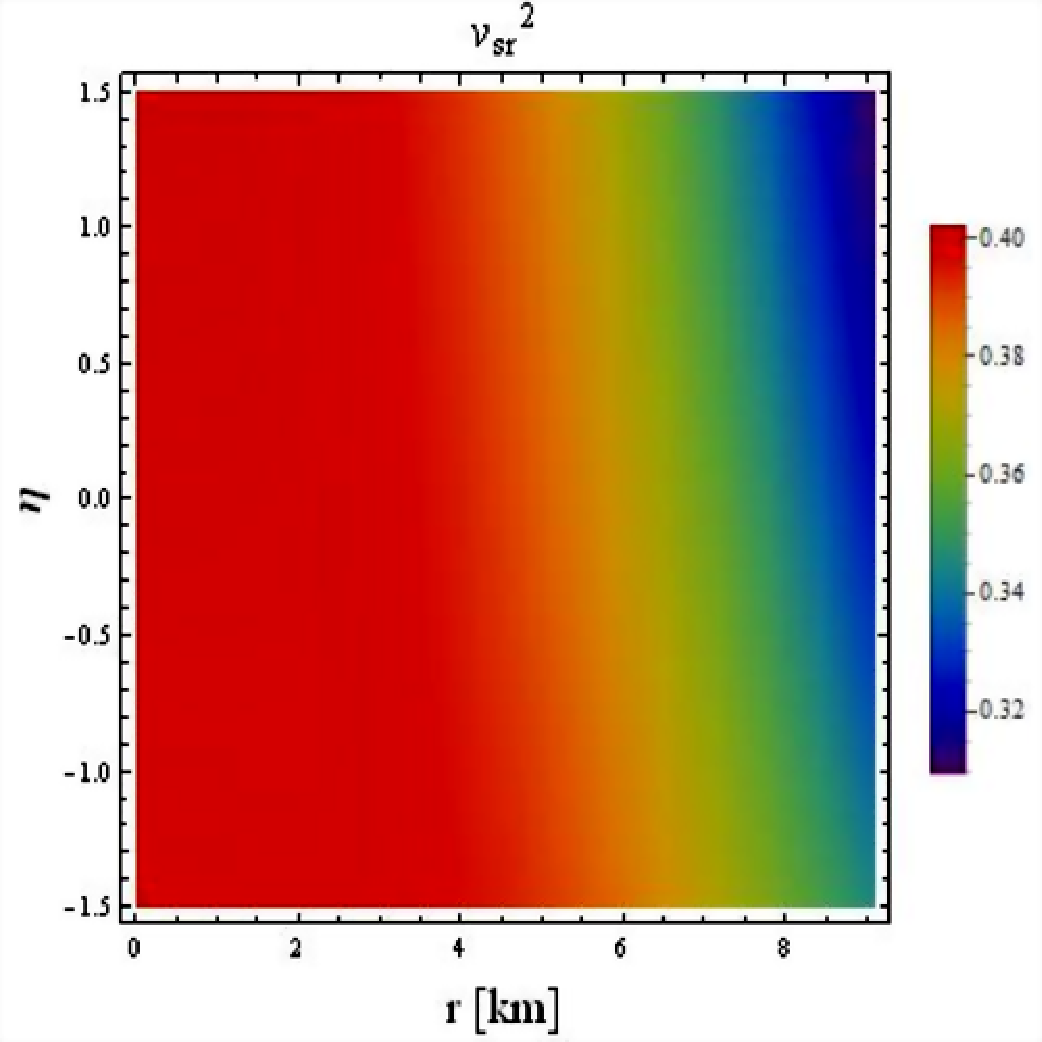,width=0.47\linewidth}
\epsfig{file=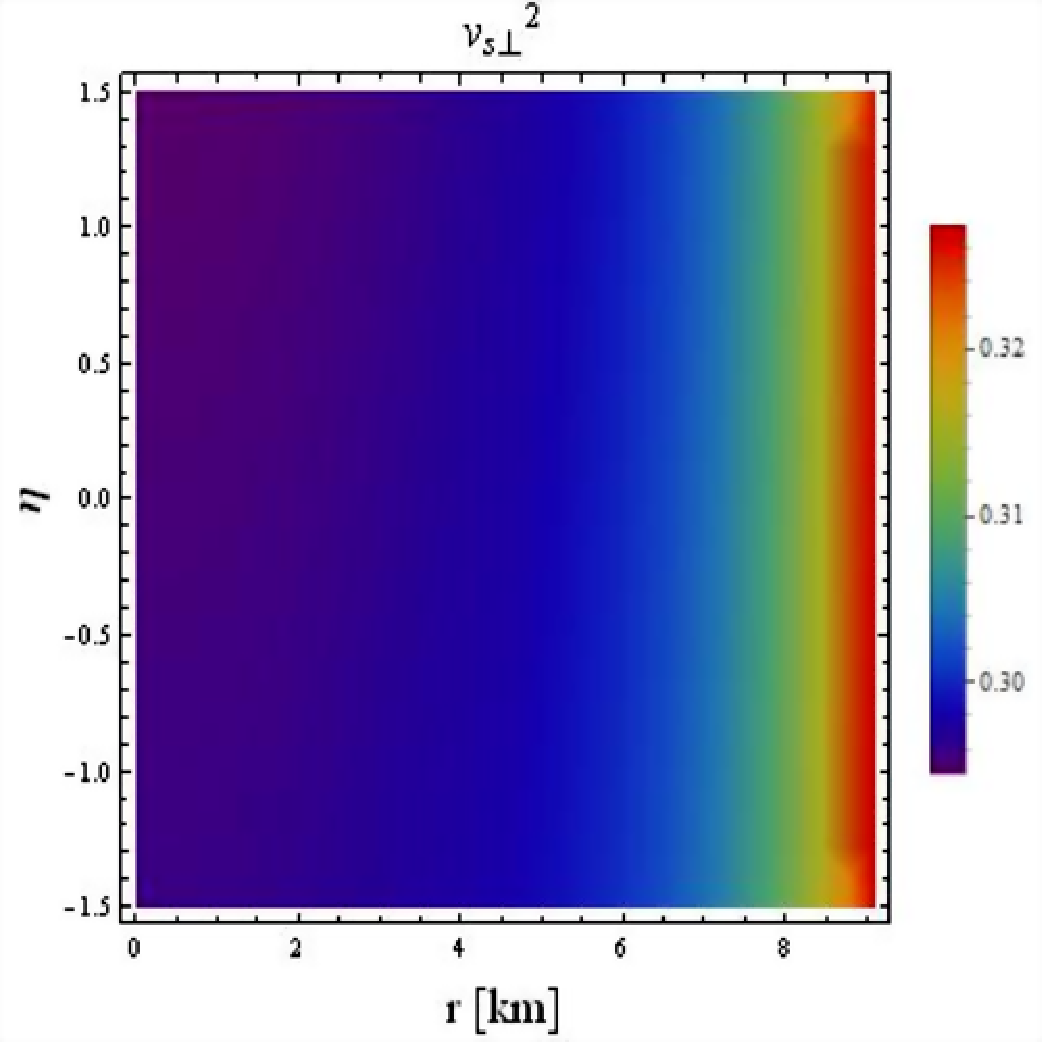,width=0.47\linewidth}\epsfig{file=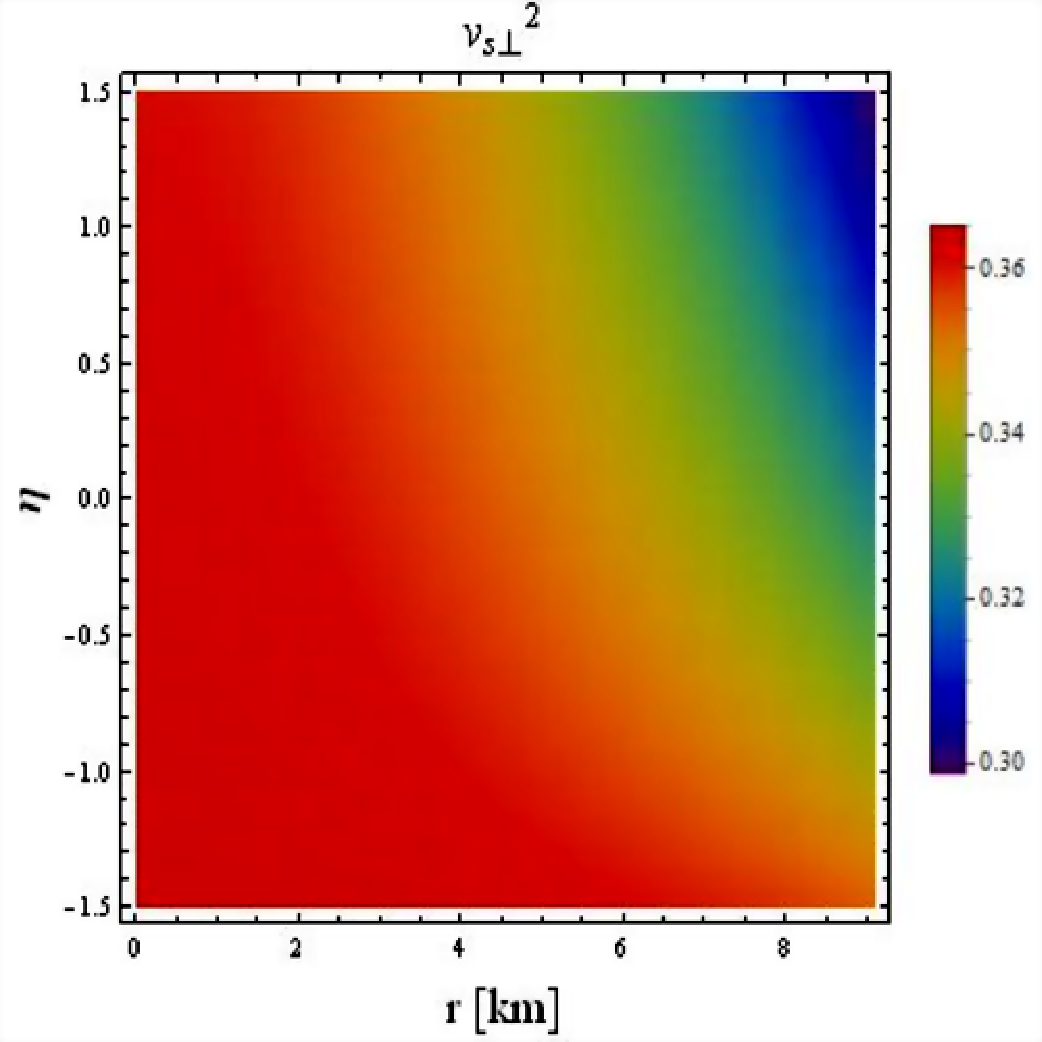,width=0.47\linewidth}
\epsfig{file=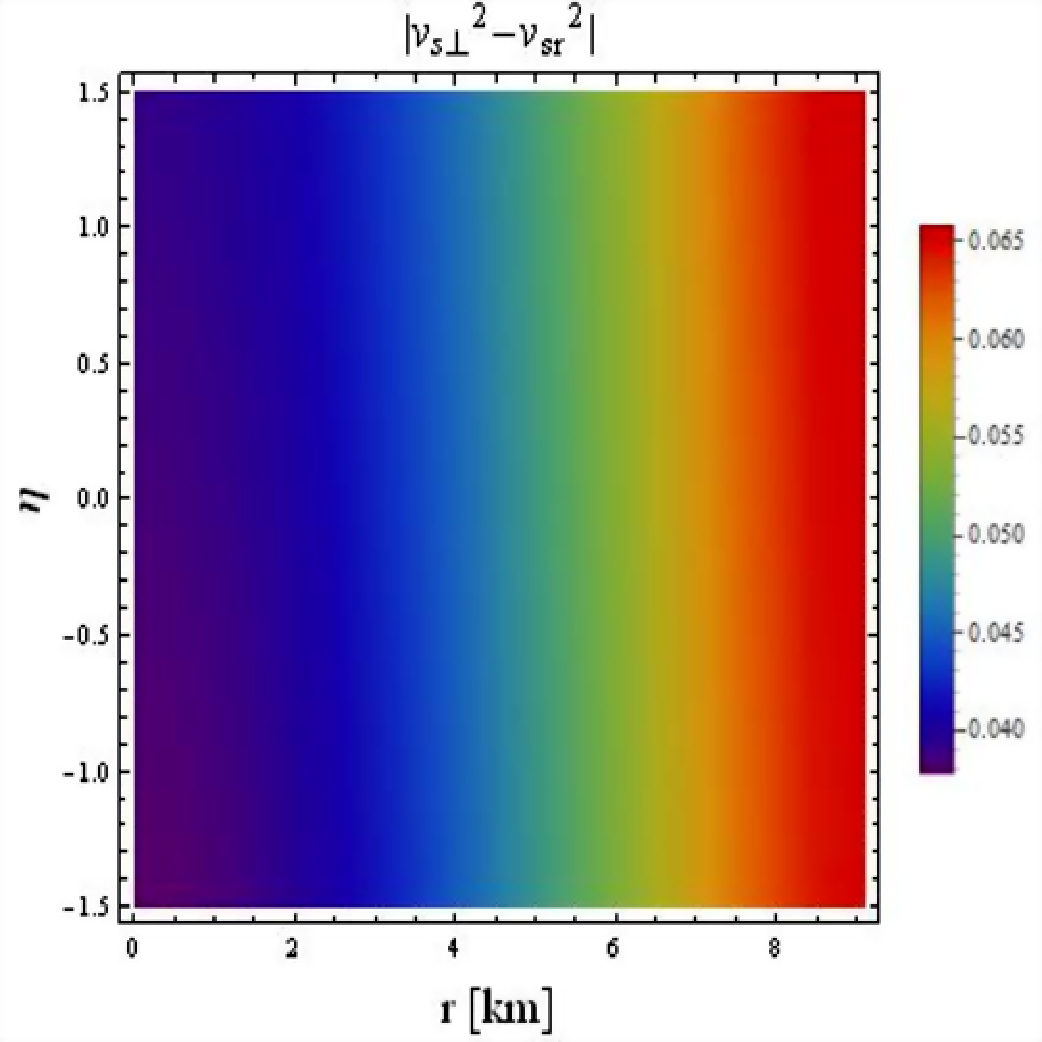,width=0.47\linewidth}\epsfig{file=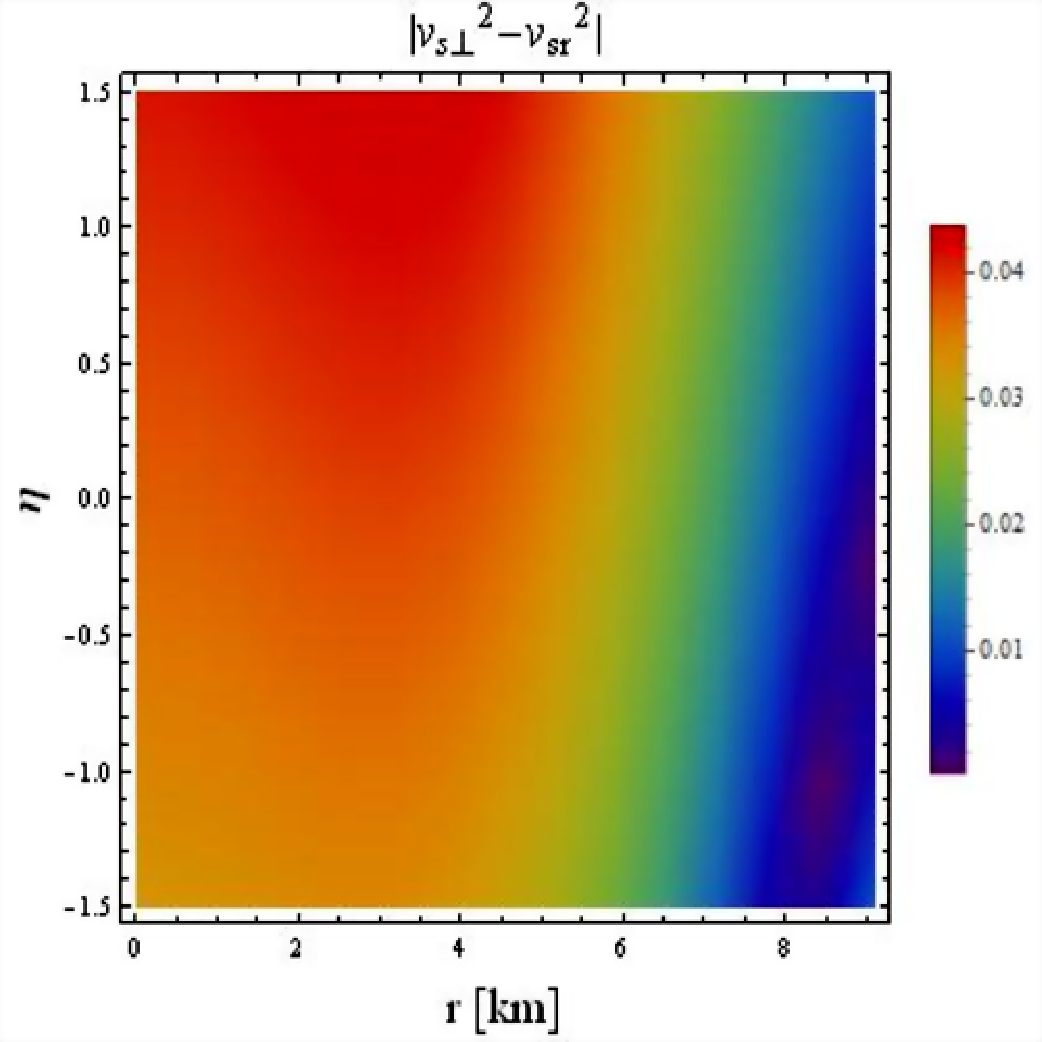,width=0.47\linewidth}
\caption{Stability checks versus $\eta$ and $r$ for $\textbf{Model 1}$ (left)
and $\textbf{Model 2}$ (right).}
\end{figure}
\begin{figure}[htp!]\center
\epsfig{file=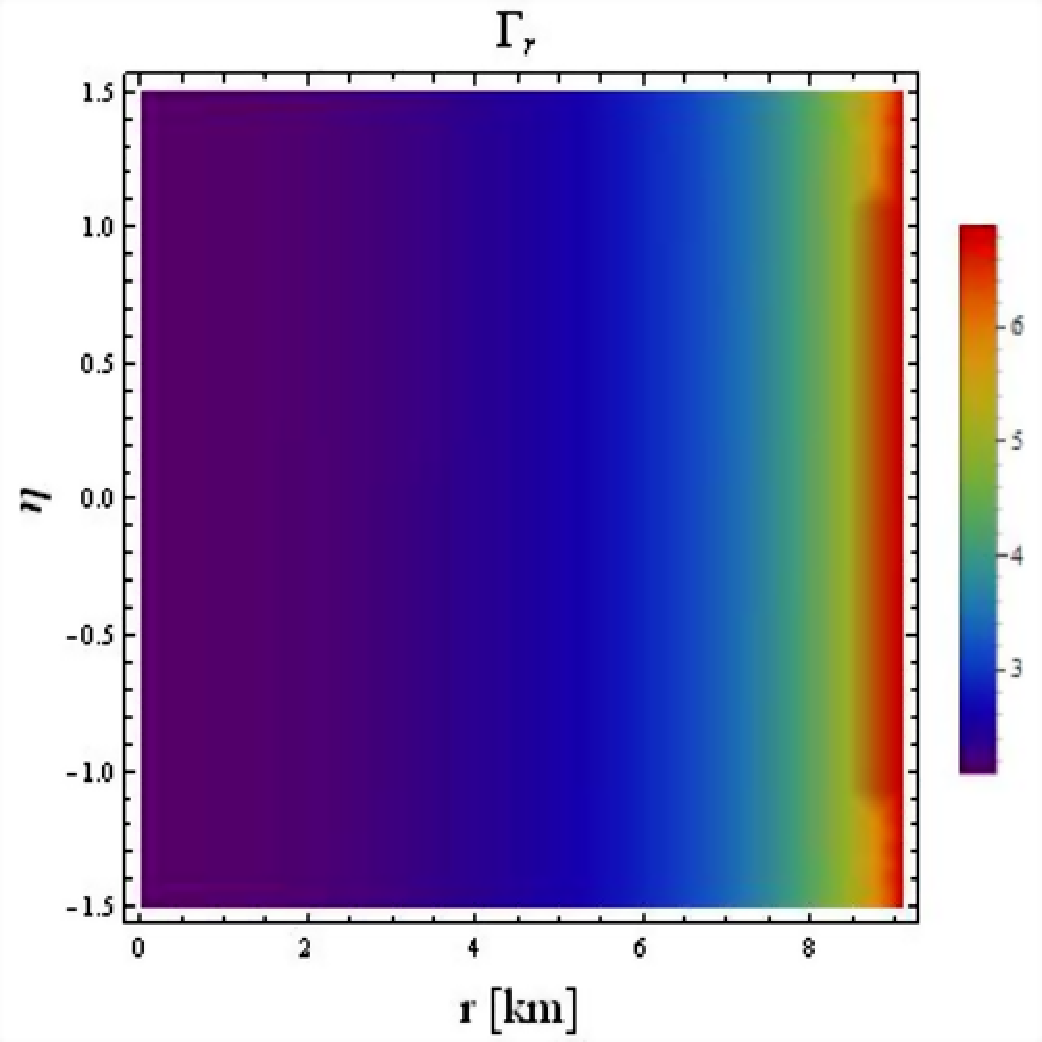,width=0.47\linewidth}\epsfig{file=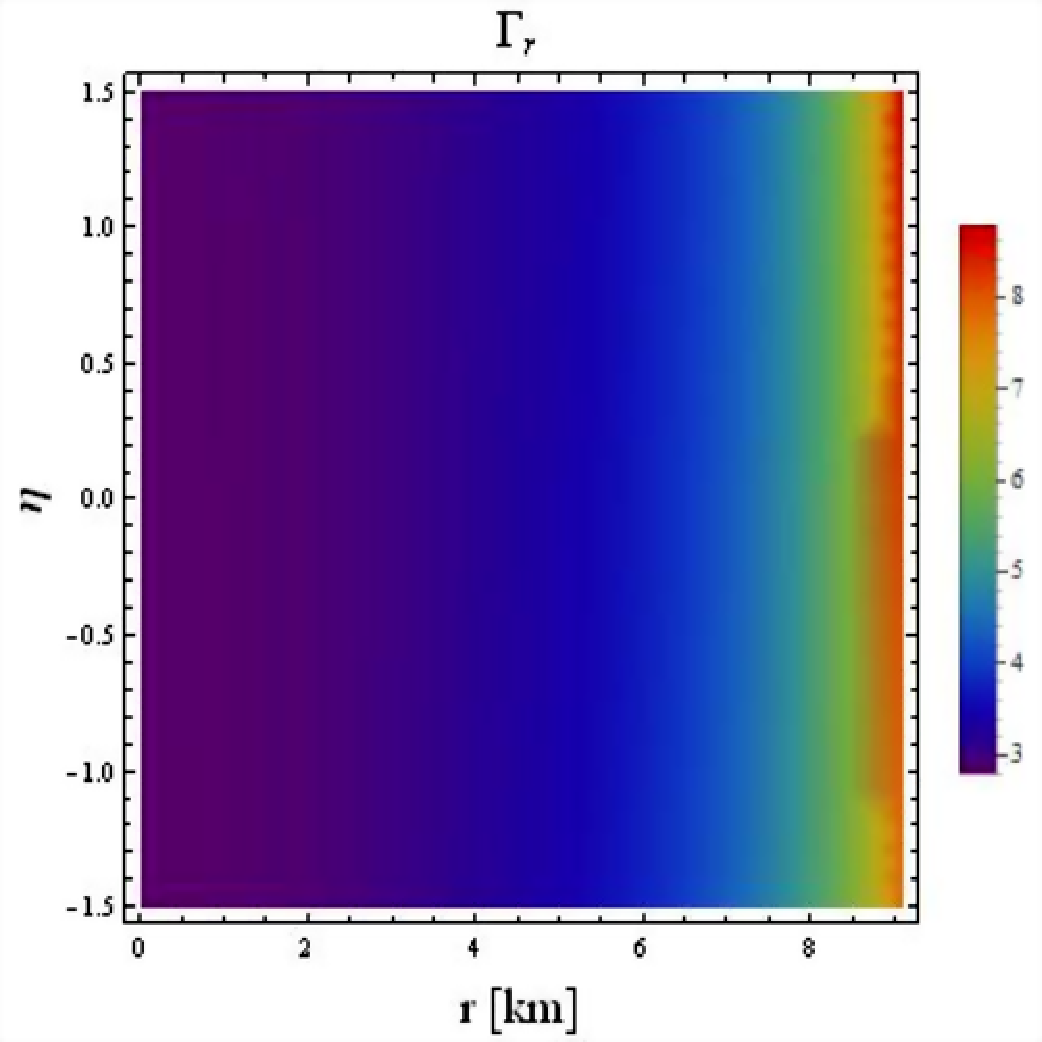,width=0.47\linewidth}
\epsfig{file=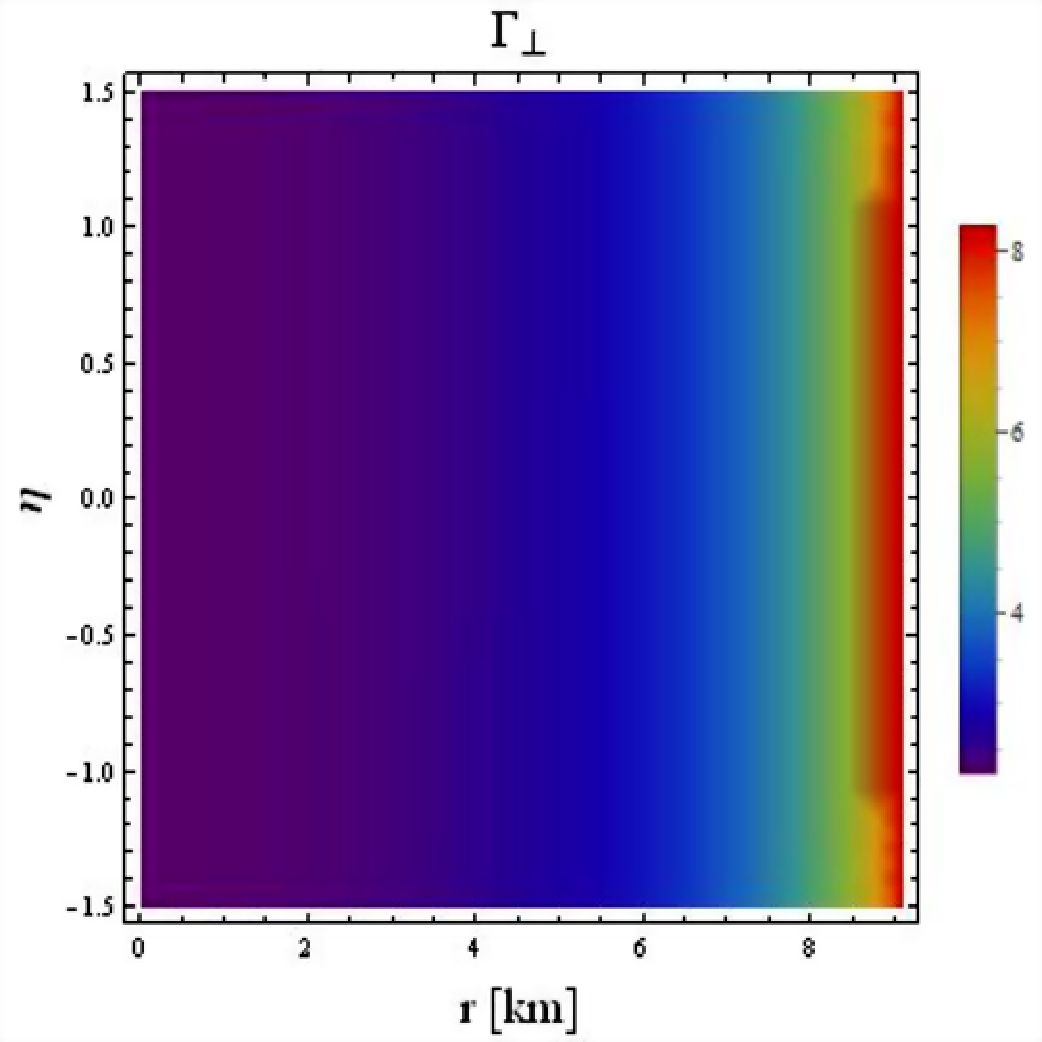,width=0.47\linewidth}\epsfig{file=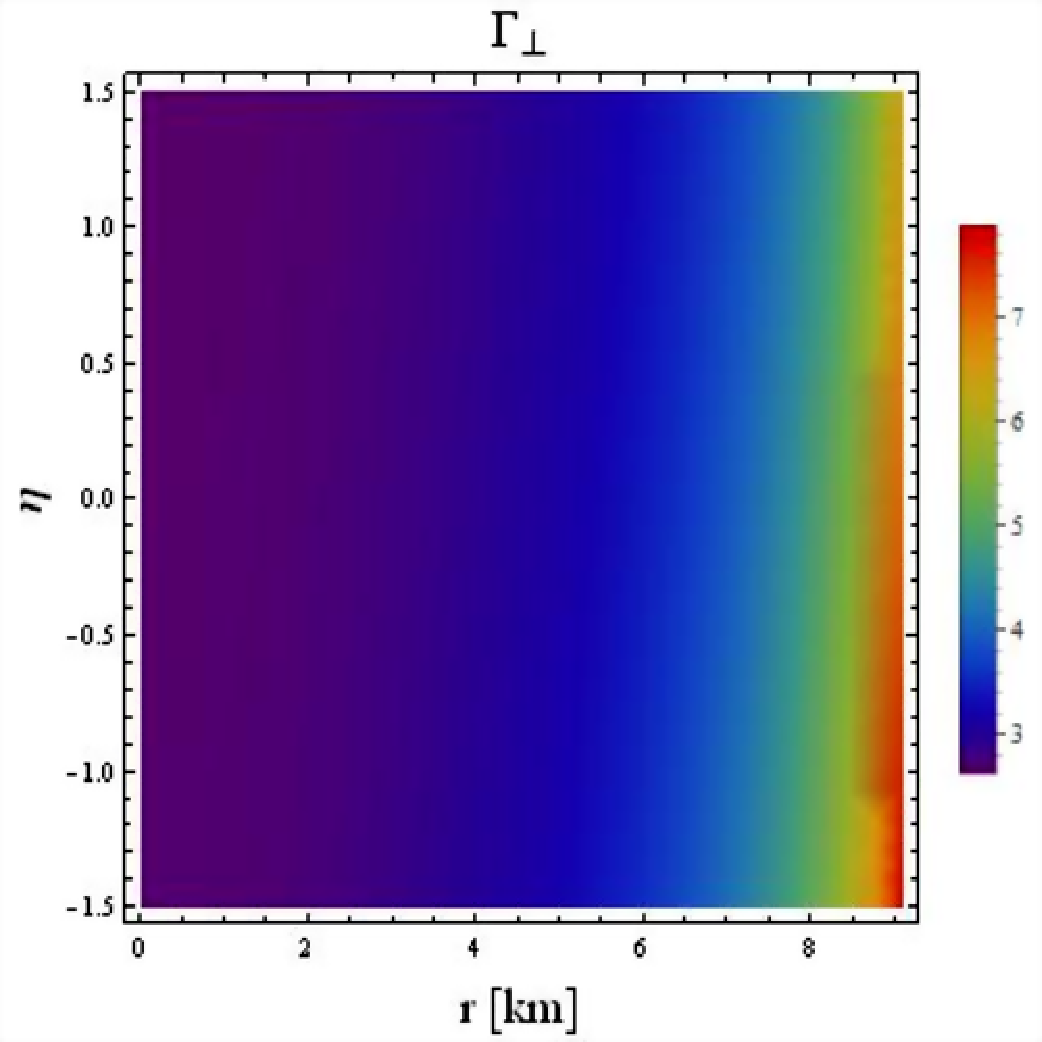,width=0.47\linewidth}
\caption{Adiabatic index versus $\eta$ and $r$ for $\textbf{Model 1}$ (left)
and $\textbf{Model 2}$ (right).}
\end{figure}

\section{Finding the Model Parameter through $P_r{_=^\Sigma}0$ Constraint}

The values of the constant $\eta$ are calculated in this section by
using preliminary information of various compact stars. We already
used first fundamental form while calculating the doublet
$(d_1,d_2)$ in the Durgapal-Fuloria metric \eqref{g15}. Now we use
the second form, resulting in the condition on the radial pressure
that becomes zero at the interface of any geometrical object, i.e.,
$P_r{_=^\Sigma}0$. Implementing this constraint on Eq.\eqref{g14c} corresponding to $\textbf{Model 1}$ as,
we have
\begin{align}\nonumber
&7 \eta \mathrm{M}^2 \emph{R} \big(79 \sqrt{\emph{R}^5 (9
\emph{R}-14 \mathrm{M})}+193 \emph{R}^3\big)-98 \eta  \mathrm{M}^3
\big(6 \sqrt{\emph{R}^5 (9 \emph{R}-14 \mathrm{M})}+11
\emph{R}^3\big)\\\nonumber &-8 B_c \emph{R}^2 \big\{56 \pi ^2
\emph{R}^8-4 \pi \eta \emph{R}^4 \big(196 \mathrm{M}^2-259
\mathrm{M} \emph{R}+93 \emph{R}^2\big)+\eta ^2 \mathrm{M} \big(2744
\mathrm{M}^3+31 \emph{R}^3\\\nonumber &-3283 \mathrm{M}^2
\emph{R}+938 \mathrm{M} \emph{R}^2\big)\big\}-40 \pi \emph{R}^5
\big(\sqrt{\emph{R}^5 (9 \emph{R}-14 \mathrm{M})}-3
\emph{R}^3\big)+\mathrm{M} \emph{R}^2 \big\{28 \pi\\\label{g26}
&\times \big(3 \emph{R}^2 \sqrt{\emph{R}^5 (9 \emph{R}-14
\mathrm{M})}-8 \emph{R}^5\big)-131 \eta \sqrt{\emph{R}^5 (9
\emph{R}-14 \mathrm{M})}-413 \eta \emph{R}^3\big\}=0.
\end{align}

On the other hand, combining the same condition with \eqref{g14f}, we can get the constraints for $\textbf{Model 2}$, which we are left due to a very lengthy expression. However, both constraints are used
to find the numerical values of $\eta$ for different choices of the
bag constant corresponsing to $\textbf{Model 1}$ and $\textbf{Model 2}$. We specify their values in Tables \textbf{III} and
\textbf{IV}, proving $\eta$ to be a real-valued parameter.

\begin{table}[H]
\scriptsize \centering \caption{Parametric values of $\eta$
corresponding to model 1.} \label{Table3} \vspace{+0.1in}
\setlength{\tabcolsep}{0.68em}
\begin{tabular}{cccccc}
\hline\hline  Values of $\mathrm{B_c}$ & & & $73~MeV/fm^3$ &
$83~MeV/fm^3$ & $93~MeV/fm^3$
\\\hline Star Candidates & Mass ($\mathrm{M}_{\bigodot}$) & Radius (km) & $\eta$ & $\eta$ & $\eta$
\\\hline Cen X-3 & 1.49 & 9.51 & -9.46 & 3.58 & 16.45
\\\hline
SMC X-4 & 1.29 & 8.83 & -235.61 & -221.391 & -207.36
\\\hline
Her X-1 & 0.85 & 8.1 & -826.024 & -525.799 & -288.779
\\\hline
4U 1820-30 & 1.58 & 9.1 & -243.39 & -222.84 & -202.57
\\\hline
4U 1608-52 & 1.74 & 9.3 & -164.96 & -154.66 & -144.51
\\\hline
PSR J 1614 2230 & 1.97 & 10.3 & 130.28 & 140.26 & 143.01
\\\hline
PSR J 1903+327 & 1.67 & 9.82 & 43.32 & 55.06 & 66.64
\\\hline
SAX J 1808.4-3658 & 0.9 & 7.95 & -447.55 & -428.58 & -209.85  \\
\hline\hline
\end{tabular}
\end{table}
\begin{table}[H]
\scriptsize \centering \caption{Parametric values of $\eta$
corresponding to model 2.} \label{Table4} \vspace{+0.1in}
\setlength{\tabcolsep}{0.68em}
\begin{tabular}{cccccc}
\hline\hline  Values of $\mathrm{B_c}$ & & & $73~MeV/fm^3$ &
$83~MeV/fm^3$ & $93~MeV/fm^3$
\\\hline Star Candidates & Mass ($\mathrm{M}_{\bigodot}$) & Radius (km) & $\eta$ & $\eta$ & $\eta$
\\\hline Cen X-3 & 1.49 & 9.51 & 0.28 & -3.36 & -6.57
\\\hline
SMC X-4 & 1.29 & 8.83 & 8.24 & 3.55 & -0.57
\\\hline
Her X-1 & 0.85 & 8.1 & 11.25 & 2.29 & -5.58
\\\hline
4U 1820-30 & 1.58 & 9.1 & 6.67 & 3.31 & 0.35
\\\hline
4U 1608-52 & 1.74 & 9.3 & 4.82 & 1.98 & -0.51
\\\hline
PSR J 1614 2230 & 1.97 & 10.3 & -3.11 & -5.34 & -7.31
\\\hline
PSR J 1903+327 & 1.67 & 9.82 & -1.17 & -4.18 & -6.82
\\\hline
SAX J 1808.4-3658 & 0.9 & 7.95 & 20.89 & 12.54 & 5.19  \\
\hline\hline
\end{tabular}
\end{table}
\section{Conclusions}

This study delves into investigating the existence of different
anisotropic compact models within the context of
$f(\mathcal{R},\mathcal{T},\mathcal{R}_{\zeta\gamma}\mathcal{T}^{\zeta\gamma})$
theory. The consequences of the non-minimal interaction between
geometry and matter have been explored by choosing two distinct
modified functional forms along with focusing on a specific range of
$\eta$. We formulated the modified field equations which correspond
to the adopted models and identified them as under-determined sets
of differential equations possessing high non-linearity. The
Durgapal-Fuloria spacetime, meeting specific criteria of
acceptability, have been employed to calculate solutions to these
equations. Furthermore, the internal structure of strange stars has
been described using the MIT bag model. In the context of the
Durgapal-Fuloria spacetime, defined by two unknowns $(d_1, d_2)$,
calculations have been performed at the hypersurface using matching
conditions to determine their values.

Figure \textbf{2} exhibited the graphical profile of the matter
triplet, ensuring validation of the resulting solutions \big(for
instance, \eqref{g14b}-\eqref{g14c} and
\eqref{g14e}-\eqref{g14f}\big). Analyzing the mass function within
the considered fluid setup indicated itself to be consistently
increasing function of the radial coordinate (Figure \textbf{6}). It
is noted that model 1 displays a less dense interior than model 2
for the specified range of $\eta$ (Tables \textbf{I} and
\textbf{II}). Additionally, the graphs
depicting compactness and redshift conformed to acceptable ranges.
Two parameters related with EoS have been illustrated in Figure
\textbf{7}, validating the practicality of the constructed models.
Throughout the interior, the constraints on EMT take positive
values, affirming the physical feasibility of our proposed solutions
depicted in Figures \textbf{8} and \textbf{9}. Furthermore, we have
plotted the TOV equations \eqref{g11} and \eqref{g11a} in Figure
\textbf{10}, demonstrating that the derived models are in the
hydrostatic equilibrium.

Finally, three distinct tests have been utilized to evaluate
stability. We have found the stability of the obtained solutions, as
evidenced by the observations presented in Figures \textbf{11} and
\textbf{12}, aligning with the findings in \cite{27a,38}. Notably,
it is evident that our solutions demonstrated superior efficacy
compared to the results obtained in \cite{25a}, suggesting that the
extra force of this extended theory may lead to more favorable
outcomes for the specified parametric range. Further, we employed
the zero radial pressure constraint at the spherical interface to
determine the parameter $\eta$ that align with the observed masses
and radii of an array of compact stars. For models 1 and 2, we have
presented these numerical values in Tables \textbf{III} and
\textbf{IV}, respectively, across various acceptable choices of the
bag constant. Notably, $\eta$ exhibits both negative/positive
values, signifying its nature as a real-valued parameter.
Ultimately, the parallel results in GR can be found when
substituting $\eta=0$.

\section*{Acknowledgements}
The author SKM is thankful for continuous support and encouragement from the administration of University of Nizwa.

\section*{Data Availability Statement}
This manuscript has no associated data or the data will not be deposited. (There is no observational data related to this article. The necessary calculations and graphic discussion can be made available on request.)

\section*{Appendix A}

The terms $\alpha_j^{'s}$ appeared in Eqs.\eqref{g8c}-\eqref{g8e}
are given by
\begin{align}\nonumber
\alpha_0&=\frac{e^{-{A_1}}}{4}\bigg(A_0'^2-A_0'{A_1}'+2A_0''+\frac{4A_0'}{r}\bigg),\\\nonumber
\alpha_1&=\frac{e^{-{A_1}}}{4}\bigg(A_0'{A_1}'-A_0'^2-2A_0''+\frac{4{A_1}'}{r}\bigg),\\\nonumber
\alpha_3&=e^{-{A_1}}\bigg(\frac{{A_1}'}{r}-\frac{A_0'}{r}+\frac{2e^{A_1}}{r^2}-\frac{2}{r^2}\bigg),\\\nonumber
\alpha_4&=\frac{e^{-{A_1}}}{4}\bigg(A_0'{A_1}'^2-A_0'^2{A_1}'-3A_0''{A_1}'-A_0'{A_1}''+2A_0'A_0''+2A_0'''-\frac{4A_0'}{r^2}\\\nonumber
&-\frac{4A_0'{A_1}'}{r}+\frac{4A_0''}{r}\bigg),\\\nonumber
\alpha_5&=\frac{e^{-{A_1}}}{4}\bigg(A_0'^2{A_1}'-A_0'{A_1}'^2+3A_0''{A_1}'+A_0'{A_1}''-2A_0'A_0''-2A_0'''-\frac{4{A_1}'}{r^2}\\\nonumber
&-\frac{4{A_1}'^2}{r}+\frac{4{A_1}''}{r}\bigg),\\\nonumber
\alpha_6&=e^{-{A_1}}\bigg(\frac{A_0'{A_1}'}{r}-\frac{4{A_1}'e^{A_1}}{r^2}-\frac{{A_1}'^2}{r}+\frac{{A_1}''}{r}+\frac{{A_1}'}{r^2}-\frac{A_0''}{r}
+\frac{A_0'}{r^2}-\frac{4e^{A_1}}{r^3}+\frac{4}{r^3}\bigg),\\\nonumber
\alpha_7&=\frac{e^{-{A_1}}}{4}\bigg(A_0'^2{A_1}'^2-A_0'{A_1}'^3+4A_0''{A_1}'^2+3A_0'{A_1}'{A_1}''-5A_0'''{A_1}'
-4A_0''{A_1}''\\\nonumber
&-A_0'^2{A_1}''-A_0'{A_1}'''-4A_0'A_0''{A_1}'+2A_0''^2+2A_0'A_0'''+2A_0''''+\frac{4A_0'{A_1}'^2}{r}\\\nonumber
&-\frac{8A_0''{A_1}'}{r}-\frac{4A_0'{A_1}''}{r}+\frac{4A_0'''}{r}-\frac{8A_0''}{r^2}+\frac{8A_0'}{r^3}\bigg),\\\nonumber
\alpha_8&=\frac{e^{-{A_1}}}{4}\bigg(A_0'{A_1}'^3-A_0'^2{A_1}'^2-4A_0''{A_1}'^2-3A_0'{A_1}'{A_1}''+4A_0'A_0''{A_1}'+5A_0'''{A_1}'\\\nonumber
&+A_0'^2{A_1}''+A_0'{A_1}'''+4A_0''{A_1}''-2A_0''^2-2A_0'A_0'''-2A_0''''+\frac{4{A_1}'^3}{r}-\frac{12{A_1}'{A_1}''}{r}\\\nonumber
&+\frac{8{A_1}'^2}{r^2}+\frac{4{A_1}'''}{r}-\frac{8{A_1}''}{r^2}+\frac{8{A_1}'}{r^3}\bigg),\\\nonumber
\alpha_9&=e^{-{A_1}}\bigg(\frac{{A_1}'^3}{r}-\frac{A_0'{A_1}'^2}{r}-\frac{3{A_1}'{A_1}''}{r}+\frac{4A_0''{A_1}'}{r}-\frac{2A_0'{A_1}'}{r^2}
-\frac{6{A_1}'}{r^3}+\frac{A_0'{A_1}''}{r}\\\nonumber
&-\frac{{A_1}''e^{A_1}}{r^2}+\frac{8{A_1}'e^{A_1}}{r^3}+\frac{{A_1}'''}{r}-\frac{A_0'''}{r}+\frac{2A_0''}{r^2}-\frac{2A_0'}{r^3}+\frac{12e^{A_1}}{r^4}
-\frac{12}{r^4}\bigg).
\end{align}
The geometric terms for the line element \eqref{g6} are
\begin{align}\nonumber
\mathcal{R}&=\frac{1}{2e^{{A_1}}}\bigg(A_0'^2-{A_1}'A_0'+2A_0''-\frac{4{A_1}'}{r}+\frac{4A_0'}{r}-\frac{4e^{{A_1}}}{r^2}
+\frac{4}{r^2}\bigg),\\\nonumber
\mathcal{R}_{00}&=\frac{1}{4e^{{A_1}-A_0}}\bigg(A_0'^2-{A_1}'A_0'+2A_0''+\frac{4A_0'}{r}\bigg),\\\nonumber
\mathcal{R}_{11}&=\frac{1}{4}\bigg({A_1}'A_0'-A_0'^2-2A_0''+\frac{4{A_1}'}{r}\bigg),
\quad
\mathcal{R}_{22}=\frac{1}{2e^{{A_1}}}\bigg({A_1}'r-A_0'r+2e^{A_1}-2\bigg).
\end{align}

\section*{Appendix B}

The anisotropic factor analogous to model 1 is
\begin{align}\nonumber
\Delta&=\bigg\{\big(2 \eta r A_0''-\eta  {A_1} ' \big(r
A_0'+4\big)+\eta r A_0'^2+4 \eta A_0'+32 \pi r e^{{A_1}}\big)
\big(2\big(8 \eta -\eta r^2 {A_1} '' \\\nonumber &+9\eta r^2
A_0''+64 \pi  r^2 e^{{A_1}}-8 \eta e^{{A_1}}\big)-\eta r^2 {A_1}
'^2+r^23 \eta  A_0 '^2-10 \eta  r {A_1} ' \big(r A_0'+4\big)
\\\nonumber &+16
\eta r A_0'\big)\bigg\}^{-1}\bigg\{3\eta r^3(6\eta
B_c+1)A_0'^4+4r\big(\eta {A_1}''\big(8 B_c\big(\eta+\big(8\pi
r^2-\eta \big)e^{{A_1}}\big)+r^2 \\\nonumber &\times (4 \eta B_c-1)
A_0''\big)+8 A_0'' \big(\eta (1+\eta  B_c )+e^{{A_1}} \big(8 \pi
r^2 (3\eta  B_c +1)-\eta (\eta B_c +1)\big)\big)\\\nonumber &+128
\pi  B_c e^{{A_1}} \big(\big(8 \pi r^2-\eta \big)e^{{A_1}}+\eta\big)
+\eta r^2 (10 \eta  B_c +9) A_0 ''^2\big)+2 r A_0'^2 \big(4 \big(2
\big(2 \eta
\\\nonumber &\times (3 \eta  B_c +1)+(2 B_c \eta +1)
\big(8 \pi r^2-\eta \big) e^{{A_1}}\big)+\eta r^2 (7 \eta B_c +3)
A_0''\big)-\eta  r^2 {A_1}''
\\\nonumber &\times (6 \eta  B_c
+1) \big)+4A_0' \big(8 \big(\eta (2 \eta  B_c +1)+e^{{A_1}} \big(4
\pi r^2 (8 \eta  B_c +1)-\eta  (2 \eta  B_c+1)\big)\big)
\\\nonumber &+ r^2 (24 \eta B_c +19)\eta A_0''-3 \eta r^2 {A_1}
''\big)+\eta r^2 {A_1} '^3 \big(88 \eta  B_c +(22 \eta B_c r+r)
A_0'+14\big)\\\nonumber &+\eta r {A_1} '^2 \big(-2 \big(8 \big( B_c
\big(\eta +24 \pi r^2\big)e^{{A_1}}-9 \eta B_c -4\big)+ (22 \eta
B_c +1) A_0 ''(r)r^2\big)+9 r^2\\\nonumber & \times A_0'^2 (2 \eta
B_c+1) +6 r (8 \eta B_c +11) A_0'\big)+2\eta r^2 (16 \eta  B_c +19)
A_0'^3-{A_1} ' \big(\eta r^3 A_0'^3
\\\nonumber &\times (34 \eta  B_c +13)+4 \big(\eta  r^2 (12 \eta  B_c +1) {A_1}
''+8 \big(\eta (2 \eta  B_c +1)-e^{{A_1}} \big(\eta  (1+2 \eta
B_c)\\\nonumber &-12 \pi r^2 (4 \eta  B_c +1)\big)\big)+\eta  r^2
(44 \eta  B_c +27) A_0''\big)+2 r A_0' \big(\eta r^2 (6 \eta B_c -1)
{A_1} ''+8\\\nonumber &\times  \big(\eta (13 \eta B_c +10)+e^{{A_1}}
\big(8 \pi r^2 (3 \eta  B_c +1)-\eta (\eta  B_c +1)\big)\big)+\eta
(22 \eta  B_c +19)
\\\nonumber &\times r^2A_0''\big)+2 \eta r^2 (92 \eta
B_c+35) A_0'^2\big)\bigg\},
\end{align}
while its value for model 2 is
\begin{align}\nonumber
\Delta&=\bigg\{2 e^{{A_1}} r^2 \eta \mathcal{R}^2+4 \eta  \big(r
{A_1}'+2 e^{{A_1}}-r A_0'-2\big) \mathcal{R}+r \big(\eta \big(r
\mathcal{R}'-2 r \alpha_6-4 \alpha_3\big) A_0'\\\nonumber &-r \eta
\alpha_3 A_0'^2+32 e^{{A_1}} \pi  r-4 r \eta \alpha_9-4 \eta
\alpha_6+\eta {A_1}' \big(r A_0' \alpha_3+4 \alpha_3+2 r \alpha_6-r
\mathcal{R}'\big)+2 r\\\nonumber &\times \eta \mathcal{R}''-2 r \eta
\alpha_3 A_0''\big)\bigg\}^{-1}\bigg\{r \big(r A_0'^2+2 A_0'-{A_1}'
\big(r A_0'+2\big)+2 r A_0''\big)\bigg\}+\bigg\{8 e^{{A_1}} r \eta
\mathcal{R}^2\\\nonumber &+\eta \big({A_1}' \big(7 r A_0'-4\big)-r
A_0'^2-20 A_0'-8 r A_0''\big) \mathcal{R}+2 \big(2 \eta \alpha_8 r-3
\eta \alpha_4 A_0' r-\eta \alpha_5 A_0'r\\\nonumber &+2 \eta
\mathcal{R}'' r+64 e^{{A_1}} \pi r-\eta (6 \alpha_1+2 \alpha_2+3 r
\alpha_4+r \alpha_5) {A_1}'-\eta \mathcal{R}' \big(r {A_1}'+r
A_0'-4\big)\\\nonumber &-6 \eta \alpha_1 A_0'-2 \eta \alpha_2
A_0'\big)\bigg\}^{-1}\bigg\{4 \big(2 e^{{A_1}} r \eta B_c
\mathcal{R}^2+\eta  B_c \big(2 {A_1}' \big(r A_0'-2\big)-r A_0'^2-8
A_0'\\\nonumber &-2 r A_0''\big) \mathcal{R}+32 e^{{A_1}} \pi r  B_c
+4 r \eta  B_c \alpha_7-4 \eta B_c \alpha_1 {A_1}'-2 r \eta
B_c\alpha_4 {A_1}'-{A_1}'-4 \eta  B_c \alpha_1\\\nonumber &\times
A_0'-2 r \eta B_c \alpha_4 A_0'-A_0'-\eta  B_c R' \big(r {A_1}'+r
A_0'-4\big)+2 r \eta  B_c \mathcal{R}''\big)\bigg\}+\bigg\{\big(8
e^{{A_1}} r \eta \mathcal{R}^2\\\nonumber &+\eta \big(-r A_0'^2-20
A_0'+{A_1}' \big(7 r A_0'-4\big)-8 r A_0''\big) \mathcal{R}+2 \big(2
\eta \alpha_8 r-3 \eta  \alpha_4 A_0' r-\eta \alpha_5 A_0'
r\\\nonumber &+2 \eta \mathcal{R}'' r+64 e^{{A_1}} \pi r-\eta  (6
\alpha_1+2\alpha_2+3 r \alpha_4+r \alpha_5) {A_1}'-6 \eta \alpha_1
A_0'-2 \eta \alpha_2 A_0'-\eta \mathcal{R}'\\\nonumber &\times
\big(r {A_1}'+r A_0'-4\big)\big)\big) \big(2 e^{{A_1}} r^2 \eta
\mathcal{R}^2+4 \eta \big(r {A_1}'+2 e^{{A_1}}-r A_0'-2\big)
\mathcal{R}-r \big(r \eta \alpha_3 A_0'^2\\\nonumber &-\eta \big(-4
\alpha_3-2 r \alpha_6+r \mathcal{R}'\big) A_0'-32 e^{{A_1}} \pi r+4
r \eta \alpha_9+4 \eta\alpha_6-\eta {A_1}' \big(r A_0' \alpha_3+2 r
\alpha_6\\\nonumber &+4 \alpha_3-r \mathcal{R}'\big)-2 r \eta
\mathcal{R}''+2 r \eta \alpha_3 A_0''\big)\big)\bigg\}^{-1}\bigg\{2
r \eta \big(\big(-r (2 \alpha_1+\mathcal{R}) A_0'^2-\big(4
\alpha_1+r4 \alpha_4\\\nonumber &-2 r \mathcal{R}'\big) A_0'-8 r
\alpha_7-8 \alpha_4+{A_1}' \big(2 r A_0' \alpha_1+4 \alpha_1+4 r
\alpha_4+\mathcal{R} \big(r A_0'+4\big)\big)-r4 \alpha_1
A_0''\\\nonumber &-2 r \mathcal{R} A_0''\big) \big(2 e^{{A_1}} r
\eta  B_c \mathcal{R}^2+\eta B_c \big(2 r A_0'^2+4 A_0'+{A_1}'
\big(r A_0'+8\big)-2 r A_0''\big) \mathcal{R}-3 {A_1}'\\\nonumber
&-32 e^{{A_1}} \pi r B_c +4 r \eta  B_c\alpha_8-4 \eta  B_c \alpha_2
{A_1}'-2 r \eta  B_c \alpha_5 {A_1}'-4 \eta  B_c \alpha_2 A_0'-2 r
\eta \alpha_5 B_c A_0'\\\nonumber &-3 A_0'+\eta B_c \mathcal{R}'
\big(r {A_1}'+r A_0'-4\big)-2 r \eta B_c \mathcal{R}''\big)+\big(2 r
\alpha_2 A_0'^2-r \mathcal{R} A_0'^2+4\alpha_2 A_0'+8
\alpha_5\\\nonumber &+4 r \alpha_5 A_0'-4 \mathcal{R} A_0'+8 r
\alpha_8+2 \mathcal{R}' \big(r {A_1}'-2 r A_0'-4\big)+ \big(
(\mathcal{R}-2 \alpha_2)rA_0'-4 (\alpha_2+r
\alpha_5)\big)\\\nonumber &\times {A_1}' -4 r \mathcal{R}''+4 r
\alpha_2 A_0''-2 r \mathcal{R} A_0''\big) \big(-2 e^{{A_1}} r \eta
B_c \mathcal{R}^2+\eta  B_c \big( A_0'^2r+8 A_0'+2 r
A_0''\\\nonumber &+{A_1}' \big(4-2 r A_0'\big)\big) \mathcal{R}-32
e^{{A_1}} \pi  r  B_c -4 r \eta  B_c \alpha_7+4 \eta  B_c \alpha_1
{A_1}'+2r \eta B_c \alpha_4 {A_1}'+{A_1}'\\\nonumber & +4 \eta B_c
\alpha_1 A_0'+2 r \eta  B_c \alpha_4 A_0'+A_0'+\eta  B_c
\mathcal{R}' \big(r {A_1}'+r A_0'-4\big)-2 r \eta  B_c
\mathcal{R}''\big)\big)\bigg\}.
\end{align}

\end{document}